\documentclass[rmp,aps,twocolumn,nofootinbib]{revtex4}

\usepackage{graphicx}
\usepackage{epsfig}

\begin{document}
\title{Physics of thin-film ferroelectric oxides}

\author{M. Dawber}
\email{matthew.dawber@physics.unige.ch} \affiliation{DPMC, University of
Geneva, 24 quai Ernest-Ansermet, CH-1211, Geneva 4, Switzerland}

\author{K.M. Rabe}
\email{rabe@physics.rutgers.edu} \affiliation{Dept of Physics and Astronomy, Rutgers University, 136 Frelinghuysen Rd, Piscataway, NJ 00854-8019, USA}

\author{J.F. Scott}
\email{jsco99@esc.cam.ac.uk} \affiliation{Dept of Earth Sciences, University of Cambridge, Downing St, Cambridge CB2 3EQ, UK}

\begin{abstract}
This review covers the important advances in recent years in the physics of thin film ferroelectric oxides, the strongest emphasis being on those aspects particular to ferroelectrics in thin film form. We introduce the current state of development in the application of ferroelectric thin films for electronic devices and discuss the physics relevant for the performance and failure of these devices.
Following this we cover the enormous progress that has been made in the first principles computational approach to understanding ferroelectrics. We then discuss in detail the important role that strain plays in determining the properties of epitaxial thin ferroelectric films. Finally, we look at the emerging possibilities for nanoscale ferroelectrics, with particular emphasis on ferroelectrics in non conventional nanoscale geometries.
\end{abstract}

\maketitle \tableofcontents

\section{INTRODUCTION}
\label{sec:intro}

The aim of this review is to provide an account of the progress in
the understanding of the physics of ferroelectric thin film oxides,
particularly the physics relevant to present and future technology
that exploits the characteristic properties of ferroelectrics. An
overview of the current state of ferroelectric devices is followed
by identification and discussion of the key physics issues that
determine device performance. Since technologically relevant films
for ferroelectric memories are typically thicker than 120 nm,
characterization and analysis of these properties can initially be
carried out at comparable length scales. However, for a deeper
understanding, as well as for the investigation of the behavior of
ultrathin films with thickness on the order of lattice constants, it
is appropriate to re-develop the analysis at the level of atomic and
electronic structure. Thus, the second half of this review is
devoted to a description of the state of the art in first principles
theoretical investigations of ferroelectric oxide thin films,
concluding with a discussion of experiment and theory of nanoscale
ferroelectric systems.

As a starting point for the discussion, it is helpful to have a
clear definition of ferroelectricity appropriate to thin films and
nanoscale systems. Here we consider a ferroelectric to be a
pyroelectric material with two or more stable states of different
nonzero polarization (unlike electrets, ferroelectrics have
polarization states that are thermodynamically stable, not
metastable). Furthermore, it must be possible to switch between the
two states by the application of a sufficiently strong electric
field, the threshold field being designated the coercive field. This
field must be less than the breakdown field of the material, or the
material is merely pyroelectric and not ferroelectric. Because of
this switchability of the spontaneous polarization, the relationship
between, the electric displacement D, and the electric field E is
hysteretic.

For thin film ferroelectrics the high fields that must be applied to
switch the polarisation state can be achieved with low voltages,
making them suitable for integrated electronics applications. The
ability to create high density arrays of capacitors based on thin
ferroelectric films has spawned an industry dedicated to the
commercialization of ferroelectric computer memories. The classic
textbooks on
ferroelectricity\cite{LinesandGlass},\cite{FatuzzoandMerz}, though
good, are now over twenty years old, and pre-date the shift in
emphasis from bulk ceramics and single crystals towards thin-film
ferroelectrics. While much of the physics required to understand
thin-film ferroelectrics can be developed from the understanding of
bulk ferroelectrics, there is also behavior specific to thin films
that cannot be readily understood in this way. This is the focus of
the present review.

One of the points that will become clear in the course of this
review is that a ferroelectric thin film cannot be considered in
isolation, but rather the measured properties reflect the entire
system of films, interfaces, electrodes and substrates. We also look
in detail at the effects of strain on ferroelectrics. All
ferroelectrics are grown on substrates which can impose considerable
strains, meaning that properties of ferroelectric thin films can
often be considerably different from those of their bulk parent
material. The electronic properties also have a characteristic
behavior in thin-film form. While bulk ferroelectric materials are
traditionally treated as good insulators, as films become thinner it
becomes more appropriate to treat them as semiconductors with a
fairly large bandgap. These observations are key to understanding
the potential and the performance of ferroelectric devices, and to
understanding why they fail when they do.

In parallel with the technological developments in the field, the
power of computational electronic structure theory has increased
dramatically, giving us new ways of understanding ferroelectricity.
Over the last fifteen years, more and more complex systems can be
simulated with more accuracy; and as the length scales of
experimental systems decrease, there is now an overlap in size
between the thinnest epitaxial films and the simulated systems. It
is therefore an appropriate and exciting time to review this work,
and to make connections between it and the problems considered by
experimentalists and engineers.

Finally we look at some issues and ideas in nano-scale
ferroelectrics, with particular emphasis on new geometries for
ferroelectric materials on the nanoscale such as ferroelectric
nanotubes and self-patterned arrays of ferroelectric nano-crystals.

We do not attempt to cover some of the issues which are of great
importance but instead refer readers to reviews by other authors.
Some of the more important applications for ferroelectrics make use
of their piezoelectric properties, for example in actuators and
microsensors; this topic has been reviewed by \textcite{Muralt00}.
Relaxor ferroelectrics in which ferroelectric ordering occurs
through the interaction of polar nanodomains induced by substitution
are also of great interest for a number of applications and have
recently been reviewed by \textcite{Samara03}.

\section{FERROELECTRIC ELECTRONIC DEVICES}

\subsection{Ferroelectric Memories}

The idea that electronic information can be stored in the electrical
polarization state of a ferroelectric material is a fairly obvious
one; however it's realization is not so straightforward. The initial
barrier to the development of ferroelectric memories was the
necessity to make them extremely thin films, because the coercive
voltage of ferroelectric materials is typically of the order of
several kV/cm, requiring sub-micron thick films to make devices that
work on the voltage scale required for computing (all Si devices
work at $\leq$ 5V). With today's deposition techniques this is no
longer a problem, and now high-density arrays of non-volatile
ferroelectric memories are commercially available. However,
reliability remains a key issue. The lack of good device models
means that design of ferroelectric memories is expensive and that it
is difficult to be able to guarantee that a device will still
operate ten years into the future. Because competing non-volatile
memory technologies exist, ferroelectric memories can succeed only
if these issues are resolved.

A ferroelectric capacitor, while capable of storing information, is
not sufficient for making a non-volatile computer memory. A
pass-gate transistor is required so that a voltage above the
coercive voltage is only applied to the capacitor when a voltage is
applied to both the word and bit line (this is how one cell is
selected from an array of memories). The current (measured through a
small load resistor in series with the capacitor) is compared to
that from a reference cell that is poled in a definite direction. If
the capacitor being read is in a different state the difference in
current will be quite large (the displacement current associated
with switching accounts for the difference). If the capacitor does
not switch because it is already in the reference state, the
difference in current between the capacitor being read and the
reference capacitor is zero.

\begin{figure}[h]
 \includegraphics[width=8.5cm]{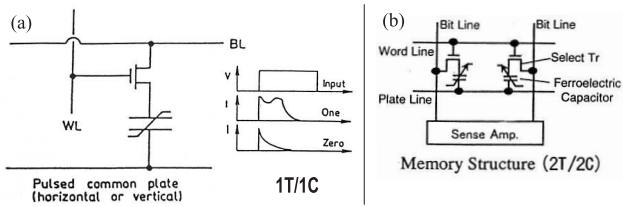}
 \caption{\textit{a)1T-1C
memory design. When a voltage is applied to both the word and bit
line, the memory cell is addressed. Shown also is the voltage
applied to the capacitor and the current output, depending on
whether a one or a zero is stored. The current for the zero state is
pure leakage current and by comparison to a reference capacitor can
be removed. (b) A 2T-2C memory cell in which the reference capacitor
is part of the memory cell}}\label{fig:1T1C2T2C}
\end{figure}

Most memories use either a 1T-1C design (1 Transistor-1 Capacitor)
or a 2T-2C (2 Transistor-2 Capacitor) design (Fig.
\ref{fig:1T1C2T2C}). The important difference is that the 1T-1C
design uses a single reference cell for the entire memory for
measuring the state of each bit, whereas in the 2T-2C there is a
reference cell per bit. A 1T-1C design is much more space-effective
than a 2T-2C design, but has some significant problems, most
significantly that the reference capacitor will fatigue much faster
than the other capacitors, and so failure of the device occurs more
quickly. In the 2T-2C design the reference capacitor in each cell
fatigues at the same rate as its corresponding storage capacitor,
leading to better device life. A problem with these designs is that
the read operation is destructive, so every time a bit is read it
needs to be written again. A ferroelectric field effect transistor,
in which a ferroelectric is used in place of the metal gate on a
field effect transistor, would both decrease the size of the memory
cell and provide a non-destructive read out; however, no commercial
product has yet been developed. Current efforts seem to run into
serious problems with data retention.

\begin{figure}[h]
\epsfig{figure=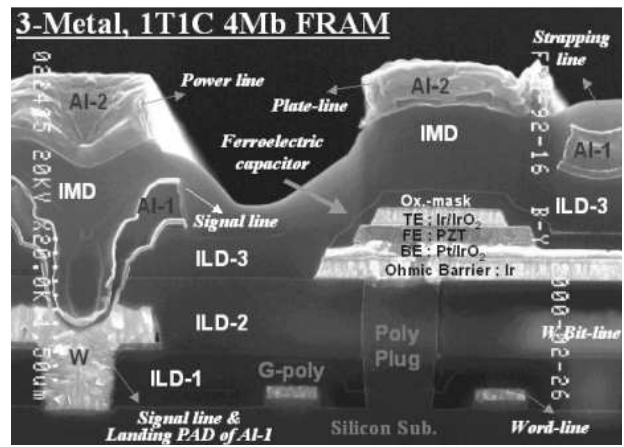,width=8.5cm}\caption{\textit{Cross-sectional
SEM image of the Samsung 4Mbit 1T-1C 3 metal
FRAM}}\label{fig:samsungcap}
\end{figure}

An example of a real commercially available memory is the Samsung
lead zirconate titanate based 4 Mbit 1T-1C ferroelectric memory. The
SEM cross-section (Fig. \ref{fig:samsungcap}) of the device gives
some indication of the complexity of design involved in a real
ferroelectric memory.

\begin{figure}[h]
\epsfig{figure=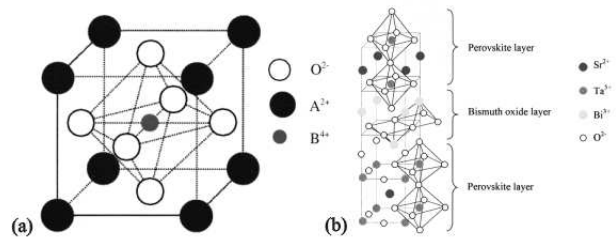,width=8.5cm}\caption{\textit{(a)ABO$_{3}$
cubic perovskite structure, (b) Strontium Bismuth Tantalate (layered
perovskite structure)}}\label{fig:perovskites}
\end{figure}

Lead zirconate titanate (PZT) has long been the leading material
considered for ferroelectric memories, though strontium bismuth
tantalate (SBT), a layered perovskite, is also a popular choice due
to its superior fatigue resistance and the fact that it is lead
free(Fig. \ref{fig:perovskites}). However it requires higher
temperature processing, which creates significant integration
problems. Some recent progress has been made in optimizing
precursors. Until recently the precursors for Sr, Bi, and Ta/Nb did
not function optimally in the same temperature range, but last year
Inorgtech developed Bi(mmp)3 -- a 2-methoxy-2-propanol propoxide
that improves reaction and lowers the processing temperature for
SBT, its traditional main disadvantage compared to PZT.  This
material also saturates the bismuth coordination number at 6.
Recently several other layered perovskites, for example bismuth
titanate, have also been considered.

As well as their applications as FRAM's,  ferroelectric materials
have potential use in DRAM's because of their high dielectric
constant in the vicinity of the ferroelectric phase transition, a
topic which has been reviewed by \textcite{Kingon2000}. Barium
strontium titanate (BST) is one of the leading materials in this
respect since by varying composition a transition temperature just
below room temperature can be achieved, leading to a high dielectric
constant over the operating temperature range.

\subsection{Future prospects for non-volatile Ferroelectric Memories}

There are two basic kinds of ferroelectric random access memories in
production today: (1) The free-standing RAMs and (2) fully embedded
devices (usually a CPU, which may be a CMOS EEPROM (complementary
metal oxide semiconductor electrically erasable programmable
read-only memory, the current generation widely used non-volatile
memory technology), plus an FRAM (ferroelectric random access
memory), and an 8-bit microprocessor). The former have reached 4
Mbit both at Samsung (using PZT) and Matsushita (using SBT). The
Samsung device is not yet, as far as the authors know, in commercial
production for real products, but the NEC FRAM is going into
full-scale production this year in Toyama (near Kanazawa). Fujitsu
clearly leads in the actual commercial use of its embedded FRAMs.
The Fujitsu embedded FRAM is that used in the SONY Playstation 2. It
consists of 64 Mbit of EEPROM plus 8 kbit of RAM, 128 kbit ROM, and
a 32-kbit FRAM plus security circuit.  The device is manufactured
with a 0.5-micron CMOS process.  The capacitor is 1.6 x 1.9 microns
and the cell size is either 27.3 square microns for the 2T-2C design
or 12.5 square microns for the 1T-1C.

The leading competing technologies in the long term for non volatile
computer memories are FRAM (ferroelectric random access memories)
and MRAM (magnetic random access memories).  These are supposed to
replace EEPROMs (electrically erasable programable read-only
memories) and "Flash" memories in devices such as digital cameras.
Flash, though proving highly commercially successful at the moment,
is not a long term technology, suffering from poor long term
endurance and scalability.   It will be difficult for Flash to
operate as the silicon logic levels decrease from 5V at present to
3.3V, 1.1V, and 0.5V in the near future. The main problem for
ferroelectrics is the destructive read operation, which means that
each read operation must be accompanied by a write operation leading
to faster degradation of the device. The operation principle of
MRAMs is that the tunneling current through a thin layer sandwiched
between two ferromagnetic layers is different depending on whether
the ferromagnetic layers have their magnetization parallel or
anti-parallel to each other. The information stored in MRAMs can
thus be read non-destructively, but their write operation requires
high power which could be extremely undesirable in high density
applications. We present a summary of the current state of
development in terms of design rule and speed of the two
technologies in the following table.

\begin{tabular}{|c|c|c|c|}
  \hline
  Company & Design Rule & Speed \\
  & (Feature Size) & (Access Time)\\
  \hline
  \textbf{MRAMs} & &\\
  NEC/Toshiba & 1Mb &  \\
  IBM & 16 Mb& \\
  Matsushita & 4Mb & \\
  Sony & 8kb 0.18 microns & \\
  Cypress & 256 kb & 70 ns \\
  State of the art & 16 Mb 0.09 microns & 25 ns \\
  \textbf{FRAMs} & & \\
  Fujitsu & 32 kb & 100 ns \\
  Samsung & 32 Mb 0.18 microns & 60 ns \\
  Matsushita & 4 Mb & 60 ns \\
  Laboratory & & 800 ps \\
  \hline

\end{tabular}

Some clarification of the numbers in this table is required. The
size of the Fujitsu FRAM memory may seem small but it is for an
actual commercial device in large scale use (in every Playstation
2), whereas the others are figures from internal sampling of
unreleased devices that have not been commercialized. No MRAMS exist
in any commercial device, giving FRAMs a substantial edge in this
regard. The most recent commercial FRAM product actually shipped is
a large-cell-area six-transistor four-capacitor (6T-4C) memory for
smart credit cards and radio frequency identification tags (RF-ID)
and features non-destructive read out \cite{Masui03}. A total of 200
million ferroelectric memories of all types have been sold
industry-wide. The Sony MRAM, though small, has sub-micron design
rules, meaning that in principle a working device could be scaled up
to Mb size.

Partly in recognition of the fact that are distinct advantages for
both ferroelectrics and ferromagnets, there has been a recent flurry
of activity in the field of multiferroics, i.e. materials that
display both ferroelectric and magnetic ordering, the hope being
that one could develop a material with a strong enough coupling
between the two kinds of ordering to realize a device that can be
written electrically and read magnetically. In general multi-ferroic
materials are somewhat rare, and certainly the conventional
ferroelectrics like PbTiO$_{3}$ and BaTiO$_{3}$ will not display any
magnetic behaviour as the Ti-O hybridisation required to stabilize
the ferroelectricity in these compounds will be inhibited by the
partially filled d-orbitals that would be required for magnetism
\cite{Hill00}. However there are other mechanisms for
ferroelectricity and in materials where ferroelecticity and
magnetism co-exist there can be coupling between the two. For
example, in BaMnF$_{4}$ the ferroelectricity is actually responsible
for changing the antiferromagnetic ordering to a weak canted
ferromagnetism \cite{Fox80}  In addition, the large magnetoelectric
coupling in these materials causes large dielectric anomalies at the
Neel Temperature and  at the in-plane spin ordering temperature
\cite{Scott77},\cite{Scott79}. More recent theoretical and
experimental efforts have focused on BiMnO$_{3}$,BiFeO$_{3}$
\cite{Seshadri01},\cite{Moreirads02},\cite{Wang03} and YMnO$_{3}$
\cite{VanAken2004},\cite{Fiebig02}.

\subsection{Ferroelectric FET's}

\begin{figure}[h]
 \includegraphics[width=5cm]{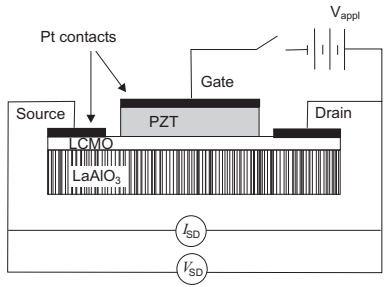}
 \caption{\textit{\textcite{Mathews97} Schematic diagram of an all-perovskite ferroelectric FET and measurement circuit}}\label{fig:FEFET}
\end{figure}

It has been known for some time that replacing the metal gate in a
field effect transistor (FET) by a ferroelectric could produce a
device with non-destructive read-out (NDRO), in which the
polarization of the gate (+ for "1" and - for "0") could be sensed
simply by monitoring the source-drain current magnitude. Thus such a
device requires no reset operation after each READ and will
experience very little fatigue in a normal frequent-read,
occasional-write usage.  The early ferroelectric FETs utilized gates
of lithium niobate (Rice Univ.) \cite{Rabson95} or BaMgF$_{4}$
(Westinghouse)
\cite{Sinharoy91},\cite{Sinharoy92},\cite{Sinharoy93}. An example of
a ferroelectric FET device as fabricated by \textcite{Mathews97} is
shown in Fig. \ref{fig:FEFET}

The optimum parameters for such a ferroelectric gate material are
extremely different from those for pass-gate switched capacitor
arrays; in particular, the latter require a remanent polarization
ca. 10 $\mu$C/cm$^{2}$, whereas the ferroelectric-gated FETs can
function well with 50x less (0.2 $\mu$C/cm$^{2}$).  However, the
switched capacitor array (FRAM) is very tolerant of surface traps in
the ferroelectric (which may be ca. $10^{20}$ cm$^{-3}$ in the
interface region near the electrode), since the ferroelectric makes
contact only with a metal (or metal-oxide) electrode. By comparison,
the ferroelectric gate in an FET contacts the Si substrate directly
(MOSFET channel - metal-oxide-semiconductor field-effect-transistor
channel).  Thus it must be buffered from the Si to prevent charge
injection.  Unfortunately, if a thin buffer layer of a
low-dielectric material such as SiO$_{2}$ is used, most of the
applied voltage will drop across the buffer layer and not the
ferroelectric gate, making it impossible to switch the gate.  As a
result, much of the ferroelectric FET research has employed buffer
layers with relatively high dielectric constants, or else rather
thick buffer layers, for example, the first BaMnF$_{4}$ FET made at
Symetrix \cite{ScottFerroRevFET} used  a buffer layer of ca. 40 nm
of SiO$_{2}$. Subsequent studies often used PZT,\cite{Kalkur94}
although the large remanent polarization in this case (ca. 40
$\mu$C/cm$^{2}$) is actually undesirable for a ferroelectric FET
gate.

As pointed out by \textcite{Yoon01}, the depolarization field in a
ferroelectric gate is inevitably generated when the gate is
grounded, and this makes it very difficult to obtain $>10$ year data
retention in an FE-FET.  Their solution is to utilize a 1T-2C
capacitor geometry in which this depolarization field is suppressed
by poling the two capacitors in opposite directions. With this
scheme  Ishiwara and his colleagues achieved an on/off source-drain
current ratio of $>$1000 for a 150 nm thick SBT film in a 5 x 50
micron MOSFET channel, with Pt electrodes on the SBT capacitor.

Note that the direct contact of the ferroelectric onto Si produces a
semiconductor junction that is quite different from the
metal-dielectric interface discussed above.  The Schottky barrier
heights for this case have been calculated by \textcite{Peacock02}.
The electron screening length in the Si will be much greater than in
the case of metal electrodes; in particular this will increase the
minimum ferroelectric film thickness required to stabilize the
device against depolarization instabilities.  Although this point
was first emphasized by Batra and Silverman (1973), it has been
neglected in the more recent context of ferroelectric FETs.  In our
opinion, this depolarization instability for thin ferroelectric
gates on FETs is a significant source of the observed retention
failure in the devices but has not yet been explicitly modeled.  If
we are correct, the retention problem in ferroelectric FETs could be
minimized by making the ferroelectric gates thicker and the Si
contacts more conducting (e.g., p+ rather than p).  See Scott
(Microelectron. Eng. 2005, in press) for a full discussion of
all-perovskite FETs.

Table I lists a number of the most promising gate materials under
recent study, together with the buffer layers employed in each case.
Studies of the I(V) characteristics of such ferroelectric FETs have
been given by \textcite{Macleod01} and a disturb-free programming
scheme described by \textcite{Ullmann01}. \scriptsize

\begin{tabular}{|c|c|c|}
  \hline
  FET Gate & Buffer Layer & Reference \\
  \hline
  LiNbO$_{3}$  & none & \textcite{Rabson95} \\
  SBT & SrTa$_{2}$O$_{6}$ & \textcite{Ishiwara93}, \\
  && \textcite{Ishiwara97},\\
  && \textcite{Ishiwara01}\\
  SBT & CeO$_{2}$ & \textcite{Shimada01}, \\
  && \textcite{Haneder01}\\
  SBT & SiO$_{2}$ & \textcite{Okuyuma01} \\
  SBT & ZrO$_{2}$ & \textcite{Park01} \\
  SBT & Al$_{2}$O$_{3}$ & \textcite{Shin01} \\
  SBT & Si$_{3}$N$_{4}$ & \textcite{Han01} \\
  SBT & Si$_{3}$N$_{4}/$SiO$_{2}$ & \textcite{Sugiyama01} \\
  SBT & poly-Si + Y$_{2}$O$_{3}$ & \textcite{Kalkur01} \\
  Pb$_{5}$Ge$_{3}$O$_{11}$ & none & \textcite{Li01} \\
  YMnO$_{3}$ & Y$_{2}$O$_{3}$ & \textcite{Cheon01}, \\
  && \textcite{Choi01} \\
  Sr$_{2}$(Ta$_{2x}$Nb$_{2-2x}$)O$_{7}$ & none & \textcite{Kato01} \\
  PZT & CeO$_{2}$ & \textcite{Xiaohua01} \\
  BST(strained) & YSZ(zirconia) & \textcite{Jun01} \\
  BaMnF$_{4}$ & SiO$_{2}$ & \textcite{ScottFerroRevFET},\\
  && \textcite{Kalkur94} \\
  \hline

\end{tabular}
\normalsize Beyond its use in modulating the current in a
semiconductor channel the ferroelectric field effect can also be
used to modify the properties of more exotic correlated oxide
systems.\cite{Ahn03}

\subsection{Replacement of gate oxides in DRAMs}

At present there are three basic approaches to solving the problem
of SiO$_{2}$ gate oxide replacement for DRAMs:  The first is to use
a high-dielectric ("high-k") material such as SrTiO$_{3}$  (k = 300
is the dielectric constant;  $\epsilon = k -1$ is the permittivity;
for k $>>$ 1 the terms are nearly interchangeable) deposited by some
form of epitaxial growth. This is the technique employed at
Motorola, but the view elsewhere is that it is too expensive to
become industry process-worthy.  The second approach is to use a
material of moderate k (of order 20), with HfO$_{2}$ favored but
ZrO$_{2}$ also a choice.  Hafnium oxide is satisfactory in most
respects but has the surprising disadvantage that it often degrades
the n-channel mobility catastrophically (by as much as x10,000).
Recently ST Microelectronics decided to use SrTiO$_{3}$ but with
MOCVD deposition from Aixtron, thus combining high dielectric
constant and cheaper processing.

The specific high-k integration problems are four: (1) depletion
effects in the polysilicon gate; (2) interface states; (3) strain
effects; and (4) etching difficulties (HfO$_{2}$ is hard to
wet-etch). The use of a poly-Si gate instead of a metal gate
produces grain boundary stress in the poly, with resultant poor
conductivity. This mobility degradation is only partly understood.
The general view is that a stable amorphous HfO$_{2}$ would be a
good strain-free solution. Note that HfO$_{2}$ normally crystallizes
into two or three phases, one of which is monoclinic
\cite{Morrison03}. Hurley in Cork has been experimenting with a
liquid injection system that resembles Isobe's earlier SONY device
for deposition of viscous precursors with flash evaporation at the
target.

\section{FERROELECTRIC THIN FILM DEVICE PHYSICS}
\label{sec:ferromems}

We now turn to some of the physics questions which are relevant to
ferroelectric thin film capacitors.

\subsection{Switching}

\begin{figure}[h]
 \includegraphics[width=8.5cm]{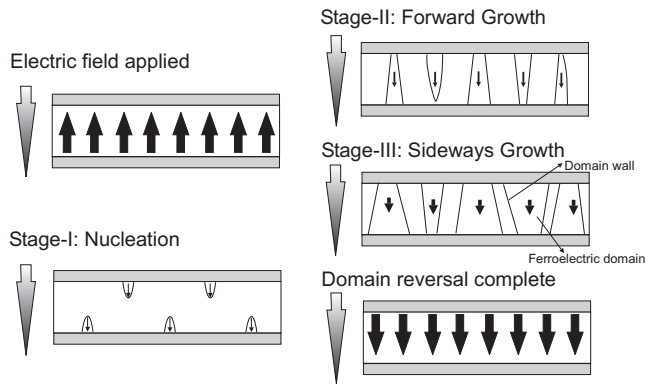}
 \caption{\textit{The
three phases of domain reversal, I. Nucleation (fast) II. Forward
growth (fast) III. Sideways growth
(slow)}}\label{fig:nucleationgrowth}
\end{figure}

In the ferroelectric phase ferroelectric materials form domains
where the polarisation is all aligned in the same direction, in an
effort to minimise energy. When a field is applied the ferroelectric
switches by the nucleation of domains and the movement of domain
walls and not by the spontaneous reorientation of all of the
polarisation in a domain at once. In contrast to ferromagnets where
switching usually occurs by the sideways movement of existing domain
walls, ferroelectrics typically switch by the generation of many new
reverse domains at particular nucleation sites, which are not
random; i.e.,  nucleation is inhomogeneous. The initial stage is
nucleation of opposite domains at the electrode, followed by fast
forward propagation of domains across the film, and then slower
widening of the domains(Fig. \ref{fig:nucleationgrowth}). In
perovskite oxides the final stage of the switching is usually much
slower than the other two stages, as first established by
\textcite{Merz54}.  In other materials nucleation can be the slowest
(rate-limiting) step.

\subsubsection{Ishibashi-Orihara Model}

For many years the standard model to describe this process has been
the Ishibashi-Orihara model\cite{Orihara94} based on
Kolomogorov-Avrami growth kinetics. In this model one considers a
nucleus formed at time t' and then a domain propagating outwards
from it with velocity V. In the Ishibashi-Orihara model the velocity
is assumed to be dependent only on the electric field E, and not on
the domain radius r(t).  This makes the problem analytically
tractable but gives rise to unphysical fitting parameters, such as
fractional dimensionality D.  The fractional D is not related to
fractals.   It is an artifact that arises because domain wall
velocity V is actually proportional to 1/r(t) for each domain and is
not a constant at constant E. The volume of a domain at time t is
given by
\begin{equation}
C(t,t')=C_{D}[\int_{t'}^{t}V(t'')dt'']^{D}
\end{equation}
where D is the dimensionality of the growth and $C_{D}$ is a
constant which depends on the dimensionality. It is also assumed
within this model that the nucleation is deterministic and occurs at
pre-defined places; i.e., this is a model of inhomogeneous
nucleation. This is an important point since some researchers still
use homogeneous nucleation models. These are completely
inappropriate for ferroelectrics (where the nucleation is
inhomogeneous, as is demonstrated by imaging
experiments\cite{Shur91,Shur96,Ganpule01}).

The result of the model is that the fraction of switched charge as a
function of field and frequency may be expressed as
\begin{equation}
Q(E,f)=1-\exp(-f^{-D}\Phi(E))
\end{equation}
where $\Phi(E)$ depends on the waveform used for switching. After
some consideration and the substitution $\Phi(E)=E^{k}$ one obtains
a useful relationship for the dependence of the field on frequency:

\begin{equation}
E_{c}=f^{\frac{D}{k}}
\end{equation}

This relationship has been used to fit data fairly well in TGS
\cite{Hashimoto94}, PZT and SBT \cite{Scott96}. More recently
however \textcite{Tsurumi01} and {\textcite{Jung02A}} have found
that over larger frequency ranges the data on several materials is
better fitted by the nucleation limited model of \textcite{Du98}.
\textcite{Tagantsev02} have also found that over large time ranges
the Ishibashi-Orihara model is not a good description of switching
current data and that a nucleation limited model is more
appropriate. It is quite possible of course that domain wall-limited
switching (Ishibashi) is operative in one regime of time and field
but that in another regime the switching is nucleation-limited.

\subsubsection{Nucleation models}

Some of the earliest detailed studies of switching in ferroelectrics
developed nucleation limited switching models where the shape of the
nucleus of the reversed domain was very important. In the work of
\textcite{Merz54} and \textcite{Wieder56},\textcite{Wieder57} a
nucleation-limited model was used in which when dagger-shaped nuclei
were assumed, the correct dependence of the switching current on
electric field could be derived. This approach leads to the concept
of an activation field for nucleation (somewhat different from the
coercive field). Activation fields in  thin film PZT capacitors were
measured by \textcite{Scott88b}; very recently \textcite{Jung04}
have studied the effects of micro-geometry on the the activation
field in PZT capacitors.

The switching model of \textcite{Tagantsev02} is a different
approach in which a number of non-interacting elementary switching
regions are considered.  These switch according to a broad
distribution of waiting times.

\subsubsection{The scaling of coercive field with thickness}

\begin{figure}[h]
 \includegraphics[width=8.5cm]{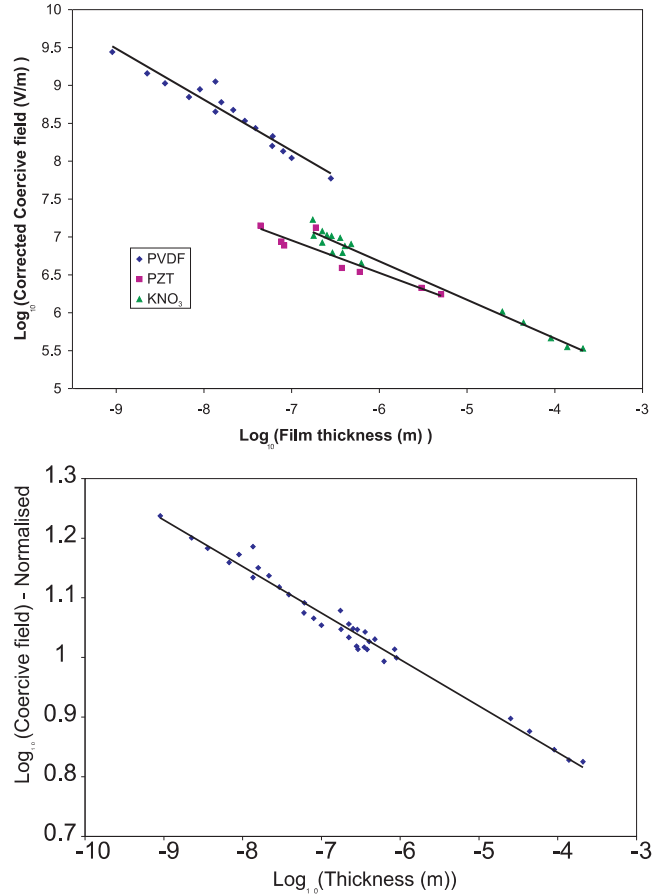}
 \caption{\textit{The scaling of coercive field with thickness in ferroelectrics; from mm to nm scale, from \textcite{Dawber03JPC}}}\label{fig:coercivescaling}
\end{figure}

For the last forty years the semi-empirical scaling
law,\cite{Janovec61,Kay62} $E_c(d) \propto d^{-{2/3}}$, has been
used successfully to describe the thickness-dependence of the
coercive field in ferroelectric films ranging from 100 microns to
200 nanometers.\cite{Scott00} In the ultrathin PVDF films of
\textcite{Bune98} a deviation from this relationship was seen for
the thinnest films\cite{Ducharme00}. Although they attribute this to
a new kind of switching taking place (simultaneous reversal of
polarisation, as opposed to nucleation and growth of domains),
\textcite{Dawber03JPC} have shown, to the contrary, that if the
effects of a finite depolarization field due to incomplete screening
in the electrode are taken into account, then the scaling law holds
over six decades of thickness and the coercive field does not
deviate from the value predicted by the scaling law (Fig.
\ref{fig:coercivescaling}). Recently \textcite{Pertsev03} measured
coercive fields in very thin PZT films.  Although they have used a
different model to explain their data, it can be seen that in fact
the scaling law describes the data very well.

\subsubsection{Mobility of 90$^{o}$ domain walls}

The mobility of domain walls, especially 90-degree walls, depends
upon their width.  In this respect the question has been
controversial, with some authors claiming very wide widths (hundreds
of angstroms) and immobile walls. Some recent papers show
experimentally that 90-degree domain walls in perovskite
ferroelectrics are extremely narrow
\cite{Foeth99},\cite{Tsai92},\cite{Stemmer95},\cite{Floquet97}. In
PbTiO$_{3}$ they are $1.0 \pm 0.3$ nm wide.  This connects the
general question of how wide they are and whether they are immobile.
The review by \textcite{Floquet99} is quite good. They make the
point that in ceramics these 90-degree walls are 14.0 nm wide (an
order of magnitude wider than in single crystals).  This could be
why theory and experiment disagree, i.e. that something special in
the ceramics makes them 10-15 times wider (and less mobile?). The
latter point is demonstrated clearly in experiments on KNbO$_{3}$,
together with a theoretical model that explains geometrical pinning
in polyaxial ferroelectrics in terms of electrostatic forces. In
this respect the first principles study of \textcite{Meyer2002} is
extremely interesting. Not only do they show that 90 degree domain
walls in PbTiO$_{3}$ are narrow and form much more easily than 180
degree domain walls, but that they should be much more mobile as
well, the barrier for motion being so low they predict thermal
fluctuation of about 12 unit cells at room temperature, which could
perhaps explain why they appear to be wide.

Some experimental studies using atomic force microscopy (AFM) have
attempted to answer the question of whether 90-degree domain walls
were mobile or not. In certain circumstances they were immobile
\cite{Ganpule00a},\cite{Ganpule00b}, but in another study
\cite{Nagarajan03} the motion of 90-degree domain walls under an
applied field was directly observed. It seems that in principle 90
degree domain walls can move, but this depends quite strongly on the
sample conditions.

\subsubsection{Imaging of domain wall motion}

The direct imaging of ferroelectric domain walls is an excellent
method for understanding domain wall motion and switching. At first
this was carried out in materials where the domains were optically
distinct such as lead germanate\cite{Shur90}, but more recently
atomic force microscopy (AFM) has become a powerful tool for
observing domain wall motion. The polarisation at a point can be
obtained from the piezoresponse detected by the tip, and the tip
itself can be used to apply a field to the ferroelectric sample and
initiate switching. It is thus possible to begin switching events
and watch their evolution over time. AFM domain writing of
ferroelectric domains can also be used to write extremely small
domain structures in high density arrays\cite{Paruch01} or other
device-like structures, such as surface acoustic wave
devices\cite{SarinKumar04}.

\begin{figure}[h]
 \includegraphics[width=8.5cm]{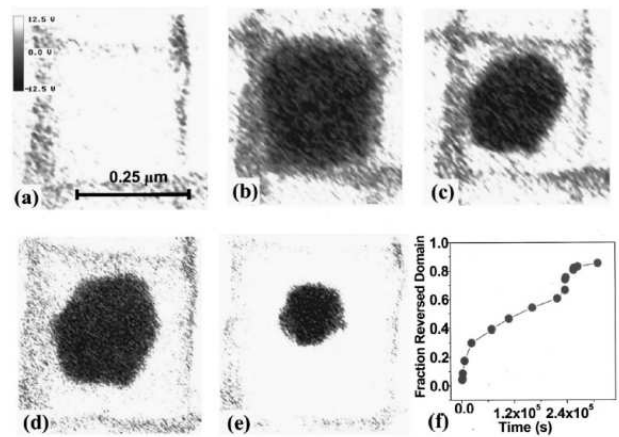}
 \caption{\textit{\textcite{Ganpule01}} Piezoresponse scans of a single cell in PbZr$_{0.2}$Ti$_{0.8}$O$_{3}$, (b)-(d) illustrate the the spontaneous reversal of polarization within this region after wait times of (b) 1.01 x 10$^3$, (c) 1.08 x 10$^5$, (d) 1.61 x 10$^5$, and (d) 2.55 x 10$^5$ s. Faceting can be seen in (c),(d),and (e). (f) Transformation-time curve for the data in (b)-(e)}\label{fig:ganpule}
\end{figure}

The backswitching studies of \textcite{Ganpule01} show two very
interesting effects (Fig. \ref{fig:ganpule}). The first is the
finding that reverse domains nucleate preferentially at antiphase
boundaries. This was studied in more detail subsequently by
\textcite{Roelofs02} who invoked a depolarisation field mediated
mechanism to explain the result. Another explanation might be that
the strain is relaxed at these antiphase boundaries,  resulting in
favorable conditions for nucleation. Secondly,  the influence of
curvature on the domain wall relaxation is accounted for within the
Kolomogorov-Avrami framework. The velocity of the domains is
dependent on curvature; and as the relaxation proceeds, the velocity
decreases and the domain walls become increasingly faceted. In Fig.
\ref{fig:ganpule} the white polarisation state is stable, whereas
the black is not. The sample is poled into the black polarisation
state and then allowed to relax back.

\begin{figure}[h]
 \includegraphics[width=8.5cm]{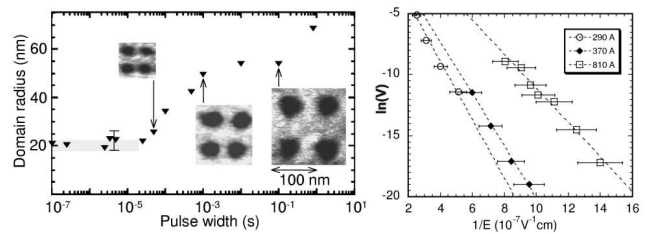}
 \caption{\textit{\textcite{Tybell02} (a) Domain size increases logarithmically with pulse widths longer than 20 $\mu$s and saturates for shorter times as indicated by the shaded area. (b) Domain wall speed as function of the inverse applied electric field for 290, 370,  and 810 ${\AA}$ thick samples. The data fit well  the characteristic velocity field relationship of a creep process. }}\label{fig:creepfig}
\end{figure}

A different kind of study was undertaken by \textcite{Tybell02}, in
which they applied a voltage pulse to switch a region of the
ferroelectric using an AFM tip and watched how the reversed domain
grew as a function of pulsewidth and amplitude. They were able to
show that the process was well described by a creep mechanism,
thought to arise due to random pinning of domain walls in a
disordered system (Fig. \ref{fig:creepfig}). Though the exact origin
of the disorder was not clear, it is suggested that it is connected
to oxygen vacancies, which also play a role in pinning the domain
walls during a fatigue process due to their ordering
\cite{Park1998},\cite{vacancyordering}.

An unsolved puzzle is the direct observation via AFM  of domain
walls penetrating grain boundaries\cite{Gruverman97}. This is
contrary to some expectations and always occurs at non-normal
incidence, i.e. at a small angle to the grain boundary.

\begin{figure}[h]
 \includegraphics[width=8.5cm]{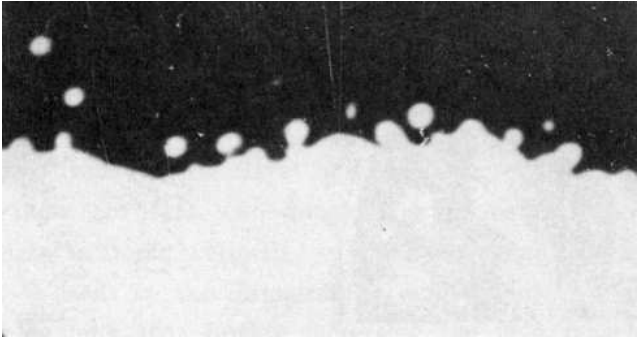}
 \caption{\textit{\textcite{Shur90} Nucleation of nanodomains in front of domain wall in lead
germanate at high electric field. Black and white represent the two
directions of polarisation.}}\label{fig:coherentnucleation}
\end{figure}

One of the interesting things to come out of the work in lead
germanate (where ferroelectric domains are optically distinct due to
electrogyration) by \textcite{Shur90} is that at high applied
electric fields (15kV cm$^{-1}$) tiny domains are  nucleated in
front of the moving domain wall (Fig. \ref{fig:coherentnucleation}).
A very similar effect is seen in ferromagnets as observed by
\textcite{Randoshkin95} in a single crystal iron garnet film.

However in ferromagnets the effect is modelled by a spin wave
mechanism.\cite{Khodenkov75} This mechanism is based on the
gyrotropic model of domain wall motion in uniaxial
materials\cite{Walker}. When a strong magnetic driving field
(exceeding the Walker Threshold) acts upon a domain wall, the
magnetization vectors in the domain wall begin to precess with a
frequency $\gamma H$, where $\gamma$ is the effective gyromagnetic
ratio. By relating the precession frequency in the domain wall with
the spin wave frequency in the domain, good predictions can be made
for the threshold fields at which the effect occurs. We note that
the domains nucleated in front of the wall may be considered as
vortex-like skyrmions. The similarity between these effects is thus
quite surprising and suggests that perhaps there is more in common
between ferroelectric domain wall motion and ferromagent domain wall
motion than is usually considered. However, whereas
\textcite{Democritov88} have shown that magnetic domain walls can be
driven supersonically (resulting in a phase-matched Cerenkov-like
bow wave of acoustic phonon emission), there is no direct evidence
of supersonic ferroelectric domains. Processes such as the
nano-domain nucleation described above seem to occur instead when
the phase velocity of the domain wall motion approaches the speed of
sound.  Of course the macroscopic electrical response to switching
can arrive at a time t $<$ v/d where v is the sound velocity and d,
the film thickness, simply from domain nucleation within the
interior of the film between cathode and anode.

\subsection{Electrical Characterization}

\subsubsection{Standard Measurement techniques}

Several kinds of electrical measurements are made on ferroelectric
capacitors. We briefly introduce them here before proceeding to the
following chapters where we discuss in detail the experimental
results obtained by using these techniques.

\paragraph{Hysteresis}

\begin{figure}[h]
 \includegraphics[width=8.5cm]{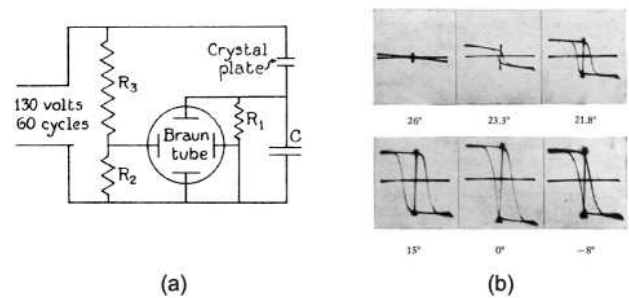}
 \caption{\textit{\textcite{SawyerTower}(a)The original Sawyer-Tower circuit, (b) Hysteresis in Rochelle salt measured using this
 circuit by Sawyer and Tower at various temperatures}}\label{fig:sawyertower}
\end{figure}

One of the key measurements is naturally the measurement of the
ferroelectric hysteresis loop. There are two measurement schemes
commonly used. Traditionally a capacitance bridge as first described
by Sawyer and Tower\cite{SawyerTower} was used (Fig.
\ref{fig:sawyertower}). Although this is no longer the standard way
of measuring hysteresis the circuit is still useful (and very simple
and cheap) and we have made several units which are now in use in
the teaching labs in Cambridge for a demonstration in which students
are able to make and test their own ferroelectric KNO$_{3}$
capacitor\cite{AJPPaper}.

This method is not very suitable in practice for many reasons, for
example the need to compensate for dielectric loss and the fact that
the film is being continuously cycled. Most testing of ferroelectric
capacitors is now carried out using commercial apparatus from one of
two companies, Radiant Technologies (http://www.ferrodevices.com/)
and AixAcct (http://www.aixacct.com/). Both companies' testers can
carry out a number of tests and measurements, and both machines use
charge or current integration techniques for measuring hysteresis.
Both machines also offer automated measurement of characteristics
such as fatigue and retention.

In measuring P(E) hysteresis loops several kinds of artifacts can
arise. Some of these are entirely instrumental, and some arise from
the effects of conductive (leaky) specimens.

Hysteresis circuits do not measure polarization P directly. Rather,
they measure switched charge Q. For an ideal ferroelectric insulator

\begin{equation}
Q=2P_{r}A
\end{equation}

where $P_{r}$ is the remanent polarization and A is electrode area
is a parallel plate capacitor. For a somewhat conductive sample

\begin{equation}
Q=2P_{r}A+\sigma E_{a}t \label{eq:hyscond}
\end{equation}

where $\sigma$ is the electrical conductivity; $E_{a}$, is the
applied field and t, the measuring time. Thus Q in a pulsed
measuring system depends on the pulse width.

\begin{figure}[h]
 \includegraphics[width=8.5cm]{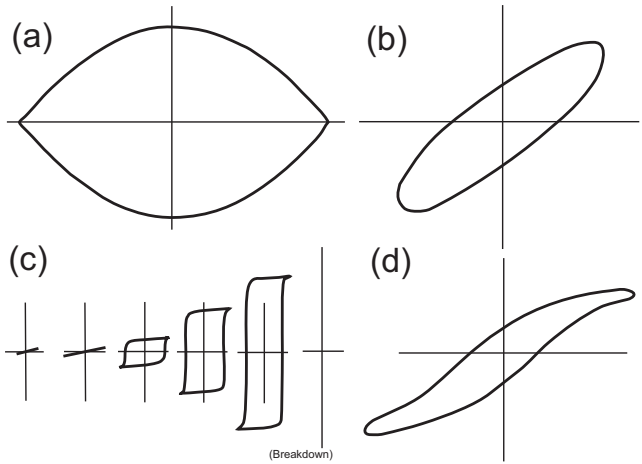}
 \caption{\textit{Common hysteresis artifacts: (a) Dead short, (b) linear lossy dielectric, (c) saturated amplifier, (d) non-linear lossy dielectric.  }}\label{fig:hysart}
\end{figure}

The four basic types of apparent hysteresis curves that are
artifacts are shown in Fig.\ref{fig:hysart}.

Fig\ref{fig:hysart} a is a dead short in a Sawyer Tower circuit or
modern variant and is discussed in the instruction documentation for
both the AixAcct\footnote{TF Analyzer 2000 FE-module instruction
manual} and
Radiant\footnote{http://www.ferroelectrictesters.com/html/specs.html$\#$tut}
testers.

Fig \ref{fig:hysart} b shows a linear lossy dielectric. The points
where the loop crosses $V_{a}=0$ are often misinterpreted as $P_{r}$
values. Actually this curve is a kind of Lissajous figure. It can be
rotated  out of the page to yield a straight line (linear dielectric
response). Such a rotation can be done electrically and give a
``compensated'' curve. Here ``compensation'' means to compensate the
phase shift caused by dielectric loss.

Fig \ref{fig:hysart} c is more subtle. Here are two seemingly
perfect square hysteresis loops, obtained on the same or nominally
equivalent specimens at different maximum fields. The smaller loop
was run at an applied voltage of $V_{a}=10 V$, and yields $P_{r}=6
\mu C cm^{-2}$ and the larger at $V_{a}=50 V$ and yields $P_{r}=100
\mu C cm^{-2}$. Note that both curves are fully saturated (flat
tops). This is impossible, if the dipoles of the ferroelectric are
saturated at $P_{r}=6 \mu C cm^{-2}$  then there are no additional
dipoles to produce $P_{r}=100 \mu C cm^{-2}$ in the larger loop at
high voltage. What actually occurs in the illustration is saturation
of the amplifier in the measuring system, not saturation of the
polarization in the ferroelectric. The figure is taken from
\textcite{JaffeJaffeCookbook} where this effect is discussed (p.
39). It will be a serious problem if conductivity is large in Eq.
\ref{eq:hyscond}. ``Large'' in this sense is $\sigma
> 10^{-6} (\Omega cm)^{-1}$ and ``small'' is $\sigma < 10^{-7}
(\Omega cm)^{-1}$. This is probably the source of $P_{r}>150 \mu C
cm^{-2}$ reports in BiFeO$_{3}$ where $\sigma$ can exceed $10^{-4}
(\Omega cm)^{-1}$.

Finally Fig \ref{fig:hysart} is a \textit{nonlinear} lossy
dielectric. If it is phase compensated it still resembles real
hysteresis. One can verify whether it is real or an artifact only by
varying the measuring frequency. Artifacts due to dielectric loss
are apt to be highly frequency dependent. Figs \ref{fig:hysart} b
and d are discussed in \textcite{LinesandGlass} (p.104).

No data resembling Figs \ref{fig:hysart} a-d should be published as
ferroelectric hysteresis.

\paragraph{Current measurements}

Another measurement of importance which is carried out in an
automated way by these machines is the measurement of the leakage
current. This is normally discussed in terms of a current-voltage
(I-V) curve, where the current is measured at a specified voltage.
It is important however that sufficient time is allowed for each
measurement step that the current is in fact true steady state
leakage current and not relaxation current, and for this reason
current-time (I-t) measurements can also be important\cite{Dietz95}.
Relaxation times in ferroelectric oxides such as barium titanate are
typically 1000 s at room temperature.

\paragraph{Dielectric permittivity}

An impedance analyser measures the real and the imaginary parts of
the impedance by use of a small-amplitude AC signal which is applied
to the sample. The actual measurement is then made by balancing the
impedance of the sample with a set of references inside the
impedance analyser. From this the capacitance and loss can be
calculated (all this is done automatically by the machine). It is
possible at the same time to apply a DC bias to the sample, so the
signal is now a small AC ripple superimposed on a DC voltage.
Ferroelectric samples display a characteristic ``butterfly loop'' in
their capacitance voltage relationships, because the capacitance is
different for increasing and decreasing voltage. The measurement is
not exactly equivalent to the hysteresis measurement. In a
capacitance voltage measurement a static bias is applied and the
capacitance measured at that bias, whereas in a hysteresis
measurement the voltage is being varied in a continuous fashion.
Therefore it is not strictly true that C(V) = (d/dV)P(V) [area taken
as unity] as claimed in some texts, since the frequency is not the
same in measurements of C and P. Impedance spectroscopy (where the
frequency of the AC signal is varied) is also a powerful tool for
analysis of films, especially as it can give information on the
timescales at which processes operate. Interpretation of these
results must be undertaken carefully, as artifacts can arise in many
circumstances and even when this is not the case many elements of
the system (e.g., electrodes, grain boundaries, leads etc.) can
contribute to the impedance in complicated ways.

\subsubsection{Interpretation of dielectric permittivity data}

\paragraph{Depletion charge vs intrinsic response}

\begin{figure}[h]
\epsfig{figure=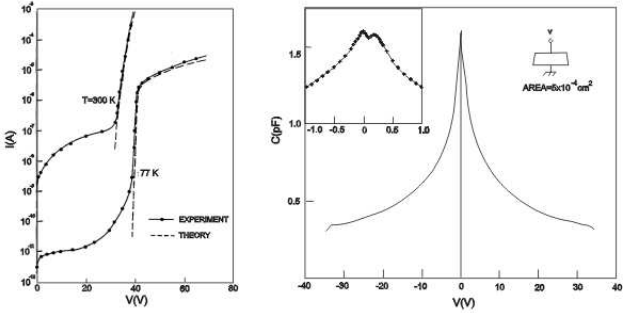,width=8.5cm}\caption{\textit{Current
voltage and capacitance voltage relationship of Pt-Si-Pt
punch-through diode \cite{Sze71},characteristics very similar to
those obtained in metal-ferroelectric-metal systems
}}\label{fig:punchthrough}
\end{figure}

Before looking for ferroelectric contributions to a system's
electrical properties one should make sure there are not
contributions due to the properties of the system unrelated to
ferroelectricity. Although much is sometimes made of the dependence
of capacitance on voltage, it is worth noting that
metal-semiconductor-metal systems have a characteristic capacitance
voltage which arises from the response of depletion layers to
applied voltage (Fig. \ref{fig:punchthrough}).

Essentially the problem boils down to the fact that there are two
possible sources of the dependence of capacitance on applied field,
either changes in depletion width or changes in the dielectric
constant of the material, i.e.

\begin{equation}
\frac{C(E)}{A}=\frac{\epsilon(E)}{d(E)}
\end{equation}

Several groups have assumed, that all the change in the capacitance
with field C(E) comes from change in depletion width d(E) and that
 (E) is changing negligibly. The first to suggest this was Joe
Evans (IEEE-ISAF Meeting, Urbana, Ill. 1990), who found d = 20 nm in
PZT.  Later Sandia claimed that there was no depletion (d=0 or
d=infinity)\cite{Miller90}. Several authors have assumed that d(E)
is responsible for C(E) \cite{Brennan92}, \cite{Mihara92},
\cite{Hwang98}, \cite{Hwang98b} ,\cite{Scott92}, \cite{Sayer92}

\begin{figure}[h]
\epsfig{figure=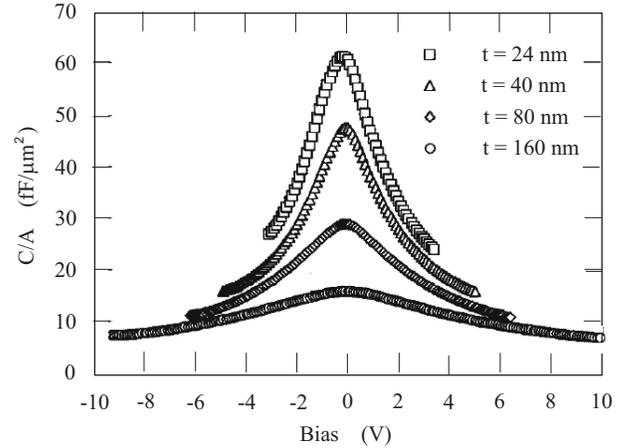,width=8.5cm}\caption{\textit{(Capacitance
v applied bias for BST thin films. The data are due to
\textcite{Basceri97}}}\label{fig:basceri}
\end{figure}

In contrast to the approach of explaining these characteristics
using semiconductor models \textcite{Basceri97} account for their
results on the basis of a Landau-Ginzburg style expansion of the
polarisation (Fig. \ref{fig:basceri}). The change in   with field
due to its nonlinearity has also been calculated by both
\textcite{Outzourhit95} and \textcite{Dietz97b}.  The real problem
is that both pictures are feasible. One should not neglect the fact
that the materials have semiconductor aspects; but at the same time
it is not unreasonable to expect that the known nonlinearity of the
dielectric response in these materials should be expressed in the
capacitance voltage characteristic. Probably the best approach is to
avoid making any conclusions on the basis of these kinds of
measurements alone, as it is quite possible that the relative sizes
of the contributions will vary greatly from sample to sample, or
even in the same sample under different experimental conditions.

\paragraph{Domain wall contributions}

Below the coercive field there are also contributions to the
permittivity from domain walls, as first pointed out by Fouskova in
1965\cite{Fouskova65},\cite{Fouskova65b}. In PZT the contributions
of domain wall pinning to the dielectric permittivity has been
studied in detail by Damjanovic and
Taylor\cite{Damjanovic97},\cite{Taylor97},\cite{Taylor98}, who
showed that the sub-coercive field contributions of the permittivity
were described by a Raleigh law with both reversible and
irreversible components, the irreversible component being due to
domain wall pinning.

\paragraph{Dielectric measurements of phase transitions}

\begin{figure}[h]
\epsfig{figure=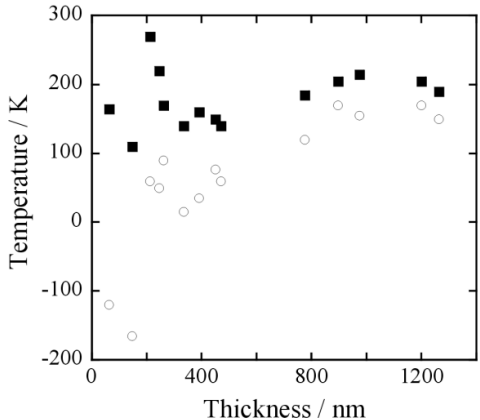,width=6.5cm}\caption{\textit{(\textcite{Lookman04}
Comparison between the \textit{apparent} Curie temperature in BST
taken from Curie-Weiss plots of raw data (empty circles) and
intrinsic data after correction for interfacial capacitance (filled
squares) had been performed. The intrinsic Curie temperature appears
to be independent of film thickness}}\label{fig:lookman}
\end{figure}

One of the most common approaches to measuring the transition
temperature of a ferroelectric material is naturally to measure the
dielectric constant and loss. However, in thin films there are
significant complications.  In bulk the maximum in the dielectric
constant is fairly well correlated with the transition temperature,
but this does not always seem to be the case in thin films. As
pointed out by \textcite{Vendik00}, a series capacitor model is
required to extract the true transition temperature, which in the
case of BST has been shown to be independent of
thickness\cite{Lookman04}, in contrast to the temperature at which
the permittivity maximum occurs, which can depend quite strongly on
thickness (Fig. \ref{fig:lookman}).

\subsubsection{Schottky barrier formation at metal-ferroelectric
junctions} \label{subsection:schottky}

In general, since ferroelectric materials are good insulator the
majority of carriers are injected from the electrode. When a metal
is attached to a ferroelectric material, a potential barrier is
formed if the metal work function is greater than the electron
affinity of the ferroelectric.  This barrier must be overcome if
charge carriers are to enter the ferroelectric, On the other hand if
the electron affinity is greater than the work function, then an
ohmic contact is formed. For the usual applications of
ferroelectrics (capacitors) it is desirable to have the largest
barrier possible. If a metal is brought into contact with an
intrinsic pure ferroelectric and surface states do not arise (i.e.,
the classic metal-insulator junction), then the barrier height is
simply

\begin{equation}
\phi_{b}=\phi_{m}-\chi
\end{equation}

In this case the fermi level of the metal becomes the fermi level of
the system, as there is no charge within the insulator with which to
change it. On the other hand, if there are dopants or surface
states, then there can be a transfer of charge between the metal and
ferroelectric, which allows the ferroelectric to bring the system
fermi level towards its own fermi level.

Although single crystals of undoped ferroelectric titanates tend to
be slightly p-type, simply because there are greater abundances of
impurities with lower valences than those of the ions for which they
substitute (Na+ for Pb+2; Fe+3 for Ti+4)
\cite{Smyth84},\cite{Smyth81},\cite{Smyth76} in reality most
ferroelectric capacitors are fine grained polycrystalline ceramics
and are almost always oxygen deficient. Typically the regions of the
capacitor near to interface are more oxygen deficient than the bulk.
Oxygen vacancies act as donor ions, and this means there can be a
transition from n-type behaviour at the interface to p-type
behaviour in the centre of the film as is evident in the kelvin
probe study of \textcite{Nowotny94}, who found that in bulk
BaTiO$_{3}$ with Pt electrodes a change in work-function from
$2.5\pm0.3$eV for surfaces and $4.4\pm0.4$eV in the bulk of the
material. The nature of the material near the surface is important
since it determines whether a blocking or ohmic contact is formed.
It has been shown by \textcite{DawberJJAP} that the defect
concentration profile as measured by capacitance voltage technique
may be explained by a model of combined bulk and grain boundary
diffusion of oxygen vacancies during the high temperature processing
of a film.

Regardless of the p-type or n-type nature of the material , in most
oxide ferroelectrics on elemental metal electrodes the barrier
height for electrons is significantly less than the barrier height
for holes\cite{Robertson1999}, and so the dominant injected charge
carriers are electrons. As the injected carriers dominate the
conduction, leakage currents in ferroelectrics are electron currents
and not hole currents, contrary to the suggestion of
\textcite{Stolichnov98b}.

The first picture of barrier formation in semiconductors is due to
\textcite{Schottky38} and \textcite{Mott38}. In this picture the
conduction band and valence band bend such that the vacuum levels at
the interface are the same and the fermi level is continuous through
the interface, but deep within the bulk of the semiconductor retains
its original value relative to the vacuum level. This is achieved by
the formation of a depletion layer which shifts the position of the
fermi level by altering the number of electrons within the
interface.

Motivated by the experimental observation that many Schottky barrier
heights seemed to be fairly independent of the metal used for the
electrode \textcite{Bardeen47} proposed a different model of
metal-semiconductor junctions. In this picture the fermi level of
the semiconductor is ``pinned'' by surface states to the original
charge neutrality level. These states, as first suggested by
\textcite{Heine65}, are not typically real surface states but rather
states induced in the band-gap of the semiconductor by the metal.

Most junctions lie somewhere between the Schottky and Bardeen
limits. The metal induced gap states (MIGS) can accommodate some but
not all of the difference in the fermi level between the metal and
the semiconductor, and so band bending still occurs to some extent.
The factor $S=\frac{d\phi_{b}}{d\phi_{m}}$ is used to define this,
with S = 1 being the Schottky limit and S = 0 representing the
Bardeen limit. The value of S is determined by the nature of the
semiconductor; originally experimental trends linking this to the
covalency or ionicity of the bonding in the material were observed
(with covalent materials developing many more MIGS. than ionic
materials) \cite{Kurtin69}. However better correlation was found
between the effective band gaps (dependent on the electronic
dielectric constant $\epsilon_{\infty}$) \cite{Schluter78}, with
\cite{Monch86}

\begin{equation}
(\frac{1}{S}-1)=0.1(\epsilon_{\infty}-1)^{2}.
\end{equation}

Although SrTiO$_{3}$ was invoked as one of the materials that
violated the electronegativity rule by \textcite{Schluter78}, it is
omitted from the plot against $(\epsilon_{\infty}-1)$. The
experimental value for S in SrTiO$_{3}$ can be measured from Dietz's
data as approximately 0.5. This does not agree well with what one
would expect from Monch's empirical relation,  which gives S = 0.28
(as used by \textcite{Robertson1999}). Note that the use of the
ionic trap-free value S=1 for BST gives a qualitative error: It
predicts that BST on Al should be ohmic, whereas in actuality it is
a blocking junction;  an S-value of approximately 0.3 predicts a 0.4
eV Schottky barrier height, in agreement with
experiment.\cite{Scott00}

We can extract the penetration depth for Pt states into BaTiO$_{3}$
from the first principles calculation of \textcite{Rao1997} by
fitting the density of states (DOS) of platinum states in the oxygen
layers to an exponential relationship to extract the characteristic
length as 1.68 angstroms.

\textcite{Cowley65} derived an expression for the barrier height for
junctions between the two extremes. In this approach the screening
charges in the electrode and the surface states are treated as delta
functions of charge separated by an effective thickness
$\delta_{eff}$. This effective thickness takes into account both the
Thomas-Fermi screening length in the metal and the penetration
length of the MIGS, and is essentially an air-gap approach.

\begin{figure}[h]
 \includegraphics[width=8.5cm]{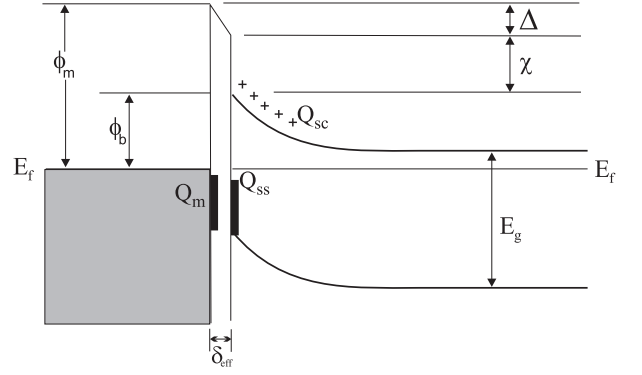}
 \caption{\textit{Energy
band diagram of a metal n-type semiconductor contact after Cowley
and Sze.}}\label{fig:cowleysze}
\end{figure}

The expression for the barrier height is

\begin{equation}
\phi_{b}=S(\phi_{m}-\chi)+(1-S)(E_{g}-\phi_{0})+\zeta
\label{eq:approxcowleysze}
\end{equation}

\lefteqn{\zeta=\frac{S^{2}C}{2}-S^{\frac{3}{2}}[C(\phi_{m}-\chi)+(1-S)(E_{g}-\phi_{0})\frac{C}{S}}
\begin{equation}
-\frac{C}{S}(E_{g}-E_{f}+kT)+\frac{C^{2}S}{4}]^{\frac{1}{2}}
\label{eq:cowleysze}
\end{equation}

\begin{figure}[h]
 \includegraphics[width=8.5cm]{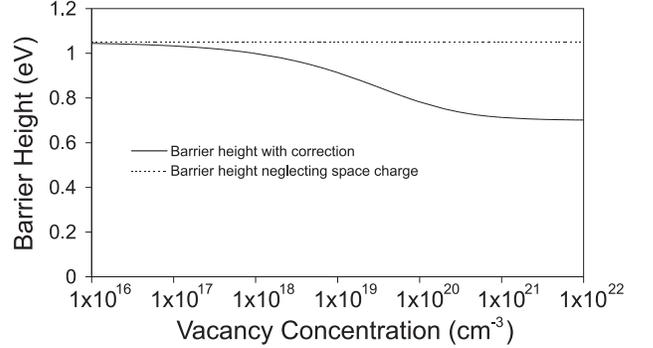}
 \caption{\textit{Schottky barrier height of Pt-SrTiO$_{3}$ as a
function of oxygen vacancy concentration, note that this may explain
the variation of experimental values from ca. 0.7 eV to 1.0
eV}}\label{fig:srtio3wvac}
\end{figure}

In the above $S=\frac{1}{1+q^{2}\delta_{eff}
D_{s}}$,$C=2q\varepsilon_{s}N_{D}\delta_{eff}^{2}$. When
$\varepsilon_{s}\approx10\varepsilon_{0}$ and $N_{D}$ $< 10^{18}$
cm$^{-3}$ C is of the order of 0.01 eV and it is reasonable to
discard the term $\zeta$ as \textcite{Cowley65} did. Neglecting this
term, as has been pointed out by \textcite{Rhoderick88}, is
equivalent to neglecting the charge in the depletion width. In the
systems under consideration here this term should not be neglected
as it can be quite large. To demonstrate the effect on the barrier
height we calculate the barrier height for a Pt-SrTiO$_{3}$ barrier
over a wide range of vacancy concentrations (Fig.
\ref{fig:srtio3wvac}).

It can be seen that the effect of vacancies on barrier height
becomes important for typical concentrations of vacancies
encountered in ferroelectric thin films. \textcite{dopants} have
addressed this issue and also the effect of introduced dopants on
barrier heights. Despite their omission of the term discussed above
the work of Robertson and Chen is valuable because of their
calculation of the charge neutrality levels for several
ferroelectric materials, an essential parameter for the calculation
of metal-ferroelectric barrier heights.

In a ferroelectric thin film this distribution of charges at the
interface manifests itself in more ways than simply in the
determination of the Schottky barrier height. Electric displacement
in the system is screened over the entire charge distribution.

In measuring the small -signal capacitance against thickness there
is always a non-zero intercept, which has been typically associated
with a "dead layer" at the metal film interface. However, in most
cases this interfacial capacitance can be understood by recognizing
that a finite potential exists across the charge at the interface.
In the simplest approximation one neglects any charge in the
ferroelectric and uses a Thomas-Fermi screening model for the metal.
This was initially considered by \textcite{Ku64} and first applied
to high k dielectrics by \textcite{Black99}. In their work they use
a large value for the dielectric constant of the oxide metal,
considering it as the dielectric response of the ions stripped of
their electrons. This may seem quite reasonable but is not however
appropriate. In general we think of metals not being able to sustain
fields, and in the bulk they certainly cannot, but the problem of
the penetration of electric fields into metals is actually well
known in a different context, that of the microwave skin depth. It
is very instructive to go through the derivation as an AC current
problem and then find the DC limit which will typically apply for
our cases of interest.

We describe the metal in this problem using the Drude Free Electron
Theory:

\begin{equation}
\sigma=\frac{\sigma_{0}}{1+i\omega \tau}
\end{equation}

There are three key equations to describe the charge distribution in
the metal.

Poisson's equation for free charges:
\begin{equation}
\rho (z)=\frac{1}{4\pi}\frac{\partial E(z)}{\partial z}
\end{equation}

The continuity equation:
\begin{equation}
-i\omega \rho(z)=\frac{\partial j(z)}{\partial z}
\end{equation}

The Einstein transport equation:
\begin{equation}
j=\sigma E - D \frac{\partial \rho}{\partial z}
\end{equation}

These are combined to give:
\begin{equation}
\frac{\partial^{2}\rho}{\partial
z^{2}}=\frac{4\pi\sigma}{D}(1+\frac{i\omega}{ 4\pi\sigma})\rho(z)
\end{equation}

This tells us that if at a boundary of the metal there exists a
charge it must decay with the metal exponentially with
characteristic screening length $\lambda$:
\begin{equation}
\lambda=(\frac{4\pi\sigma}{D}(1+\frac{i\omega}{ 4\pi\sigma}))^{-1/2}
\end{equation}

In the DC limit (which applies for most frequencies of our interest)
this length is the Thomas-Fermi screening length;

\begin{equation}
\lambda_{0}=(\frac{4\pi\sigma_{0}}{D})^{-1/2}
\end{equation}

So it becomes clear that the screening charge in the metal may be
modelled by substituting a sheet of charge displaced from the
interface by the Thomas-Fermi screening length, but that in
calculating the dielectric thickness of this region the effective
dielectric constant that must be used is 1, consistent with the
derivation of the screening length. Had we used a form of the
Poisson equation that had a non-unity dielectric constant, i.e.,

\begin{equation}
\rho (z)=\frac{\epsilon}{4\pi}\frac{\partial E(z)}{\partial z}
\end{equation}

then our screening length would be

\begin{equation}
\lambda_{0}=(\frac{4\pi\sigma_{0}}{\epsilon D})^{-1/2}
\end{equation}

which is not the Thomas-Fermi screening length. Thus the use of a
non-unity dielectric constant for the metal is not compatible with
the use of the Thomas-Fermi screening length.

Measurements on both sol-gel and CVD lead zirconate-titanate (PZT)
films down to ca. 60 nm thickness show that reciprocal capacitance
1/C(d) versus thickness d extrapolates to finite values at d=0,
demonstrating an interfacial capacitance. However whereas the value
for the sol-gel films is consistent with the Thomas Fermi screening
approach (.05nm), the value of interfacial thickness (0.005 nm) for
the CVD films is only $10\%$ of the interfacial capacitance that
would arise from the known Fermi-Thomas screening length of 0.05 nm
in the Pt electrodes \cite{Dawber03JPC}. That is, if this result
were interpreted in terms of a "dead layer", the dead layer would
have negative width. This result may arise from a compensating
"double-layer" of space charge inside the semiconducting PZT
dielectric; the Armstrong-Horrocks \cite{Armstrong97} semiconductor
formalism form of the earlier Helmholtz and Gouy-Chapman
polar-liquid models of the double layer can be used. Such a double
layer is unnecessary in PVDF because that material is highly
insulating \cite{Moreira02}. This explains quantitatively the
difference (x8) of interfacial capacitance in sol-gel PZT films
compared with CVD PZT films of the same thickness. The magnitude of
the electrokinetic potential (or zeta-potential) $\zeta =
\frac{\sigma d'}{\epsilon \epsilon_{0}} $ that develops from the
Helmholtz layer can be estimated without adjustable parameters from
the oxygen vacancy gradient data of Dey for a typical oxide
perovskite, SrTiO3; using Dey's surface charge density $\sigma$ of
2.8 x 10$^{18}$ e / m$^{2}$, a Gouy screening length in the
dielectric d' = 20 nm, and a dielectric constant of $\epsilon$ =
1300 yields $\zeta$ = 0.78 eV; since this is comparable to the
Schottky barrier height, it implies that much of the screening is
provided internally by mobile oxygen vacancies. [Here $\sigma(\tau ,
\mu)$ is a function of time $\tau$ and mobility $\mu$ for a bimodal
(a.c.) switching process.]

\subsubsection{Conduction mechanisms}

In general conduction is undesirable in memory devices based on
Capacitors, and so the understanding and minimization of conduction
has been a very active area of research over the years. Many
mechanisms have been proposed for the conduction in ferroelectric
thin films.

\paragraph{Schottky injection} Perhaps the most commonly observed
currents in ferroelectrics are due to thermionic injection of
electrons from the metal into the ferroelectric. The current-voltage
characteristic is determined by the image force lowering of the
barrier height when a potential is applied. A few points should be
made about Schottky injection in ferroelectric thin films. The first
is about the dielectric constant appropriate for use. In
ferroelectrics the size of the calculated barrier height lowering
depends greatly on which dielectric constant, the static or the
electronic, is used. The correct dielectric constant is the
electronic one ($\sim5.5$), as discussed by Scott and used by Dietz,
and by Zafar. \textcite{Dietz97} used the more general injection law
of \textcite{Murphy56} to describe charge injection in SrTiO$_{3}$
films. They found that for lower fields the Schottky expression was
valid, but at higher fields numerical calculations using the general
injection law were required. They did not however find that
Fowler-Nordheim Tunneling was a good description of any of the
experimental data.

It has been shown by \textcite{Zafar98b} that in fact the correct
form of the Schottky equation that should be used for ferroelectric
thin films is the diffusion limited equation of Simmons.
Furthermore, very recently \textcite{Dawber04} have shown that when
one considers the ferroelectric capacitor as a metal-insulator-metal
system with diffusion-limited current (as opposed to a single metal-
insulator junction),  the leakage current is explained well; in
addition, a number of unusual effects,,such as the negative
differential resistivity observed by \textcite{Watanabe98} and the
PTCR effect observed by \textcite{Hwang97},\textcite{Hwang98} are
accounted  for.

\paragraph{Poole-Frenkel} One of the standard ways of identifying
a Schottky regime is to plot $\log(\frac{J}{T})$ against
$V^{\frac{1}{2}}$. In this case the plot will be linear if the
current injection mechanism is Schottky injection. Confusion can
arise because carriers can also be generated from internal traps by
the Poole-Frenkel effect, which on the basis of this plot is
indistinguishable from Schottky injection. However, if the I-V
characteristic is asymmetric with respect to positive and negative
voltages (as is usually the case) then the injection process is most
probably Schottky injection. There are however some papers that show
symmetrical I-V curves and  correctly explain their data on the
basis of a Poole-Frenkel conduction mechanism \cite{Chen98}.

\paragraph{Fowler-Nordheim Tunneling}

\begin{figure}[h]
 \includegraphics[width=8.5cm]{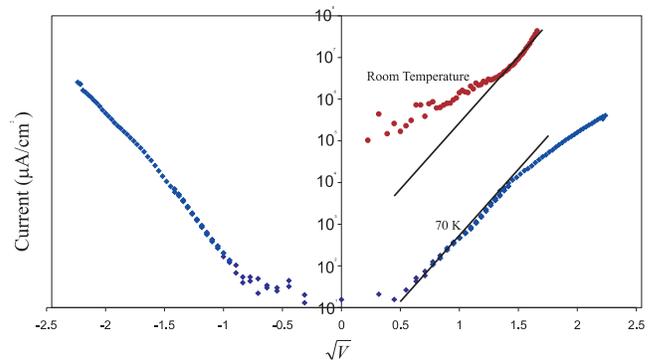}
 \caption{\textit{Leakage current data from Au-BST-SrRuO$_{3}$ film at room temperature and at T=70K}}\label{fig:lowTlc}
\end{figure}

Many researchers have discussed the possibility of tunnelling
currents in ferroelectric thin film capacitors. For the most part
they are not discussing direct tunneling through the film, which
would be impossible for typical film thicknesses, but instead
tunneling through the potential barrier at the electrode. The chief
experimental evidence that it might indeed be possible is due to
\textcite{Stolichnov98} who have seen currents that they claim to be
entirely tunneling currents in PZT films 450 nm thick at
temperatures between 100-140 K. It should be noted however that they
only observed tunneling currents above 2.2MV/cm, below which they
were unable to obtain data. The narrowness of the range of fields
for which they have collected data is a cause for concern, since the
data displayed in their paper go from 2.2-2.8 MV/cm. We conducted
leakage current measurements on a 70 nm BST thin film at 70 K and
found that the leakage current, while of much lower magnitude, was
still well described by a Schottky injection relationship, although
if one fitted this data to a similarly narrow field region it did
appear to satisfy the Fowler-Nordheim relationship well (Fig.
\ref{fig:lowTlc}).

The effective masses for tunneling obtained in the studies of
\textcite{Stolichnov99} and \textcite{Baniecki01} also seem to be at
odds with the normal effective masses considered for these
materials. Whereas they use effective masses of 1.0,  the effective
masses in perovskite oxides seem to be somewhat larger, from 5-7
m$_{e}$ for barium titanate and strontium titanate.\cite{Scott03}.
Although the tunneling mass and the effective (band) mass need not
be the same in general, if the tunneling is through thicknesses of
$>$ 2 nm, they are nearly so. [\textcite{Schnupp67};
\textcite{Conley67}, also find that the tunnelling mass due to light
holes in GaAs fits the band mass very well. ]."

\paragraph{Space Charge Limited Currents}

The characteristic quadratic relationship between current and
voltage that is the hallmark of space charge limited currents are
often seen in ferroelectrics. Sometimes it is observed that space
charge limited currents are seen when a sample is biased in one
direction, whereas for the opposite bias Schottky injection
dominates.

\paragraph{Ultra-Thin Films - Direct Tunneling}

Very recently \textcite{Contreras03} have succeeded in producing
metal-PZT-metal junctions sufficiently thin (6nm) that it appears
that direct tunneling  or phonon assisted tunneling (in contrast to
Fowler Nordheim tunneling)through the film may occur, though this
result requires more thorough investigation, since the authors note
the barrier heights extracted from their data using a direct
tunelling model are much smaller than expected. The principal result
of this paper is resistive switching,  which may be of considerable
interest in device applications, but also requires more thorough
investigation. This very interesting experimental study raises
important questions about the way that metal wave functions
penetrating from the electrode and ferroelectric polarisation
interact with each other in the thinnest ferroelectric junctions.

\paragraph{Grain boundaries}

Grain boundaries are often considered to be important in leakage
current because of the idea that they will provide conduction
pathways through the film.

Gruverman's results suggest that this is not the case in SBT. In his
experiment an AFM tip is rastered across the surface of a
polycrystalline ferroelectric film. The imaged pattern records the
leakage current at each point: white areas are high-current spots;
dark areas, low current.  If the leakage were predominantly along
grain boundaries, we should see dark polyhedral grains surrounded by
white grain boundaries, which become brighter with increasing
applied voltage.  In fact, the opposite situation obtains:  This
indicates that the grains have relatively low resistivity, with
high-resistivity grain boundaries.  The second surprise is that the
grain conduction comes in a discrete step; an individual grain
suddenly ``turns on'' (like a light switch). Smaller grains
generally conduct at lower voltages (in accord with Maier's theory
of space charge effects being larger in small grains with higher
surface-volume ratios\cite{Lubomirsky02}).

\subsection{Device Failure}

\subsubsection{Electrical Breakdown}

The process of electrical shorting in ferroelectric PZT was first
shown by Plumlee (1967) to arise from dendrite-like conduction
pathways through the material, initiated at the anodes and/or
cathodes.  These were manifest as visibly dark filamentary paths in
an otherwise light material when viewed through an optical
microscope.  They have been thought to arise as ``virtual cathodes''
via the growth of ordered chains of oxygen-deficient material. This
mechanism was modeled in detail by Duiker et al
\cite{Duiker90a},\cite{Duiker90b},\cite{Duiker90c}.

To establish microscopic mechanisms for beakdown in ferroelectric
oxide films one must show that the dependences of breakdown field
E$_{B}$ upon film thickness d, ramp rate, temperature, doping, and
electrodes are satisfied.  The dependence for PZT upon film
thickness is most compatible with  a low power-law dependence or
possibly logarithmic \cite{Scott03}. The physical models compatible
with this include avalanche (logarithmic), collision ionization from
electrons injected via field emission from the cathode
\cite{Forlani64}, which gives

\begin{equation}
E_{B}=Ad^{-w}
\end{equation}

with $\frac{1}{4} < w < \frac{1}{2}$, or the linked defect model of
\textcite{Gerson59}, which has d = 0.3. The dependence on electrode
material arises from the electrode work function and the
ferroelectric electron affinity through the resultant Schottky
barrier height. Following \textcite{VonHippel35} we have
(\textcite{Scott00}, p62.)

\begin{equation}
e E_{B}\lambda=h(\Phi_{M}-\Phi_{FE})
\end{equation}

where $\Phi_{M}$ and $\Phi_{FE}$ are the work functions of the metal
and of the semiconducting ferroelectric,;   is electron mean free
path; and h is a constant of order unity.

Even in films for which there is considerable Poole-Frenkel
limitation of current (a bulk effect), the Schottky barriers at the
electrode interfaces will still dominate breakdown behavior.

In general electrical breakdown in ferroelectric oxides is a hybrid
mechanism (like spark discharge in air) in which the initial phase
is electrical but the final stage is simple thermal run-away.  This
makes the dependence upon temperature complicated.

There are at least three different contributions to the temperature
dependence. The first is the thermal probability of finding a
hopping path through the material.  Following Gerson and Marshall
and assuming a random isotropic distribution of traps, Scott
\cite{ScottKluwerBook} showed that

\begin{equation}
E_{B}=G-\frac{k_{B}T}{B}\log{A} \label{eq:rightTdep}
\end{equation}

which gives both the dependence on temperature T and electrode area
A in agreement with practically all experiments on PZT, BST, and
SBT.

In agreement with this model the further assumption of exponential
conduction (non-ohmic) estimated to occur for applied field $E > 30$
MV/m \cite{Scott00}

\begin{equation}
\sigma(T)=\sigma_{0}\exp{\frac{-b}{k_{B}T}}
\end{equation}

in these materials yields the correct dependence of breakdown time
$t_{B}$ upon field

\begin{equation}
\log{t_{B}}=c_{1}-c_{2}E_{B}
\end{equation}

as well as the experimentally observed dependence of $E_{B}$ on rise
time $t_{c}$ of the applied pulse:

\begin{equation}
E_{B}=c_{3}t_{c}^{-\frac{1}{2}}
\end{equation}

Using the same assumption of exponential conduction, which is valid
for

\begin{equation}
aeE\ll k_{B}T
\end{equation}

where a is the lattice nearest neighbor oxygen-site hopping distance
(approximately a lattice constant) and e, the electron charge, Scott
\cite{Scott00} shows that the general breakdown field expression

\begin{equation}
C_{V}\frac{dT}{dt}-\nabla(K\cdot\nabla T)=\sigma E_{B}^{2}
\end{equation}

in the impulse approximation (in which the second term in the above
equation is neglected) yields

\begin{equation}
E_{B}(T)=[\frac{3C_{V}K}{\sigma_{0}bt_{c}}]^{\frac{1}{2}}T\exp(\frac{b}{2k_{B}T})
\end{equation}

which suffices to estimate the numerical value of breakdown field
for most ferroelectric perovskite oxide films; values approximating
800 MV/m are predicted and measured.

A controversy has arisen regarding the temperature dependence of
$E_{B}(T)$ and the possibility of avalanche \cite{Stolichnov01} In
low carrier concentration single crystals, especially Si, avalanche
mechanisms give a temperature dependence that is controlled by the
mean free path of the injected carriers.  This is physically because
at higher temperatures the mean free path $\lambda$ decreases due to
phonon scattering and thus one must apply a higher field E$_{B}$ to
achieve avalanche conditions.

\begin{equation}
\lambda=\lambda_{0}\tanh(\frac{E_{B}}{kT}) \label{eq:wrongTdep}
\end{equation}

However, this effect is extremely small even for low carrier
concentrations (10\% change in $E_{B}$ between 300K and 500K for n =
10$^{16}$ cm$^{-3}$) and negligible for higher concentrations. The
change in $E_{B}$ in BST between 600K and 200K is $>500\%$ and
arises from Eq. \ref{eq:rightTdep}, not Eq. \ref{eq:wrongTdep}).
Even if the ferroelectrics were single crystals, with $10^{20}$
cm$^{-3}$ oxygen vacancies near the surface, any T-dependence from
Eq. \ref{eq:wrongTdep} would be unmeasurably small; and for the
actual fine-grained ceramics (40 nm grain diameters), the mean free
path is ca. 1 nm and limited by grain boundaries (T-independent).
Thus the conclusion of \textcite{Stolichnov01} regarding avalanche
is qualitatively and quantitatively wrong in ferroelectric oxides.

\subsubsection{Fatigue}

Polarisation fatigue, which is the process whereby the switchable
ferroelectric polarisation is reduced by repetitive electrical
cycling is one of the most serious device failure mechanisms in
ferroelectric thin films. It is most commonly a problem when Pt
electrodes, desirable because of their high work-functions, are
used.

\begin{figure}[h]
\epsfig{figure=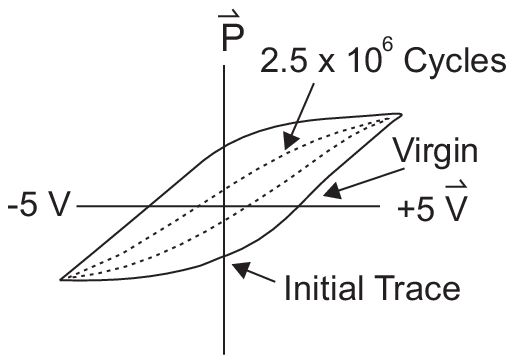,width=6cm}\caption{\textit{Change in
the polarisation hysteresis loop with fatigue \textcite{Scott88}
}}\label{fig:fatiguedloop}
\end{figure}

Importantly fatigue occurs through the pinning of domain walls,
which pins the polarization in a particular direction, rather than
any fundamental reduction of the polarisation. \textcite{Scott88}
demonstrated in KNO$_{3}$ via Raman spectroscopy that only a very
small part of the sample was converted from the ferroelectric to non
ferroelectric phase with fatigue, thus implying that fatigue must be
caused by pinning of the domain walls. They also demonstrated that
the domain walls could be de-pinned via the application of a large
field, as shown in Fig.17.  The pinning of domain walls has also
been observed directly by Atomic Force microscopy by
\textcite{Gruverman96} and by \textcite{Colla98}.

\begin{figure}[h]
\epsfig{figure=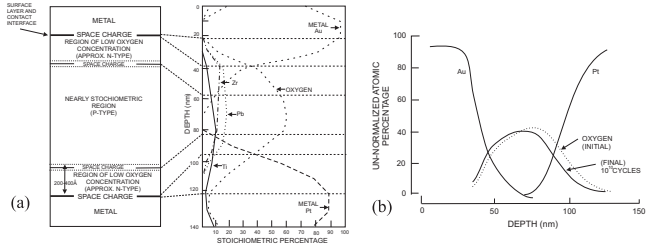,width=8.5cm}\caption{\textit{\textcite{Scott91}(a)
Auger depth profile of PZT thin film capacitor (b) Effect of fatigue
on oxygen concentration near the
electrode}}\label{fig:depthprofiles}
\end{figure}

There is a fairly large body of evidence that oxygen vacancies play
some key part in the fatigue process. Auger data of
\textcite{Scott91} show areas of low oxygen concentration in a
region near to the metal electrodes, implying a region of greater
oxygen vacancy data. Scott et al also reproduced Auger data from
\textcite{Troeger} for a film that had been fatigued by $10^{10}$
cycles showing an increase in the width of the region with depleted
oxygen near the platinum electrode (Fig. \ref{fig:depthprofiles}).
There is however no corresponding change at the gold electrode. Gold
does not form oxides, this might be an explanation of the different
behaviour at the two electrodes. Although some researchers believe
that platinum also does not form oxides, the adsorption of oxygen
onto Pt surfaces is actually a large area of research, because of
the important role platinum plays as a catalyst in fuel cell
electrodes. Oxygen is not normally adsorbed onto gold surfaces but
can be if there is significant surface roughness \cite{Mills03}.

It has also been found experimentally that films fatigue differently
in atmospheres containing different oxygen partial pressures.
\cite{Brazier99}. \textcite{Pan96} claim to have seen oxygen
actually leaving a ferroelectric sample during switching, though we
note that \textcite{Nuffer01} claim this to be an experimental
artefact. The results of \textcite{Schloss02} are very interesting
in that they show directly by O$^{18}$ tracer studies that the
oxygen vacancies redistribute themselves during voltage cycling. In
their original paper they concluded that the redistribution of
oxygen vacancies this is not the cause of fatigue, because they did
not see redistribution of O$^{18}$ when the sample has been
annealed, though the sample still fatigues, however in a more recent
publication they conclude the reason they could not see the oxygen
tracer distribution was more probably due to a change in the oxygen
permeability of the electrode after annealing \cite{Schloss04}.

It has been known for some time that the fatigue of PZT films can be
improved by the use of oxide electrodes, such as iridium oxide or
ruthenium oxides. \textcite{Araujo95} explain the improved fatigue
resistance by the fact that oxides of iridium and platinum can
reduce or re-oxidise reversibly and repeatedly without degradation.
It is for the same reason that iridium is preferred to platinum an a
electrode materials for medical applications where this property was
originally studied by \textcite{Robblee}. This property does make
the leakage current properties of these electrodes more complicated,
and generally films with Ir/IrO$_{2}$ or Ru/RuO$_{2}$ electrodes
have higher leakage currents than those with platinum electrodes.
Unless carefully annealed at a certain temperature RuO$_{2}$
electrodes will have elemental Ru metallic islands.  Since the work
function for Ru is 4.65 eV and that for RuO2 is 4.95 eV, almost all
the current will pass through the Ru islands, producing hot spots
and occasional shorts \cite{Hartmann00}.  By contrast, although one
expects that there will be similar issues with mixtures of Ir and
IrO$_{2}$ in iridium based electrodes, when metallic Ir is oxidized
to IrO$_{2}$ its workfunction decreases to 4.23 eV
\cite{Chalamala99}.

\begin{figure}[h]
\epsfig{figure=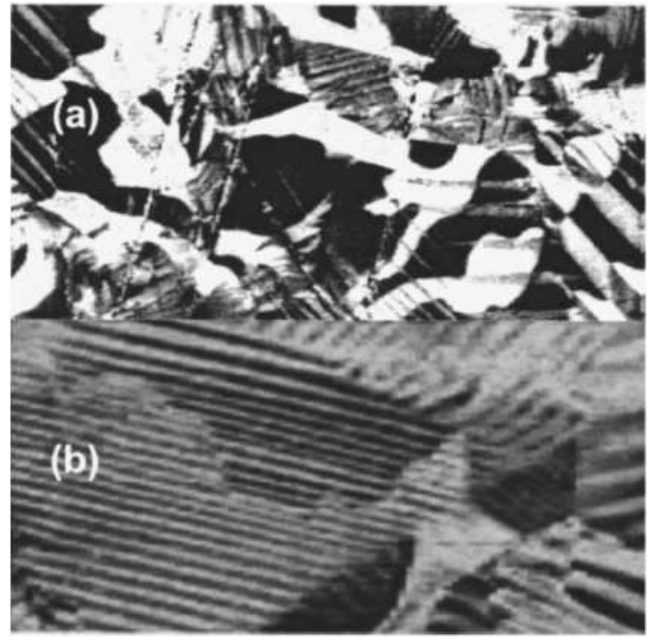,width=8.5cm}\caption{\textit{Atomic force
microscopy images (Lupascu) (a) before and (b) after, cycling
showing evidence of planes of oxygen vacancies in the fatigued
sample. Image width = $10\mu m$.}}\label{fig:Lupascu}
\end{figure}

The idea that planes of oxygen vacancies perpendicular to the
polarisation direction could pin domain walls is originally due to
\textcite{Brennan93}. Subsequently, in a theoretical microscopic
study of oxygen vacancy defects in PbTiO$_{3}$ \textcite{Park1998}
showed that planes of vacancies are much more effective at pinning
domain walls than single vacancies. In bulk ferroelectrics
\textcite{Arlt88} have discussed how under repetitive cycling the
vacancies can move from their originally randomly distributed sites
in the perovskite structure to sites in planes parallel to the
ferroelectric-electrode interface. We suspect that while this may
account for fatigue in bulk ferroelectrics it is not the operative
mechanism in thin films. \textcite{vacancyordering} have suggested
that in thin films the vacancies can reach sufficiently high
concentrations that they order themselves into planes in a similar
way as occurs in Fe-doped bulk samples and on the surfaces of highly
reduced specimens.  Direct evidence that this occurs in bulk PZT was
found using Atomic force microscope imagery of PZT grains by
\textcite{Lupascu02} (Fig. \ref{fig:Lupascu}). Recently evidence of
oxygen vacancy ordering has also been found in barium titanate
reduced after an accelerated life test \cite{Woodward04}.

While most researchers acknowledge that oxygen vacancies play a role
in fatigue, it should be noted that \textcite{Tagantsev01} has
aggressively championed a model of charge injection; however since
this model is not developed into a quantitative form it is very hard
to verify or falsify it. Charge injection probably does play a role
in fatigue, an idea at least in part supported by the detailed
experimental study of \textcite{Du98b}, but in the model of
\textcite{FatigueModel}, (which draws upon the basic idea of
\textcite{Yoo92} that fatigue is due to the electromigration of
oxygen vacancies) it is not included, nevertheless most of the
experimental results in the literature may be accounted for. The
model of \textcite{FatigueModel} basically shows that in a
ferroelectric thin film under an AC field there is in fact a net
migration of vacancies towards the interface and it is the high
concentration of vacancies in this region that results in ordering
of the vacancies and pinning of domain walls. The interfacial nature
of fatigue in thin films has been demonstrated by \textcite{Colla98}
and by \textcite{Dimos94}.

To really understand fatigue better what is needed is more
experiments that try to look at the problem in novel ways, standard
electrical measurements alone probably cannot shed a great deal of
additional light on the problem, especially given the problems
between comparing samples grown in different labs using different
techniques. Very recently a very interesting study was undertaken by
\textcite{Do04} using X-Ray microdiffraction to study fatigue in
PZT. Using this technique they were able to see how regions of the
film stopped switching as it fatigued. One of the key findings of
this study was that there appears to be two fatigue effects
operative, a reversible effect that occurs when low to moderate
fields are used for switching and an irreversible effect which
occurs under very high fields.

\subsubsection{Retention failure}

Clearly a non-volatile memory that fails to retain the information
stored in it will not be a successful device. Furthermore producers
of memories need to be able to guarantee retention times over much
longer periods of time then they can possibly test. A better
understanding of retention failure is thus required so that models
can used that allow accelerated retention tests to be carried out,
the work of \textcite{Kim01b} is a step in this direction.

It seems that imprint and retention failure are closely linked
phenomena, ie. if a potential builds up across the system of the
time it can destabilize the ferroelectric polarisation state and
thus cause loss of information. Further comparison of retention-time
data and fatigue data suggest quite strongly a link between the two
effects. DC degradation of resistance in BST seems also to be a
related effect \cite{Zafar99}. The electromigration of oxygen
vacancies under an applied field in a Fe-doped SrTiO$_{3}$ single
crystal has been directly observed via electro-chromic effects
\cite{Waser90}. Oxygen vacancy redistribution under applied field
has also been invoked to explain a slow relaxation of the
capacitance in BST thin films \cite{Boikov01}. It would seem to make
sense that whereas fatigue relates to the cumulative motion of
oxygen vacancies under an AC field, resistance degradation is a
result of their migration under an applied DC field and retention
failure is a result of their migration under the depolarisation
field or other built-in fields in the material.

\section{FIRST PRINCIPLES}
\label{sec:firstp}

With continuing advances in algorithms and computer hardware, first
principles studies of the electronic structure and structural
energetics of complex oxides can now produce accurate,
material-specific information relevant to the properties of
thin-film ferroelectrics. In this section, we focus on
first-principles studies that identify and analyze the
characteristic effects specific to thin films. First, we briefly
review the relevant methodological progress and the application of
these methods to bulk ferroelectric materials. Next, we will survey
the first-principles investigations of ferroelectric thin films and
superlattices reported in the literature. It will be seen that the
scale of systems that can be studied directly by first principles
methods is severely limited at present by practical considerations.
This can be circumvented by the construction of nonempirical models
with parameters determined by fitting to the results of selected
first-principles calculations. These models can be parameterized
interatomic potentials, permitting molecular dynamics studies of
nonzero temperature effects, or first-principles effective
Hamiltonians for appropriate degrees of freedom (usually local
polarization and strain). The form of the latter strongly resembles
that of a Landau-Devonshire theory, providing a connection between
first-principles approaches and the extensive literature on
phenomenological models for the behavior of thin film
ferroelectrics. The advantages and disadvantages of using first
principles results rather than experimental data to construct models
will be considered. In addition to allowing the study of systems far
more complex than those that can be considered by first principles
alone, this modelling approach yields physical insight into the
essential differences between bulk and thin film behavior, which
will be discussed at greater length in subsection
\ref{subsection:lessons}. Finally, it will be seen that despite
practical limitations, the complexity of the systems for which
accurate calculations can be undertaken has steadily increased in
recent years, to the point where films of several lattice constants
in thickness can be considered. While this is still far thinner than
the films of current technological interest, concommitent
improvements in thin film synthesis and characterizaton have made it
possible to achieve a high degree of atomic perfection in comparable
ultrathin films in the context of research. This progress has led to
a true relevance of calculational results to experimental
observations, opening a meaningful experimental-theoretical
dialogue. However, this progress in some ways only serves to
highlight the full complexity of the physics of real ferroelectric
films: as questions get answered, more questions, especially about
phenomena at longer length scales and about dynamics, are put forth.
These challenges will be discussed in subsection
\ref{subsubsection:challenges}.

\subsection{DFT studies of bulk ferroelectrics}

In parallel with advances in laboratory synthesis, the past decade
has seen a revolution in the atomic-scale theoretical understanding
of ferroelectricity, especially in perovskite oxides, through
first-principles density-functional-theory (DFT) investigations. The
central result of a DFT calculation is the ground state energy
computed within the Born-Oppenheimer approximation; from this the
predicted ground state crystal structure, phonon dispersion
relations, and elastic constants are directly accessible. The latter
two quantities can be obtained by finite-difference calculations, or
more efficiently, through the direct calculation of derivatives of
the total energy through density-functional perturbation theory
(DFPT) \cite{Baroni2001}.

For the physics of ferroelectrics, the electric polarization and its
derivatives, such as the Born effective charges and the dielectric
and piezoelectric tensors, are as central as the structural
energetics, yet proper formulation in a first-principles context
long proved to be quite elusive. Expressions for derivatives of the
polarization corresponding to physically measurable quantities were
presented and applied in DFPT calculations in the late
1980's\cite{deGironcoli1989}. A key conceptual advance was
establishing the correct definition of the electric polarization as
a bulk property through the Berry phase formalism of King-Smith,
Vanderbilt and Resta \cite{Kingsmith93,Resta94}. With this and the
related Wannier function expression \cite{Kingsmith94}, the
spontaneous polarization and its derivatives can be computed in a
post-processing phase of a conventional total-energy calculation,
greatly facilitating studies of polarization-related properties.

For perovskite oxides, the presence of oxygen and first row
transition metals significantly increases the computational demands
of density functional total energy calculations compared to those
for typical semiconductors. Calculations for perovskite oxides have
been reported using essentially all of the available
first-principles methods for accurate representation of the
electronic wavefunctions: all-electron methods, mainly linearized
augmented plane wave (LAPW) and full-potential linearized augmented
plane wave (FLAPW), linear muffin-tin orbitals (LMTO),
norm-conserving and ultrasoft pseudopotentials, and
projector-augmented wavefunction (PAW) potentials. The effects of
different choices for the approximate density functional have been
examined; while most calculations are carried out with the
local-density approximation (LDA), for many systems the effects of
the generalized-gradient approximation (GGA) and weighted-density
approximation (WDA)\cite{Singh2004} have been investigated, as well
as the alternative use of the Hartree-Fock approach. Most
calculations being currently reported are performed with an
appropriate standard package, mainly VASP\cite{VASPa,VASPb}, with
ultrasoft pseudopotentials and PAW potentials, ABINIT \cite{ABINIT},
with norm-conserving pseudopotentials and PAW potentials, PWscf
\cite{PWscf}, with norm-conserving and ultrasoft pseudopotentials,
SIESTA \cite{SIESTA}, with norm-conserving pseudopotentials, WIEN97
(FLAPW) \cite{WIEN97} and CRYSTAL (Hartree-Fock)\cite{CRYSTAL}.

To predict ground state crystal structures, the usual method is to
minimize the total energy with respect to free structural parameters
in a chosen space group, in a spirit similar to that of a Rietveld
refinement in an experimental structural determination. The space
group is usually implicitly specified by a starting guess for the
structure. For efficient optimization, the calculation of forces on
atoms and stresses on the unit cell is essential and is by now
included in every standard first-principles implementation following
the formalism of Hellmann and Feynman \cite{HellmannFeynman} for the
forces and Nielsen and Martin \cite{NielsenMartin} for stresses.

The accuracy of DFT for predicting the ground state structures of
ferroelectrics was first investigated for the prototypical cases of
BaTiO$_3$ and PbTiO$_3$ \cite{Cohen1990,Cohen1992,Cohen92a}, and
then extended to a larger class of ferroelectric
perovskites\cite{Kingsmith94}. Extensive studies of the structures
of perovskite oxides and related ferroelectric oxide structures have
since been carried out \cite{Resta2003}. The predictive power of
first-principles calculations is well illustrated by the results of
Singh for PbZrO$_3$ \cite{Singh1995}, in which the correct energy
ordering between ferroelectric and antiferroelectric structures was
obtained and furthermore, comparisons of the total energy resolved
an ambiguity in the reported space group and provided an accurate
determination of the oxygen positions.

It is important to note, however, that there seem to be limitations
to the accuracy to which structural parameters, particularly lattice
constants, can be obtained. Most obvious is the underestimate of the
lattice constants within the LDA, typically by about 1\% (GGA tends
to shift lattice constants upward, sometimes substantially
overcorrecting). Considering that the calculation involves no
empirical input whatsoever, an error as small as 1\% could be
regarded not as a failure, but as a success of the method. Moreover,
the fact that the underestimate varies little from compound to
compound means that the relative lattice constants and thus the type
of lattice mismatch (tensile/compressive) betwen two materials in a
heterostructure is generally correctly reproduced when using
computed lattice parameters. However, for certain questions, even a
1\% underestimate can be problematic. The ferroelectric instability
in the perovskite oxides, in particular, is known to be very
sensitive to pressure \cite{Samara} and thus to the lattice
constant, so that 1\% can have a significant effect on the
ferroelectric instability. In addition, full optimization of all
structural parameters in a low symmetry space group can in some
cases, PbTiO$_3$ being the most well studied example, lead to an
apparently spurious prediction, though fixing the lattice constants
to their experimental values leads to good agreement for the other
structural parameters \cite{Cohen1998}. Thus, it has become
acceptable, at least in certain first-principles contexts, to fix
the lattice parameters or at least the volume of the unit cell, to
the experimental value, when this value is known.

In a first-principles structural prediction, the initial choice of
space group may appear to limit the chance that a true ground state
structure will be found. In general, once a minimum is found, it can
be proved (or not) to be a local minimum by computation of the full
phonon dispersion and of the coupling, if allowed by symmetry,
between zone-center phonons and homogeneous strain
\cite{Garcia1996}. Of course, this does not rule out the possibility
of a different local minimum with lower energy with an unrelated
structure.

For ferroelectrics, the soft-mode theory of ferroelectricity
provides a natural conceptual framework for the identification of
low-symmetry ground state structures and for estimating response
functions. The starting point is the identification of a
high-symmetry reference structure. For perovskite compounds, this is
clearly the cubic perovskite structure, and for layered perovskites,
it is the nonpolar tetragonal structure. The lattice instabilities
of the reference structure can be readily identified by the
first-principles calculation of phonon dispersion relations
\cite{Waghmare1997,Ghosez1999,Sai2000a}, this being especially
efficient within density functional perturbation theory. In simple
ferroelectric perovskites, the ground state structure is obtained to
lowest order by freezing in the most unstable mode (a zone-center
polar mode). This picture can still be useful for more complex
ground state structures that involve the freezing in of two or more
coupled modes (e.g. PbZrO$_3$) \cite{Waghmare1997a,Cockayne2000}, as
well as for identifying low-energy structures that might be
stabilized by temperature or pressure \cite{Stachiotti2000,Fennie}
The polarization induced by the soft polar mode can be obtained by
computation of the Born effective charges, yielding the mode
effective charge. The temperature-dependent frequency of the soft
polar mode and its coupling to strain are expected largely to
determine the dielectric and piezoelectric response of ferroelectric
and near-ferroelectric perovskite oxides; this idea has been the
basis of several calculations \cite{Cockayne1998,Garcia1998}.

Minimization of the total energy can similarly be used to predict
atomic arrangements and formation energies of point defects
\cite{Park1998,Poykko2000,Betsuyaku2001,Man2002,Park2003,
Robertson2003,Astala2004}, domain walls
\cite{Poykko2000,Meyer2002,He2003} and non-stoichiometric planar
defects such as antiphase domain boundaries
\cite{Suzuki2001,HaoLi2002},in bulk perovskite oxides. The supercell
used must accommodate the defect geometry, and generally must
contain many bulk primitive cells to minimize the interaction of a
defect with its periodically-repeated images. Thus, these
calculations are extremely computationally intensive, and many
important questions remain to be addressed.

While much of the essential physics of ferroelectrics arises from
the structural energetics, the polarization and the coupling between
them, there has been increasing interest in ferroelectric oxides as
electronic and optical materials, for which accurate calculations of
the gap and dipole matrix elements are important. Furthermore, as we
will discuss in detail below, the bandstructures enter in an
essential way in understanding the charge transfer and dipole layer
formation of heterostructures involving ferroelectrics, other
insulators, and metals. While density functional theory provides a
rigorous foundation only for the computation of Born-Oppenheimer
ground-state total energies and electronic charge densities, it is
also often used for investigation of electronic structure. In the
vast majority of density functional implementations, calculation of
the ground state total energy and charge density involves the
computation of a bandstructure for independent electron states in an
effective potential, following the work of Kohn and Sham
\cite{KohnSham1965}. This bandstructure is generally regarded as a
useful guide to the electronic structure of materials, including
perovskite and layered perovskite oxides
\cite{Cohen1992,Robertson1996,Tsai2003}. It should be noted that it
is consistently found that with approximate functionals, such as the
LDA, the fundamental bandgaps of insulators and semiconductors,
including perovskite ferroelectrics, are substantially
underestimated. While for narrow gap materials, the system may even
be erroneously found to be metallic, for wider gap systems such as
most of the simple ferroelectric perovskite compounds considered
here, the band gap is still nonzero and thus the structural
energetics in the vicinity of the ground state structure is
unaffected. While this error might be considered to be an
insuperable stumbling block to first-principles investigation of
electronic structure and related properties, there is at present no
widely available, computationally tractable, alternative (there are,
though, some indications that the use of exact-exchange functionals
can eliminate much of this error \cite{Piskunov2004}; it has long
been known that Hartree-Fock, i.e. exact exchange only, leads to
overestimates of the gaps). The truth is that, as will be discussed
further below, these results can, with care, awareness of the
possible limitations and judicious use of experimental input, be
used to extract useful information about the electronic structure
and related properties in individual material systems.

At present, the computational limitations of full first-principles
calculations to 70-100 atoms per supercell have stimulated
considerable interest in the development and use in simulations of
effective models, from which a subset of the degrees of freedom have
been integrated out. Interatomic shell-model potentials have been
developed for a number of perovskite oxide systems, eliminating most
of the electronic degrees of freedom except for those represented by
the shells \cite{Tinte1999,Heifets2000,Sepliarsky2004}. A more
dramatic reduction in the number of degrees of freedom is performed
to obtain effective Hamiltonians, in which typically one vector
degree of freedom decribes the local polar distortion in each unit
cell. This approach has proved useful for describing finite
temperature structural transitions in compounds and solid solutions
\cite{Zhong1994,Rabe1995,Waghmare1997,Bellaiche2000, Leung2002}. To
the extent that the parameters appearing in these potentials are
determined by fitting to selected first-principles results
(structures, elastic constants, phonons), these approaches can be
regarded as approximate first-principles methods. They allow
computation of the polarization as well as of the structural
energetics, but not, however, of the electronic states. Most of the
effort has been focused on BaTiO$_3$, though other perovskites,
including PbTiO$_3$ and KNbO$_3$, SrTiO$_3$ and (Ba,Sr)TiO$_3$ and
Pb(Zr,Ti)O$_3$ have been investigated in this way, and useful
results obtained.

\subsection{First-principles investigation of ferroelectric thin films}

The fascination of ferroelectric thin films and superlattices lies
in the fact that the properties of the system as a whole can be so
different from the individual properties of the constituent
material(s). Empirically, it have been observed that certain
desirable bulk properties, such as a high dielectric response, can
be degraded in thin films, while in other investigations there are
signs of novel interesting behavior obtained only in thin film form.

Theoretical analysis of the observed properties of thin films
presents a daunting challenge. It is well known that the process of
thin film growth itself can lead to nontrivial differences from the
bulk material, as observed in studies of homoepitaxial oxide films
such as SrTiO$_3$ \cite{Stemmer2004}. However, as synthetic methods
have developed, the goal of growing nearly ideal, atomically
ordered, single crystal films and superlattices is coming within
reach, and the relevance of first-principles results for perfect
single crystal films to experimental observations emerging.

As will be clear from the discussion in the rest of this section, an
understanding of characteristic thin film behavior can best be
achieved by detailed quantitative examination of individual systems
combined with the construction of models incorporating various
aspects of the physics, from which more general organizing
principles can be identified. In first principles calculations,
there is a freedom to impose constraints on structural parameters
and consider hypothetical structures that goes far beyond anything
possible in a real system being studied experimentally. This will
allow us to isolate and examine various influences on the state of a
thin film: epitaxial strain, macroscopic electric fields, surfaces
and interfaces, characteristic defects associated with thin-film
growth, and ``true" finite size effects, and how they change the
changes in atomic arrangements, electronic structure, polarization,
vibrational properties, and responses to applied fields and
stresses. While the main focus of this review is on thin films, this
approach also applies naturally to multilayers and superlattices.
Extending our discussion to include these latter systems will allow
us to consider the effects of the influencing factors in different
combinations, for example the changing density of interfaces, the
degree of mismatch strain, and the polarization mismatch. These
ideas also are relevant to investigating the behavior of bulk
layered ferroelectrics, which can be regarded as natural
short-period superlattices.

Within this first-principles modelling framework, we can more
clearly identify specific issues and results for investigation and
analysis. The focus on modelling is also key to the connection of
first-principles results to the extensive literature on
phenomenological analysis and to experimental observations. This
makes the most effective use of first-principles calculations in
developing a conceptual and quantitative understanding of
characteristic thin film properties, as manifested by thickness
dependence as well as by the dependence on choice of materials for
the film, substrate and electrodes.

\subsubsection{First principles methodology for thin films}
\label{subsection:methodology}

The fundamental geometry for the study of thin films, surfaces and
interfaces is that of an infinite single-crystalline planar slab.
Since three-dimensional periodicity is required by most
first-principles implementations, in those cases the slab is
periodically repeated to produce a supercell; a few studies have
been carried out with electronic wavefunction basis sets that permit
the study of an isolated slab. The variables to be specified include
the orientation, the number of atomic layers included, choice of
termination of the surfaces, and width of vacuum layer separating
adjacent slabs. As in the first-principles prediction of bulk
crystal structure, choice of a space group for the supercell is
usually established by the initial structure; relaxations following
forces and stresses do not break space group symmetries. The
direction of the spontaneous polarization is constrained by the
choice of space group, allowing comparison of unpolarized
(paraelectric) films with films polarized along the normal or in the
plane of the film.

As we will see below, in most cases the slabs are very thin (ten
atomic layers or fewer). It is possible to relate the results to the
surface of a semi-infinite system or a coherent epitaxial film on a
semi-infinite substrate by imposing certain constraints on the
structures considered. In the former case, the atomic positions for
interior layers are fixed to correspond to the bulk crystal
structure. In the latter, the in-plane lattice parameters of the
supercell are fixed to the corresponding bulk lattice parameters of
the substrate, which is not otherwise explicitly included in the
calculation. More sophisticated methods developed to deal with the
coupling of vibrational modes at the surface with bulk modes of the
substrate\cite{Lewis1996} could also be applied to ferroelectric
thin films, though this has not yet been done.

The LDA underestimate of the equilibrium atomic volume will in
general also affect slab calculations, and similar concerns arise
about the coupling of strain and the ferroelectric instability. As
in bulk crystal structure prediction, it may in some cases be
appropriate to fix certain structural parameters according to
experimental or bulk information. In the case of superlattices and
supercells of films on substrates, it may on the other hand be a
good choice to work consistently at the (compressed) theoretical
lattice constant, since the generic underestimate of the atomic
volume ensures that the lattice mismatch and relative
tensile/compressive strain will be correctly reproduced. This
applies for example to the technique mentioned in the previous
paragraph, in which the effects of epitaxial strain are investigated
by performing slab calculations with an appropriate constraint on
the in-plane lattice parameters.

As in first-principles predictions of bulk crystal structure, the
initial choice of space group constrains, to a large extent, the
final ``ground-state" structure. If the supercell is constructed by
choosing a bulk termination, the energy minimization based on forces
and stresses will preserve the initial symmetry, yielding
information about surface relaxations of the unreconstructed
surface.  A lower-energy structure might result from breaking
additional point or translational symmetries to obtain a surface
reconstruction. This type of surface reconstruction could be
detected by computing the Hessian matrix (coupled phonon dispersion
and homogeneous strain) for the relaxed surface. More complex
reconstructions involving adatoms, vacancies or both would have to
be studied using appropriate starting supercells. Information
regarding the existence and nature of such reconstructions might be
drawn from experiments and/or from known reconstructions in related
materials.

One very important consideration in the theoretical prediction of
stable surface orientations and terminations and of favorable
surface reconstructions is that these depend on the relative
chemical potential of the constituents. Fortunately, since the
chemical potential couples only to the stoichiometry, the prediction
of the change of relative stability with chemical potential can be
made with a single total-energy calculation for each structure (see,
for example, \cite{Meyer1999}). Because of the variation in
stoichiometry for different (001) surface terminations, what is
generally reported is the average surface energy of symmetric AO and
BO$_2$-terminated slabs.

\begin{figure}
\centerline{\epsfig{file=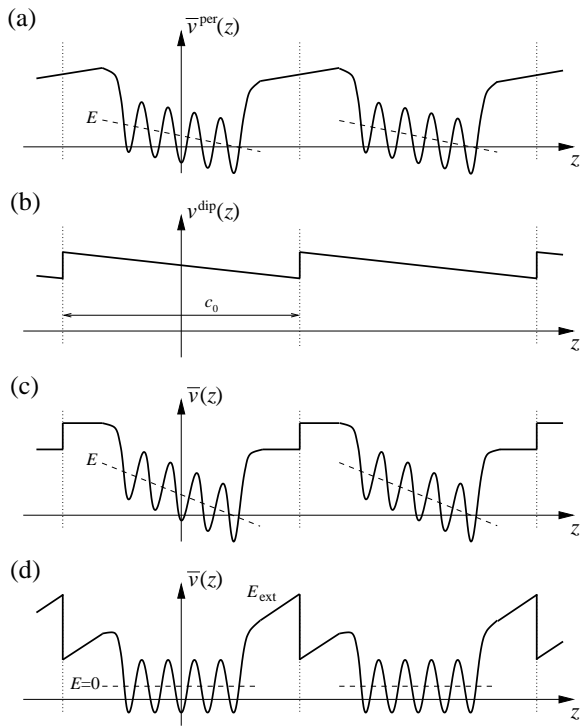,width=3.00in,angle=0}}
\caption{Schematic picture of the planar-averaged potential
$\overline v(z)$ for periodically repeated slabs: (a) with periodic
boundary conditions, (b) potential of the dipole layer, (c)
dipole-corrected slabs with vanishing external electric field, and
(d) dipole-corrected slabs with vanishing internal electric field.
From  \protect\onlinecite{Meyer2001}.} \label{fig:slabpotential}
\end{figure}

A problem peculiar to the study of periodically repeated slabs with
polarization along the normal is the appearance of electric fields
in the vacuum. As shown in Fig. \ref{fig:slabpotential}, this occurs
because there is a nonzero macroscopic depolarizing field in the
slab and thus a nonzero potential drop between the two surfaces of
the slab. As the potential drop across the entire supercell must be
zero, this inevitably leads to a nonzero electric field in the
vacuum. Physically, this can be interpreted as an external field
applied uniformly across the unit cell which acts partially to
compensate the depolarizing field. An analogous situation arises for
an asymmetrically terminated slab when the two surfaces have
different work functions. To eliminate the artificial field in the
vacuum, one technique is to introduce a dipole layer in the
mid-vacuum region far away from the slab \cite{Bengtsson1999}. This
can accommodate a potential drop up to a critical value, at which
point electrons begin to accummulate in an artificial well in the
vacuum region (see Fig. \ref{fig:slabpotential}). This approach can
be also be used to compensate the depolarization field in a
perpendicular polarized film, though it may happen that the maximum
field that can be applied is smaller than that needed for full
compensation. Alternatively, by using a first-principles
implementation with a localized basis set, it is possible to perform
computations for isolated slabs and thus avoid not only the spurious
electric fields, but also the interaction between periodic slab
images present even for symmetric nonpolar slabs. Comparison between
results obtained with the two approaches is presented in
\cite{Fu1999}.

In determining the properties of an ultrathin film, the
film-substrate and/or film electrode interface plays a role at least
as important as the free surface. In first-principles calculations,
the atomic and electronic structure of the relevant interface(s) is
most readily obtained by considering a periodic multilayer geometry
identical to that used for computing the structure and properties of
superlattices. To simulate a semi-infinite substrate, the in-plane
lattice constant should be fixed to that of the substrate bulk. The
relaxation of a large number of structural degrees of freedom
requires substantial computer resources, and some strategies for
efficiently generating a good starting structure will be described
in the discussion of specific systems in Section
\ref{subsection:individual}.

Calculations of quantities characterizing the electronic structure
are based on use of the Kohn-Sham one-electron energies and
wavefunctions. Band structures for 1x1 (001) slabs are generally
displayed in the 2D surface Brillouin zone for  $k_{supercell,z}$=0.
One way to identify surface states is by comparison with the
appropriately folded-in bulk bandstructure. Another analysis method
is to compute the partial density of states projected onto each atom
in the slab. From these plots, the dominant character of a state at
a particular energy can be found. In addition, an estimate of
valence band offsets and Schottky barriers at an interface can be
obtained by analyzing the partial density of states for a
superlattice of the two constituents. This is done comparing the
energies of the highest occupied states in the interior of the
relevant constituent layers (because of the band gap problem, the
positions of the conduction bands are computed by using the
experimental bulk bandgaps). This estimate can be refined, as
described in  \onlinecite{Junquera2003}, by computation of the
average electrostatic energy difference between the relevant
constituent layers.

As the computational resources required for full first-principles
calculations even of the simplest slab-vacuum system are
considerable, there is a strong motivation to turn to interatomic
potentials and effective Hamiltonians. Interatomic potentials based
on shell models fitted to bulk structural energetics are generally
directly transferred to the isolated slab geometry, with no changes
for the undercoordinated atoms at the surface. As will be discussed
further below, this approach seems to be successful in reproducing
the relaxations observed in full first-principles studies and has
been applied to far larger supercells and superlattices. In the case
of effective Hamiltonians, it is, at least formally, possible simply
to perform a simulation by removing unit cells and use bulk
interaction parameters for the unit cells in the film. For a more
accurate description, modification of the effective Hamiltonian
parameters for the surface layers is advisable to restore the charge
neutrality sum rule for the film \cite{Ruini1998,Ghosez2000}. In
addition, the effects of surface relaxation would also result in
modified interactions at the surface.

\subsubsection{Overview of systems}

In this section, we present a list of the materials and
configurations that have been studied, followed by a brief overview
of the quantities and properties that have been calculated in one or
more of the reported studies. A more detailed description of the
work on individual systems is provided in the following subsection.

The configurations that have been considered to date in first
principles studies can be organized into several classes. The
simplest configuration is a slab of ferroelectric material
alternating with vacuum; this can be used to investigate the free
surface of a semi-infinite crystal, an unconstrained thin film, or
an epitaxial thin film constrained to match the lattice constant of
an implicit substrate. Specific materials considered included
BaTiO$_3$ ((001) surfaces
\cite{Cohen1996,Cohen1997,Fu1999,Meyer2001,Heifets2001,Krcmar2003}
and (110) surfaces\cite{Heifets2001}) SrTiO$_3$
\cite{Padilla1998,Heifets2001,Heifets2002a,Heifets2002b,Kubo2003},
PbTiO$_3$ \cite{Ghosez2000,Meyer2001,Bungaro2004u}, KNbO$_3$
\cite{Heifets2000b} and KTaO$_3$ \cite{Li2003}. Another type of
configuration of comparable complexity is obtained by replacing the
vacuum by a second material. If this is another insulating
perovskite oxide, the calculation can yield information about
ferroelectric-dielectric (e.g BaTiO$_3$/SrTiO$_3$ \cite{Neaton2003}
and KNbO$_3$/KTaO$_3$ \cite{Sepliarsky2001,Sepliarsky2002}) or
ferroelectric-ferroelectric interfaces and superlattices. This
configuration can also be used to study the interface between the
ferroelectric and a dielectric (nonferroelectric) oxide (e.g.
BaTiO$_3$/BaO and SrTiO$_3$/SrO \cite{Junquera2003}). Replacement of
the vacuum by a conductor simulates a film with symmetrical top and
bottom electrodes, e.g. BaTiO$_3$/SrRuO$_3$\cite{Junquera2003n}.
More complex multilayer geometries including two or more different
materials as well as vacuum layers have been used to simulate
ferroelectric thin film interactions with the substrate (e.g.
PbTiO$_3$/SrTiO$_3$/PbTiO$_3$/vacuum \cite{Johnstonpreprint}, and
with realistic electrodes (e.g. Pt/BaTiO$_3$/Pt/vacuum\cite{Rao1997}
and Pt/PbTiO$_3$/Pt/vacuum \cite{Saipreprint}, as well as the
structure of epitaxial alkaline-earth oxide on silicon, used as a
buffer layer for growth of perovskite oxide films \cite{McKee2003}.

\begin{figure}
\centerline{\epsfig{file=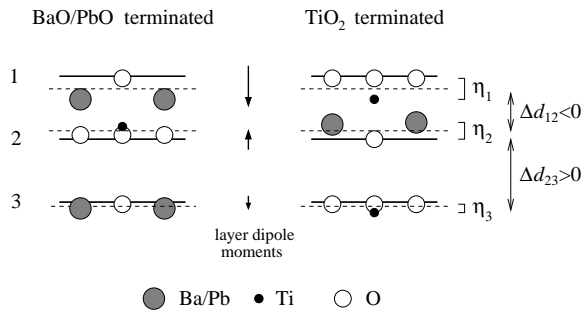,width=3.00in,angle=0}}
\caption{Schematic illustration of the structure of the first three
surface layers. From  \protect\onlinecite{Meyer2001}.}
\label{fig:slabrelax}
\end{figure}

In each class of configurations, there are corresponding quantities
and properties that are generally calculated. For the single
slab-vacuum configuration, for each orientation and surface
termination the surface energy is obtained. While in some of the
initial studies the atomic positions were fixed according to
structural information from the bulk \cite{Cohen1996}, in most
current studies relaxations are obtained by energy minimization
procedures. For the most commonly studied perovskite (001) slab, the
relaxation geometry is characterized by changes in interplane
spacings and rumplings quantified as shown in Fig.
\ref{fig:slabrelax}
Most studies assume bulk periodicity in the plane. For the study of
surface reconstructions, it is necessary to expand the lateral unit
cell, leading to a substantial additional cost in computational
resources. Most attention has been focused on the paraelectric
SrTiO$_3$, although recently studies have been carried out as well
for PbTiO$_3$. A byproduct of the total energy calculation that is
often though not universally presented is the bandstructure and or
density of states; the local density of states at the surface is of
particular interest.

The two-component superlattice configuration can be taken to model a
film on a substrate and/or with electrode layers, or to model an
actual superlattice such as that which can be obtained by techniques
such as MBE. In these studies, the interface and coupling between
the two constituents is of primary interest; combinations most
considered to date are ferroelectric + paraelectric, ferroelectric +
dielectric, and ferroelectric + metal, while the combination of two
ferroelectrics or ferroelectric + ferromagnetic material has been
less intensively investigated. The main questions of interest are
the structural rearrangements at the interface and the change in the
structure, polarization, and related properties of individual layers
relative to the bulk resulting from the interaction with the other
constituents. Analysis of the trends with varying thickness(es) of
the ferroelectric film and, in the superlattice, other constituents
is particularly useful. The electronic structure of these systems
can be most readily characterized by the band offset between the two
materials, which should also control the charge transfer across the
interface, formation of a dipole layer, and the potential difference
between the two constituents. In the case of a metal, this will
determine the type of contact. The existence of interface states is
also very relevant to the physical behavior of the system.

At a considerable increase in computational expense but also in
realism, a system with three or more components can be studied; e.g.
the combination of a substrate, a film, and vacuum. The main
questions of interest in the few such studies to date are the
analysis of the ferroelectric instability in the film, and the
film-induced changes in the substrate layers closest to the
interface. As in two-component heterostructures, the partial density
of states and the layer-average electrostatic potential are also
useful in extracting the electronic behavior of the system.

In all of these studies, one of the main questions is that of the
structure and polarization of the ferroelectric layer compared to
that of the bulk. Certainly, the change of environment (electrical
and mechanical boundary conditions) and the finite dimensions (film
thickness, particle size) are expect strongly to affect the
structure and perhaps to eliminate the ferroelectric instability
entirely. Relevant quantities to examine include the relative
stability of lower and higher symmetry phases, spatial variation in
polarization, changes in the average polarization magnitude and
direction, and the depth and shape of the ferroelectric double-well
potential. These changes can also be expected to lead to changes in
the dielectric and piezoelectric response of thin films and
superlattices, which can be studied theoretically and compared with
experiments.
The implications of the various first-principles studies included in
this review will be described at further length below.

\subsubsection{Studies of individual one component-systems}
\label{subsection:individual}

In this section, we describe a representative sampling of first
principles studies and their results. Most of the literature has
concentrated on BaTiO$_3$, providing a useful comparative test of
various first-principles implementations, as well as a benchmark for
evaluation and analysis of results on other systems. We first
consider the calculations for single slabs of pure material that
focus on the properties of surfaces: surface relaxation, surface
reconstructions and surface states. Depending on the structural
constraints, these calculations are relevant either to the surface
of a semi-infinite bulk, for a free-standing thin film, or for a
thin film epitaxially constrained by a substrate. This will be
followed by discussions of studies of systems with two or more
material components.

\paragraph{BaTiO$_{3}$}

For BaTiO$_3$, full first-principles results have been reported
primarily for the (001) orientation, with a few results for the
(110) and (111) orientations. In the slab-vacuum configuration,
systems up to 10 atomic layers have been considered. The unpolarized
slab is compared with slabs with nonzero polarization, in the plane
and/or along the normal. After reviewing the results on structures,
we will describe the results of extension to larger-scale systems
through the use of interatomic potentials. The discussion of
BaTiO$_3$ will be concluded by description of the first-principles
results for surface electronic structure.

First-principles FLAPW calculations for the BaTiO$_3$ (001) and
(111) slabs were first presented in 1995 \cite{Cohen1996}, and later
extended using the LAPW+LO method \cite{Cohen1997}. The supercells
contained 6 and 7 atomic layers, corresponding to asymmetric
termination and two symmetric terminations (BaO and TiO$_2$), and an
equal vacuum thickness. The central mirror plane symmetry
z$\rightarrow$ -z symmetry is broken for the asymmetric termination
even in the absence of ferroelectric distortion. The primitive cell
lattice constant was fixed at the experimental cubic value 4.01 \AA.
Total energies of several selected paraelectric and ferroelectric
structures were computed and compared: the ideal paraelectric slab
and ferroelectric slabs with displacements along $\hat z$
corresponding to the experimental tetragonal structure. For the
asymmetrically terminated slab, both the ferroelectric structure
with polarization towards the BaO surface (+) and the other towards
the TiO$_2$ surface (-) were considered. The surface layers were
relaxed for the ideal and (+) ferroelectric
asymmetrically-terminated slabs and the ferroelectric BaO-BaO slab.
It was found that the depolarization field of the ferroelectric
slabs indeed strongly destabilizes the ferroelectric state, as
expected, even taking into account the energy lowering due to the
surface relaxation. In all slabs considered, this consists of an
inward-pointing dipole arising from the relative motion of surface
cations and oxygens. The average surface energy of the ideal BaO and
TiO$_2$ surfaces is 0.0574 eV/\AA$^2$=0.923 eV per surface unit
cell.

In  \onlinecite{Padilla1997}, ultrasoft pseudopotential calculations
with fully relaxed atomic coordinates were reported for
symmetrically terminated (both BaO and TiO$_2$) 7-layer BaTiO$_3$
(001) slabs separated by two lattice constants of vacuum. The
in-plane lattice constant was set equal to the theoretical
equilibrium lattice constant $a$ computed for the bulk tetragonal
phase ($a$ = 3.94 \AA). The average surface energy of the ideal BaO
and TiO$_2$ surfaces is 1.358 eV per surface unit cell; at least
part of the difference relative to  \onlinecite{Cohen1996} could be
due to the different lattice constant. Relaxations were reported for
unpolarized slabs and for polarized slabs with polarization along
(100) (in the plane of the slab). Deviations from the bulk structure
were confined to the first few atomic layers. The surface layer
relaxes substantially inwards, and rumples such that the cation (Ba
or Ti) moves inward relative to the oxygens, as in Fig.
\ref{fig:slabrelax}. While the relaxation energy was found to be
much greater than the ferroelectric double well depth, the in-plane
component of the unit-cell dipole moment was seen to be relatively
insensitive to the surface relaxation, with a modest enhancement at
the TiO$_2$ terminated surface and a small reduction at the BaO
terminated surface. The relative stability of BaO and TiO$_2$
terminations were compared and both found to be stable depending on
whether the growth was under Ba-rich or Ti-rich conditions.

This investigation was extended in  \onlinecite{Meyer2001} to
7-layer and 9-layer polarized slabs with polarization along the
normal. The problem of the artificial vacuum field in this
periodically repeated slab calculation was addressed by the
techniques of introducing an external dipole layer in the vacuum
region of the supercell described above in Section
\ref{subsection:methodology}. This technique can also be used to
generate an applied field that partially or fully compensates the
depolarization field for BaTiO$_3$ slabs. As a function of applied
field, the change in structure can be understood as arising from
oppositely-directed electrostatic forces on the positively charged
cations and negatively charged anions, leading to corresponding
changes in the rumplings of the atomic layers and field-induced
increases of the layer dipoles. Analysis of the internal electric
field as a function of the applied field allows a determination of
whether the slab is paraelectric or ferroelectric. The
BaO-terminated slab is clearly ferroelectric, with vanishing
internal electric field at an external field of 0.05 a.u. and a
polarization of 22.9 $\mu C cm^{-2}$, comparable to the bulk
spontaneous polarization. The ferroelectric instability is
suppressed in the TiO$_2$-terminated slab, which appears to be
marginally paraelectric.

The Hartree-Fock method was used in \textcite{Cora99} and
\textcite{Fu1999}. In \textcite{Cora99}, a detailed analysis of the
bonding was performed using tightbinding parametrization. For the
7-layer BaO-terminated slab, the reported displacement of selected
Ti and O atoms is in good agreement with the results of
\textcite{Padilla1997}, and these calculations were extended to
slabs of up to 15 layers. Fu et al. \cite{Fu1999} performed
Hartree-Fock calculations for slabs of two to eight atomic layers,
with symmetric and asymmetric terminations. Using a localized basis
set, they were able to perform calculations for isolated slabs as
well as periodically repeated slabs. Calculations of the
macroscopically averaged planar charge density, surface energy and
surface dynamical charges were reported as a function of thickness
and termination for a cubic lattice constant of 4.006 \AA. The
relative atomic positions were fixed to their bulk tetragonal
structure values (note that this polarized structure in both
isolated and periodic boundary conditions has a very high
electrostatic energy and is not the ground state structure). This
would significantly affect the comparison of the computed surface
properties with experiment. In particular, it is presumably
responsible for the high value of the average surface energy
reported (1.69 eV per surface unit cell). However, a useful
comparison between isolated and periodic slabs is possible. It was
found that the surface charge and surface dipoles of isolated slabs
converge quite rapidly as a function of slab thickness and can be
used, combined with a value of $\epsilon_\infty$ taken from the
bulk, to extract a spontaneous polarization of 0.245 C/m$^2$
(corrected to zero field using the electronic dielectric constant
$\epsilon_\infty = 2.76$), to be compared with 0.240 C/m$^2$ from a
Berry-phase calculation. This is only slightly less than the bulk
value of 0.263 C/m$^2$ taken from experiment. The average surface
energy for the two terminations of symmetrically terminated slabs is
0.85 eV per surface unit cell. Surface longitudinal dynamical
charges differ considerably from bulk values, satisfying a sum rule
that the dynamical charges at the surface planes add up to half of
the corresponding bulk value \cite{Ruini1998}. Convergence of all
quantities with slab and vacuum thickness of periodically repeated
slabs was found, in comparision, to be slow, with significant
corrections due to the fictitious field in the vacuum (for polarized
slabs) and the interaction between slab images.

The isolated slab was also the subject of an FLAPW study
\cite{Krcmar2003}. The symmetric TiO$_2$-TiO$_2$ (9-layer) and
asymmetric TiO$_2$-BaO (10-layer) slabs were considered in a
paraelectric structure with $a$ fixed to 4.00 \AA~ and a polar
tetragonal structure with $a$ and $c$ = 4.00 and 4.04 \AA,
respectively. For the cubic TiO$_2$-terminated slab, displacements
in units of $c$ for the surface Ti, surface O, subsurface Ba and
subsurface O are -0.021, +0.007, +0.022 and -0.009 c, to be compared
with the results\cite{Padilla1997} (-0.0389,-0.0163,+0.0131,-0.0062)
for a periodically repeated 7-layer slab with lattice constant 3.94
\AA. The tetragonal phase was relaxed to a convergence criterion of
0.06 eV/\AA~ on the atomic forces; the rumplings of the layers
follow overall the same pattern as that reported in
\onlinecite{Meyer2001}, with an inward-pointing surface dipole
arising from surface relaxation, though the reduction of the
rumpling in the interior is not as pronounced as for the
zero-applied field case in  \onlinecite{Meyer2001}. The energy
difference between the paraelectric and ferroelectric slabs was not
reported.

\begin{figure}
\centerline{\epsfig{file=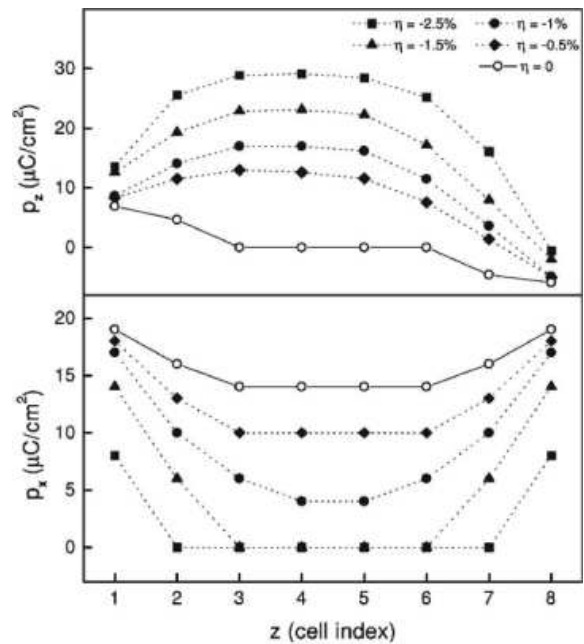,width=3.00in,angle=0}}
\caption{Cell-by-cell out-of-plane (top panel) and in-plane (bottom
panel) polarization profiles of a randomly chosen chain
perpendicular to the slab surface for different misfit strains
$\eta$ at T = 0 K. In the in-plane polarization profiles $p_y=p_x$.
From  \protect\onlinecite{Tinte2001}.} \label{fig:profile}
\end{figure}

With interatomic potentials, it is possible to study additional
aspects of surface behavior in BaTiO$_3$ thin films and
nanocrystals. The most important feature of interatomic studies of
thin films relative to full first-principles calculations is the
relative ease of extending the supercell in the lateral direction,
allowing the formation of 180 degree domains and molecular dynamics
studies of finite temperature effects. In \textcite{Tinte2001}, a
15-layer TiO$_2$-terminated slab periodically repeated with a vacuum
region of 20 \AA~was studied, using interatomic potentials that had
previously been benchmarked against first-principles surface
relaxations and energies\cite{Tinte2000}. The unconstrained
(stress-free) slab is found to undergo a series of phase transitions
with decreasing temperature, from a paraelectric phase to
ferroelectric phases, first with polarization in the plane along
(100), and then along (110). Enhancement of the surface polarization
at low temperatures appears to be linked to the existence of an
intermediate temperature regime of surface ferroelectricity. For
slabs with a strongly compressive epitaxial strain constraint
($\eta$ = -2.5\%), there is a transition to a ferroelectric state
with 180 degree domains and polarization along the surface normal.
At the surface, the polarization has a nonzero x component and a
reduced z-component, giving a rotation at the surface layer. The
width of the stripe domains cannot be determined, as it is limited
by the lateral supercell size at least up to 10x10. Reducing the
compression to $\eta$ = -1.0\% also gives a 180 degree domain
structure in the z component of the polarization, combined with a
nonzero component along [110] in the interior of the film as well as
the surface, as can be see from examination of Fig.
\ref{fig:profile}.
The thickness dependence of the transition temperature to this
ferroelectric domain phase was studied at $\eta$ = -1.5\%, with
T$_c$ decreasing from the 15-layer film to the 13-layer and lowest
T$_c$ 11-layer film. At and above a critical thickness of 3.6 nm,
stress-free films exhibit the same ferroelectric-domain ground-state
structure.

Heifets and coworkers \cite{Heifets2000,Heifets2001} studied BTO
(001) and (110) surfaces of an isolated slab using the shell model
of  \onlinecite{Heifets2000}. Between 1 and 16 atomic planes were
relaxed in the electrostatic potential of a a rigid slab of 20
atomic planes whose atoms were fixed in their perfect (presumably
cubic) lattice sites. The relaxed structures of the two (001)
terminations are in good agreement with other calculations, except
for the sign of the surface dipole in the relaxation of the BaO
terminated surface, which is found to be positive (though small).
The (110) surfaces are found to have much higher surface energy,
except for the relaxed ``asymmetric O-terminated" surface where
every second surface O atom is removed and the others occupy the
same sites as in the bulk structure. In this structure,
displacements of cations parallel to the surface are found
substantially to lower the surface energy. A similar study of the
KNbO$_3$ (110) surface \cite{Heifets2000b}, which is the surface of
most experimental interest for this 1+/5+ perovskite, showed very
strong relaxations extending deep below the surface, consistent with
suggestions that this surface has a complicated chemistry.

\begin{figure}
\centerline{\epsfig{file=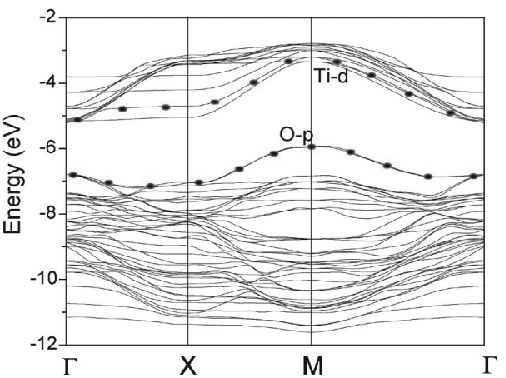,width=3.00in,angle=0}}
\centerline{\epsfig{file=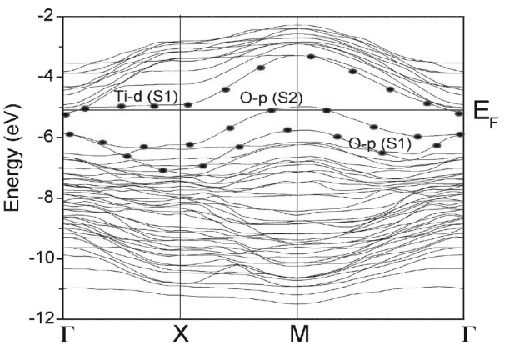,width=3.00in,angle=0}}
\caption{(a) Paraelectric phase energy-band structure of nine-layer
slab of BaTiO$_3$. The surface states in (or near) the band gap are
highlighted. (b) As in (a), for the ferroelectric nine-layer slab.
From  \protect\onlinecite{Krcmar2003}.} \label{fig:chargetransfer}
\end{figure}

Next, we consider results on the electronic structure of BaTiO$_3$
films, particularly the surface states. In the FLAPW study of Cohen,
the band gap of the ideal slab is is found to be reduced from the
bulk. A primarily O $p$ occupied surface state on the TiO$_2$
surface was identified at M \cite{Cohen1996}, with a primarily Ti
$d$ surface state near the bottom of the conduction band. Analysis
of the ferroelectric BaO-terminated slab showed that the macroscopic
field resulted in a small charge transfer to the subsurface Ti $d$
states from the O $p$ and Ba $p$ states at the other surface, making
the surfaces metallic. Further study of this effect in
\onlinecite{Krcmar2003} showed that for symmetric TiO$_2$ 9-layer
slab the ferroelectric distortion similarly shifts the top surface
Ti states and bottom surface O state toward the bulk midgap as in
Fig. \ref{fig:chargetransfer}, resulting in a small charge transfer
and a metallic character for the surfaces.

\paragraph{PbTiO$_{3}$}

In contrast to the numerous papers on calculations on the surfaces
of BaTiO$_3$, there are relatively few for the related material
PbTiO$_3$. Regarding surface relaxations and energies, it was found
in  \onlinecite{Meyer2001} that the two compounds are quite similar.
In perpendicularly polarized films, it seems that both terminations
give ferroelectric films if the depolarization field is compensated,
consistently with the stronger ferroelectric instability of
PbTiO$_3$ and the microscopic model analysis of
\onlinecite{Ghosez2000}, the latter not including the effects of
surface relaxation. Because of the larger spontaneous polarization
of PbTiO$_3$, it is not possible fully to compensate the
depolarization field using the dipole-layer technique.

There are important differences between A-site Ba and Pb, which are
evident even for the bulk. While the polarization in BaTiO$_3$ is
dominated by the Ti displacements, Pb off-centering contributes
substantially to the spontaneous polarization of PbTiO$_3$; this can
be linked to the much richer chemistry of Pb oxides compared to
alkaline-earth oxides. One downside is that it is more challenging
to construct accurate interatomic potentials for perovskites with Pb
than with alkaline earth A cations \cite{Sepliarsky2004}. In the
surface, the characteristic behavior of Pb leads to an
antiferrodistortive surface reconstruction of the (001)
PbO-terminated surface \cite{Munkholm2002,Bungaro2004u}.
Specifically, first principles calculations \cite{Bungaro2004u} show
that the reconstruction in the subsurface TiO$_2$ layer occurs only
for the PbO termination and not for TiO$_2$ termination, and also
that if the Pb in the surface layer is replaced by Ba, the
reconstruction is suppressed.

\paragraph{SrBi$_{2}$Ta$_{2}$O$_{9}$}

A first-principles study of an isolated BiO$_2$-terminated slab of
SrBi$_2$Ta$_2$O$_9$ (SBT) one lattice constant thick (composition
SrBi$_2$Ta$_2$O$_{11}$ was reported in  \onlinecite{Tsai2003}.
Spin-polarized calculations showed that such a film would be
ferromagnetic as well as ferroelectric. This intriguing possibility
suggests further investigation.

\paragraph{SrTiO$_{3}$ and KTaO$_{3}$}

In addition to the work on true ferroelectrics, there has been quite
a lot of interest in first-principles studies of surfaces and
heterostructures of incipient ferroelectrics, mainly SrTiO$_3$ and,
to a lesser extent, KTaO$_3$. As these have closely related
properties that can illuminate issues in the ferroelectric
perovskites, we include a description of a few representative
results drawn from the very extensive literature on this subject.

Many of the studies of BaTiO$_3$ discussed above included analogous
calculations for SrTiO$_3$. As already noted in the discussion of
PbTiO$_3$, the (001) surface relaxations and energies of the
nonpolar slab are very similar for all three materials. First
principles surface relaxation for the SrO surface is reported (in
units of a = 3.86\AA) for surface Sr, surface O, subsurface Ti and
subsurface O as -0.057, 0.001, 0.012 and 0.0, respectively
\cite{Padilla1998}, to be compared with the values -0.071, 0.012,
0.016, 0.009 (computed using a shell model with the experimental
lattice constant 3.8969 \AA). Similarly, first principles surface
relaxation for the TiO$_2$ surface is reported for surface Ti,
surface O, subsurface Sr and subsurface O as -0.034, -0.016, +0.025
and -0.005, respectively \cite{Padilla1998}, to be compared with the
values -0.030, -0.017, +0.035 and -0.021. The average energy of the
two surfaces is found to be 1.26 eV per surface unit cell. A
detailed comparison of various Hartree-Fock and density-functional
implementations showed generally good agreement for the surface
relaxation \cite{Heifets2001b}. Inward surface dipoles due to
relaxation are found for both terminations, with the TiO$_2$
termination to be smaller in magnitude than for BaTiO$_3$
\cite{Heifets2000}. The possibility of an in-plane ferroelectric
instability at the surface was examined and it was found to be quite
weak \cite{Padilla1998}. In these studies, the antiferrodistortive
instability exhibited by bulk SrTiO$_3$ at low temperatures was
suppressed by the choice of a 1x1 in-plane unit cell. The surface
electronic band structures show a behavior highly similar to that of
BaTiO$_3$ described above.

For SrTiO$_3$, there is considerable evidence for a wide variety of
surface reconstructions of varying stoichiometry, depending on
conditions such as temperature and oxygen partial pressure as well
as the relative chemical potentials of TiO$_2$ and SrO. Candidate
structures can be obtained by creating vacancies on the surface (for
example, missing rows of oxygen) and adding adatoms \cite{Kubo2003}.
More drastic rearrangements of the surface atoms have also been
proposed, for example a (2x1) Ti$_2$O$_3$ reconstruction
\cite{Castell2002a} and a (2x1) double-layer TiO$_2$ reconstruction
with edge-sharing TiO$_6$ octahedra \cite{Erdman2002a}. As in the
case of semiconductor surface reconstructions, first-principles
calculations of total energies are an essential complement to
experimental structural determination, and can also be used to
predict scanning tunnelling microscope images for comparison with
experiment \cite{Johnston2004a}. Even so, there are still many open
questions about the atomic and electronic structure of SrTiO$_3$
surfaces under various conditions. The same applies to KTaO$_3$; the
structures and lattice dynamics of a variety of (1x1) and (2x1)
surface structures were studied in  \cite{Li2003}).

As previously mentioned, the theoretical and experimental literature
on SrTiO$_3$ is so extensive that it would require a review paper in
its own right to cover it fully. Since, if anything, we expect the
ferroelectric instability in related systems such as BaTiO$_3$ and
PbTiO$_3$ to make the physics more, not less, complicated, this
suggests that we have only scratched the surface in developing a
complete understanding of the surfaces of perovskite ferroelectrics
and the resulting effects on thin film properties.

\subsubsection{Studies of individual heterostructures}

Now we turn to the description of studies of systems with two or
more material components. The main structural issues are the
rearrangements at the interface, the change in electrical and
mechanical boundary conditions felt by each constituent layer, and
how these changes modify the ferroelectric instability exhibited by
the system as a whole. This geometry also allows the calculation of
band offsets and/or Schottky barriers, crucial in principle to
understanding the electronic behavior (though with the caveat that
the measured Schottky barrier in real systems will be strongly
influenced by effects such as oxidation of the electrodes that are
not included in the highly idealized geometries studied
theoretically). The current state of knowledge, derived from
experimental measurements, is described in Section.
\ref{subsection:schottky}.

The first combination we discuss is that of a ferroelectric thin
film with metallic electrodes. Transition metal interfaces with
nonpolar BaO-terminated layers of BaTiO$_3$ were studied in
\onlinecite{Rao1997}, specifically systems of 3 and 7 atomic layers
of BaTiO$_3$, with lattice constant set to the bulk value of 4.00
\AA, combined with top and bottom monolayers of Ta, W, Ir and Pt
representing the electrodes. The preferred absorption site for the
metal atoms was found to be above the O site, with calculated metal
oxygen distances ranging from 2.05 \AA~for Ta to 2.11 \AA~for Pt.
The BaTiO$_3$ slabs were assumed to retain their ideal cubic
structure. Analysis of the partial density of states of the
heterostructure shows that the Pt and Ir Fermi energies lie in the
gap of the BaTiO$_3$ layer at 0.94 eV and 0.64 eV, respectively,
above the top of the valence band (this is, fortunately, smaller
than the underestimated computed gap of 1.22 eV for the BaTiO$_3$
slab). Using the experimental gap of 3.13 eV, a Schottky barrier
height of 2.19 eV for Pt and 2.49 eV for Ir is thus obtained.
Experimentally, however, the Schottky barrier is known to be
substantially {\it lower} for Ir than for Pt, illustrating the
limitations mentioned in the previous paragraph.

In  \onlinecite{Robertson1999}, first principles calculations of the
charge neutrality levels were combined with experimental values of
the band gap, electronic dielectric constant $\epsilon_\infty$, the
electron affinity, and the empirical parameter S, described above in
Section III.B 3. Values for SrTiO$_3$ and PZT were reported for Pt,
Au, Ti, Al and the conduction and valence bands of Si.

To explore how the electrodes affect the ferroelectric instability
of the film, Junquera and Ghosez considered a supercell of 5 unit
cells of metallic SrRuO$_3$ and 2-10 unit cells of BaTiO$_3$, with a
SrO/TiO$_2$ interface between the SrRuO$_3$ and the BaTiO$_3$. A
SrTiO$_3$ substrate was treated implicitly by constraining the
in-plane lattice constant of the supercell to that of bulk
SrTiO$_3$. For each BaTiO$_3$ thickness, the system was relaxed
assuming a nonpolar state for the BaTiO$_3$, and then the energy of
the bulk-like tetragonal distortion was computed as a function of
overall amplitude of the distortion. Above a critical thickness of
six unit cells, this distortion lowered the energy, demonstrating
the development of a ferroelectric instability. As will be discussed
further in section \ref{subsection:lessons}, this finite size effect
can be largely understood by considering the imperfect screening in
the metal layers.

Considerable first-principles effort has been devoted to
investigating various aspects of epitaxial ferroelectric thin films
on Si. As perovskite oxides cannot be grown directly on Si, an
approach developed by McKee and Walker \cite{McKee1998,McKee2001} is
to include a AO buffer layer, which apparently also results in the
formation of a silicide interface phase. The constituent layers of
this heterostructure should thus be considered to be
Si/ASi$_2$/AO/ABO$_3$. The full system has not been simulated
directly, but first-principles approaches have been used to
investigate individual interfaces. The importance of relaxations,
the additional role of the buffer layer in changing the band offset,
and the analysis of electronic structure within the LDA are
illustrated by the following.

\begin{figure}
\centerline{\epsfig{file=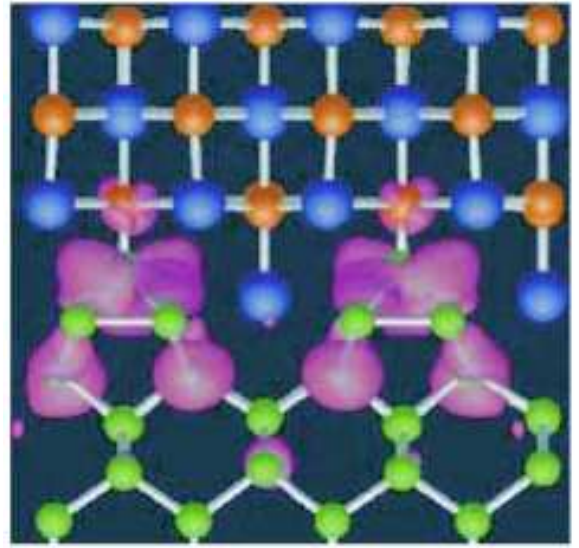,width=3.00in,angle=0}}
\caption{The image illustrates three layers of the alkaline earth
oxide on the (001) face of silicon observed in a cross section at
the [110] zone axis (blue, alkaline earth metal; yellow, oxygen; and
green, silicon). A distinct interface phase can be identified as a
monolayer structure between the oxide and the silicon in which the
charge density in interface states is strongly localized around the
silicon atoms in the interface phase. The dipole in the ionic A-O
bond between the alkaline earth metal in the silicide and the oxygen
in the oxide buffers the junction against the electrostatic
polarization of the interface states localized on silicon. The
electron density of this valence surface state at the center of the
Brillouin zone is shown with the purple isosurface (0.3 10$^{-3}$
e). From  \protect\onlinecite{McKee2003}} \label{fig:AObond}
\end{figure}

In \cite{McKee2003}, first-principles results are presented for the
atomic arrangements and electronic structure in the Si/ASi$_2$/AO
system, in conjunction with an experimental study. A strong
correlation is found between the valence band offset and the dipole
associated with the A-O bond linking the A atom in the silicide to
the O atom in the oxide, shown in Fig. \ref{fig:AObond}. It is thus
seen that the structural rearrangements in the interface are a key
determining factor in the band offset.

\begin{figure}
\centerline{\epsfig{file=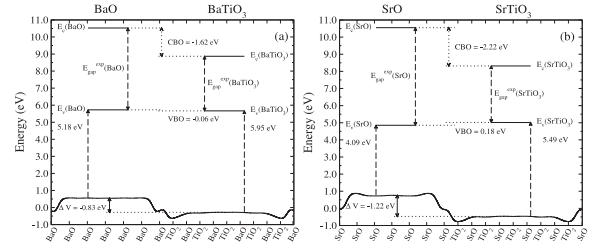,width=3.00in,angle=0}}
\caption{Schematic representation of the valence-band offset (VBO)
and the conduction-band offset (CBO) for BaO/BaTiO$_3$ (a) and
SrO/SrTiO$_3$ (b) interfaces. E$_v$, E$_c$, and E$_{gap}^{expt}$
stand for the top of the valence band, the bottom of the conduction
band, and the experimental band gap, respectively. Values for E$_v$
, measured with respect to the average of the electrostatic
potential in each material, are indicated. The solid curve
represents the profile of the macroscopic average of the total
electrostatic potential across the interface. $\Delta V$ stands for
the resulting lineup. The in-plane lattice constant was set up to
the theoretical one of Si (5.389 Å). The size of the supercell
corresponds to n=6 and m=5. From  \onlinecite{Junquera2003}}
\label{fig:bandoffset}
\end{figure}

A detailed examination of the interface between the perovskite oxide
and the alkaline oxide buffer layer, specifically BaO/BaTiO$_3$ and
SrO/SrTiO$_3$, was carried out in \cite{Junquera2003}. A periodic
1x1x16 supercell was chosen with stacking of (001) atomic layers:
(AO)$_n$-(AO-TiO$_2$)$_m$, with $n$ = 6 and $m$ = 5. Two mirror
symmetry planes were fixed on the cental AO and BO$_2$ layers, and
the in-plane lattice constant chosen for perfect matching to the
computed LDA lattice constant of Si (this epitaxial strain
constraint is the only effect of the Si substrate included in the
calculation). Relaxations within the highest-symmetry tetragonal
space group consistent with this supercell were performed. Analysis
of the partial density of states showed no interface-induced gap
states. The main effect observed for relaxations was to control the
size of the interface dipole, which in turn was found to control the
band offsets, shown here in Figure \ref{fig:bandoffset}.
As in the studies described above, the conduction band offset is
obtained from the computed valence band offset by using the bulk
experimental band gap. These results were combined with offsets
reported for the other relevant interfaces to estimate band
alignments for Si/SrO/SrTiO$_3$/Pt and Si/BaO/BaTiO$_3$/SrRuO$_3$
heterostructures, confirming that the AO layer introduces an
electrostatic barrier of height greater than 1 eV. This is
sufficient to eliminate the carrier injection from the Si into the
conduction band states of the perovskite that would be occur if the
two were in direct contact \cite{Robertson1999}.

\subsubsection{First principles modelling: methods and lessons}
\label{subsection:lessons}

As discussed above, the analysis and prediction of the behavior of
ferroelectric thin films and heterostructures can be carried out
with direct first-principles simulations only for highly idealized
configurations. However, it is possible to consider more complex and
realistic situations by constructing models that incorporate certain
physical ideas about the nature of these systems, with
material-specific parameters determined by fitting to the results of
first-principles calculations carefully selected for a combination
of informativeness and tractability. This modelling approach also
has the advantage of providing a conceptual framework for organizing
the vast amount of microscopic information in large-scale
first-principles calculations, and communicating those results,
particularly to experimentalists. This will not diminish in
importance even as such calculations become easier with continuing
progress in algorithms and computer hardware.

For successful modeling of measurable physical properties, the film
must be considered as part of a system (substrate+film+electrode) as
all components of the system and their interaction contribute to
determine properties such as the switchable polarization and the
dielectric and piezoelectric response. We start by considering the
class of first-principles models in which the constituent layers
(film, superlattice layers, electrodes and substrate) are assumed to
be subject to macroscopic electric fields and stresses resulting
from the combination of applied fields and stresses and the effects
of the other constituents, with the responses of the layers being
given to lowest order by the bulk responses. For systems with
constituent layer thicknesses as low as one bulk lattice constant,
it seems at first unlikely that such an approach could be useful,
but in practice it has been found to be surprisingly successful.

One simple application of this approach has been used to predict and
analyze the strain in nonpolar thin films and multilayers.  In the
construction of the reference structure for the AO/ABO$_3$
interfaces in \cite{Junquera2003}, macroscopic modelling of the
structure with bulk elastic constants for the constituent layers
yielded accurate estimates for the lattice constants along the
normal direction. In cases of large lattice mismatch, very high
strains can be obtained in very thin films and nonlinear
contributions to the elastic energy can become important. These can
be computed with a slightly more sophisticated though still very
easy-to-implement method that has been developed to study the
effects of epitaixial strain more generally on the structure and
properties of a particular material, described next.

As is discussed further in Section \ref{section:strain}, the effects
of epitaxial strain in ultrathin films and heterostructures have
been identified as a major factor in determining
polarization-related properties, and have been the subject of
intense interest in both phenomenological and first principles
modelling. In particular, for ferroelectric perovskite oxides it has
long been known that there is a strong coupling between strain (e.g.
pressure-induced) and the ferroelectric instability, as reflected by
the frequency of the soft mode and the transition temperature. In
both phenomenological and first-principles studies, it has become
common to study the effects of epitaxial strain induced by the
substrate by studying the structural energetics of the strained
bulk. Specifically, two of the lattice vectors of a bulk crystal are
constrained to match the substrate and other structural degrees of
freedom are allowed to relax. as described in the previous
paragraph. In most cases, these calculations are performed for zero
macroscopic electric field, as would be the case for a film with
perfect short-circuited electrodes. Indeed, it is often the case
\cite{Pertsev1998,Junquera2003,Neaton2003} that the strain effect is
considered to be the dominant effect of the substrate, which is
otherwise not included (thus greatly simplifying the calculation).
At zero temperature, the sequence of phases and phase boundaries can
be readily identified as a function of in-plane strain directly
through total-energy calculations of the relaxed structure subject
to the appropriate constraints. Atomic-scale information can be
obtained for the precise atomic positions, bandstructure, and phonon
frequencies and eigenvectors. The temperature axis in the phase
diagram can be included by using effective Hamiltonian (or
interatomic potential) simulations. Results for selected perovskite
oxides are discussed in Section \ref{section:strain}; a similar
analysis was reported for TiO$_2$ in  \cite{Montanari2004}.

This modelling is based on the assumption that the layer stays in a
single-domain state. As discussed in Section \ref{section:strain},
the possibility of strain relaxation through formation of
multidomain structures must be allowed for. While this cannot be
readily done directly in first-principles calculations,
first-principles data on structural energetics for large misfit
strains could be used to refine Landau parameters for use in
calculations such as those in \onlinecite{Speck1994, Alpay1998,
Bratkovsky2001, Li2003}. The effects of inhomogeneous strain due to
misfit dislocations that provide elastic relaxation in thicker films
have been also been argued to be significant.

Next we consider the application of these ``continuum" models to
analyzing structures in which the macroscopic field is allowed to be
nonzero. Macroscopic electrostatics is applied to the systems of
interest by a coarse-graining over a lattice-constant-scale window
to yield a value for the local macroscopic electric potential.
Despite the fact that this is not strictly within the regime of
validity of the classical theory of macroscopic electrostatics,
which requires slow variation over many lattice constants, this
analysis turns out to be remarkably useful in the analysis of first
principles results. In the simplest example, the polarization of an
polar BaTiO$_3$ slab (periodically repeated in a supercell with
vacuum) is accurately reproduced using bulk values for the bulk
spontaneous polarization and electronic dielectric constant even for
slabs as thin as two lattice constants \cite{Junqueraunpub}.

We have already mentioned in the previous section that perpendicular
(to the surface) polarization can lead to a nonzero macroscopic
field that opposes the polarization (the depolarizing field). Unless
compensated by fields from electrodes or applied fields, this
strongly destabilizes the polarized state. In systems with two or
more distinct constituent layers, this condition in the absence of
free charge favors states with $\nabla \cdot \vec P$ = 0. For
example, in the first-principles calculations of short-period
BaTiO$_3$/SrTiO$_3$ superlattices, the local polarization along
[001] is found to be quite uniform in the two layers
\cite{Neaton2003,Johnstonpreprint}, though the in-plane component
can be very different \cite{Johnstonpreprint}. To the extent that
layered ferroelectrics such as SBT can be treated in this
macroscopic framework, one similarly expects that polarization along
c will tend to be energetically unfavorable, since the layers
separating the polarized perovskite-like layers typically have low
polarizability \cite{Fennie}. This observation provides a
theoretical framework for evaluating claims of large ferroelectric
polarization along c in layered compounds \cite{Chon2002}; unless
there is an unusually high polarizability for the non-perovskite
layers, or a strong competing contribution to the energy due, for
example, to the interfaces to help stabilize a high c polarization,
other reasons for the observations need to be considered
\cite{Garg2003}.

These considerations become particularly important for ultrathin
films with metal electrodes. The limiting case of complete screening
of the depolarizing field by perfect electrodes is never realized in
real thin film systems. The screening charge in real metal
electrodes is spread over a characteristic screening length and is
associated with a voltage drop in the electrode. For thick films,
this can be neglected, but the relative size of the voltage drop
increases as the film thickness decreases. This has been identified
as a dominant contribution to the relation between the applied field
and the true field in the film for the thinnest films
\cite{Dawber03JPC}. One way this shows up is in the thickness
dependence of the apparent coercive field; it is found that the true
coercive field scales uniformly down to the thinnest films. Effects
are also expected on the structure and polarization. While films
with partial compensation of the depolarization field may still
exhibit a ferroelectric instability, the polarization and the energy
gain relative to the nonpolar state are expected to decrease. This
simple model was developed and successfully used in
\cite{Junquera2003n} to describe the thickness dependence of the
ferroelectric instability in a BaTiO$_3$ film between SrRuO$_3$
electrodes. This analysis identified the thickness dependence of the
residual depolarization field as the principal source of thickness
dependence in this case. In  \onlinecite{Trisconepreprint}, the
reduction of the uniform polarization by the residual field and its
coupling to tetragonal strain was suggested to be the cause of the
decrease in tetragonality with decreasing thickness of PbTiO$_3$
ultrathin films.

It is well known that 180 degree domain formation provides an
effective mechanism for compensating the depolarization field, and
is expected to be favored when the screening available from
electrodes is poor or nonexistent (for example, on an insulating
substrate \cite{Streiffer2002}). Instability to domain formation is
discussed in  \onlinecite{Bratkovsky2000} as the result of a nonzero
residual depolarization field due to the presence of a passive
layer. Similarly, a phase transition from a uniform polarized state
to a 180 degree domain state with zero net polarization is expected
to occur with decreasing thickness \cite{JGRunpub}.

Despite the usefulness of macroscopic models, it should not be
forgotten that they are being applied far outside the regime of
their formal validity (i.e length scales of many lattice constants)
and that atomistic effects can be expected to play an important
role, especially at the surfaces and interfaces. The structural
energetics could be substantially altered by relaxations and
reconstructions (atomic rearrangements) at the surfaces and
interfaces. These relaxations and reconstructions are also expected
to couple to the polarization \cite{Meyer2001,Bungaro2004u}, with
the possibility of either enhancing or suppressing the switchable
polarization. The surfaces and the interfaces will also be primarily
responsible for the asymmetry in energy between up and down
directions for the polarization. For ultrathin films, the surface
and interface energy can be important enough to dominate over
elastic energy, leading to a possible tradeoff between lattice
matching and atomic-scale matching for favorable bonding at the
interface. These surface and interface energies could even be large
enough to stabilize non-bulk phases with potentially improved
properties. This should be especially significant for interfaces
between unlike materials.

Electronic states associated with surfaces and interfaces will also
contribute to determining the equilibrium configuration of electric
fields and polarizations. In the simple example of periodically
repeated slabs separated by vacuum, as the slab gets thicker, a
breakdown is expected where the conduction band minimum on one
surface of the slab falls below the valence band maximum on the
other. In this case, charge will be transferred across the slab,
with the equilibrium charge (for fixed atomic positions) being
determined by a combination of the macroscopic electrostatic energy
and the single particle density of states. This tends partially to
screen the depolarization field. The role of interface states in
screening the depolarization field in the film has been discussed in
a model for BaTiO$_3$ on Ge \cite{Reiner2004}. The presence of
surface and interface states can be established by examination of
the bandstructure and PDOS, as discussed in the previous section.

Finally, we turn our attention to an analysis of what the discussion
above tells us about finite-size effects in ferroelectric thin
films. We have seen that many factors contribute to the thickness
dependence of the ferroelectric instability: the thickness
dependence of the depolarization field, the gradual relaxation of
the in-plane lattice constant from full coherence with the substrate
to its bulk value and the changing weight of the influence of
surfaces and interfaces. The ``true" finite-size effect, i.e the
modification of the collective ferroelectric instability due to the
removal of material in the film relative to the infinite bulk. could
possibly be distangled from the other factors by a
carefully-designed first principles calculation, but this has not
yet been done. We speculate that this effect does not universally
act to suppress ferroelectricity, but could, depending on the
material, enhance ferroelectricity \cite{Ghosez2000}.

\subsubsection{Challenges for first principles modelling}
\label{subsubsection:challenges}

First principles calculations have advanced tremendously in the last
decade, to the point where systems of substantial chemical and
structural complexity can be addressed, and a meaningful dialogue
opened up between experimentalists and theorists. With these
successes, the bar gets set ever higher, and the push is now to make
the theory of ferroelectrics truly realistic. The highest long-term
priorities include making finite temperature calculations routine,
proper treatment of the effects of defects and surfaces, and the
description of structure and dynamics on longer length and time
scales. In addition, there are specific issues that have been raised
that may be addressable in the shorter term through the interaction
of theory and experiment, and the rest of this section will
highlight some of these.

Many applications depend on the stability of films with a uniform
switchable polarization along the film normal. This stability
depends critically on compensation of the depolarization field.
Understanding and controlling the compensation mechanism(s) are thus
the subjects of intense current research interest. There are two
main classes of mechanism: compensation by ``free" charges (in
electrodes/substrate or applied fields) and compensation by the
formation of polarization domains. On insulating substrates, this
latter alternative has been observed and characterized in ultrathin
films \cite{Streiffer2002,Fong2004}; it has been proposed that
domain formation will occur in films on conducting substrates at
very low thicknesses as well where the finite screening length in
realistic electrodes inhibits that mechanism of compensation
\cite{JGRunpub}. The critical thickness for this instability depends
on the domain wall energy. This is expected to be different in thin
films than in bulk, one factor being that the bulk atomic plane
shifts across the domain walls.

Compensation of the depolarization field by free charges appears to
be the dominant mechanism in films on conducting substrates (even
relatively poor conductors), with or without a top electrode. In the
latter case there must be free charge on the surface; the challenge
is to understand how the charge is stabilized. There are also
unresolved questions about how the charge is distributed at the
substrate-film interface, and how this couples to local atomic
rearrangements. Asymmetry of the compensation mechanism may prove to
be a significant contribution to the overall up-down asymmetry in
the film discussed in the previous section. A better understanding
could lead to the identification of system configurations with more
complete compensation and thus an enhancement of stability.

The study of the behavior of ferroelectrics in applied electric
fields also promises progress in the relatively near future.
Recently with the solution of long-standing questions of principle,
it has become possible to perform DFT calculations for crystalline
solids in finite electric fields \cite{Souza2002}. In
ferroelectrics, this allows the investigation of nonlinearities in
structure and polarization at fields relevant to experiments, and
the possibility of more accurate modelling of constituent layers of
thin film and superlattice systems subject to nonzero fields. It is
also of interest to ask what the intrinsic breakdown field would be
in the absence of defects, though the question is rather academic
with respect to real systems.

The nonzero conductivity of real ferroelectrics becomes particularly
important for thinner films, both since a higher concentration of
free carriers is expected to be associated with characteristic
defects in the film, and because a given concentration of free
carriers will have a more significant impact as thickness decreases.
Free carriers can at least partially screen macroscopic electric
fields. At the macroscopic level, the concepts of band bending and
space charge arising in semiconductor physics can be applied to thin
film ferroelectrics, while a correct atomic-scale treatment of this
effects could be important to describing the behavior of ultrathin
films.

The physics of switching presents a significant challenge, requiring
description of structure and dynamics on long length and time
scales. The questions of what changes, if any, occur in switching as
films become thinner continue to be debated. The possibility of
switching as a whole rather than via a domain-wall mechanism has
been raised for ultrathin films of PVDF \cite{Bune98}, while a
different interpretation has been offered in
\onlinecite{Dawber03JPC}. Some progress has been made using
interatomic potentials for idealized defect-free films, though real
systems certainly are affected by defects responsible for such
phenomena as imprint and fatigue. Ongoing comparison of
characteristics such as coercive fields, time scales, material
sensitivity, and thickness dependence of domain wall nucleation,
formation energy, and motion with experimental studies promise that
at least some of these issues will soon be better understood.

To conclude this section, we emphasize that it is not very realistic
to expect first-principles calculations quantitatively to predict
all aspects of the behavior of chemically and structurally complex
systems such as ferroelectric thin films, although successful
predictions should continue to become increasingly possible and
frequent. Rather, the quantitative microscopic information and the
development of a useful conceptual framework contribute in a close
interaction with experiment to build an understanding of known
phenomena and to propel the field into exciting new directions.

\section{STRAIN EFFECTS} \label{section:strain}
Macroscopic strain is a important factor in determining the
structure and behavior of very thin ferroelectric films. The primary
origin of homogeneous film strain is lattice mismatch between the
film and the substrate. In addition, defects characteristic of thin
films can produce inhomogeneous strains that can affect the
properties of thicker relaxed films of technological relevance.
Because of the strong coupling of both homogeneous and inhomogeneous
strains to polarization, these strains have a substantial impact on
the structure, ferroelectric transition temperatures and related
properties such as the dielectric and piezoelectric responses, which
has been the subject of extensive experimental and theoretical
investigation.

The largest effects are expected in coherent epitaxial films. These
films are sufficiently thin that the areal elastic energy density
for straining the film to match the substrate at the interface is
less than the energy cost for introducing misfit dislocations to
relax the lattice parameters back towards their unconstrained
equilibrium values. (We note that for ultrathin films, the relaxed
in-plane lattice constant will not in general not be the same as the
bulk lattice constant, and the former is more appropriate for
computing lattice mismatch \cite{Rabeunpub}.) Very high homogeneous
strains, of the order of 2\%, are achievable. For example, BTO films
on STO, with a bulk mismatch of 2.2\%, remain coherent in
equilibrium up to a critical thickness of 2-4 nm \cite{Sun2004}.
With low-temperature growth techniques, the formation of misfit
dislocations is kinetically inhibited and coherent films can be
grown to thicknesses of two to three times the critical thickness
\cite{Choi2004}. Even these films, however, are much thinner than
the minimum 120 nm thickness for films used in contemporary
applications, and thus much of the discussion in this section is at
present primarily of fundamental rather than technological interest.

The structure of a coherent film can be a single-crystal monodomain
structure, a polydomain structure, or even possibly multiphase. We
discuss the simplest single-crystal monodomain case first. The phase
diagram as a function of in-plane strain will in general include
lower symmetry phases due to the symmetry-breaking character of the
epitaxial constraint. A nomenclature for these phases of perovskites
on a surface with square symmetry (e.g. a perovskite (001) surface))
has been established in  \onlinecite{Pertsev1998}. For example, a
ferroelectric perovskite rhombohedral phase will be lowered to
monoclinic symmetry (called the r-phase). For highly compressive
in-plane strains, coupling between strain and the polarization tends
to favor the formation of a tetragonal phase with polarization along
c (the c-phase) for highly compressive strains. Conversely, highly
tensile strains lead to an orthorhombic phase with polarization
along the cube face diagonal perpendicular to the normal (the
aa-phase) or along the in-plane cartesian direction (the a phase).
As a result of the added constraint, it is possible in principle to
stabilize perovskite-derived phases not observed in bulk, for
example, the monoclinic r-phase and the orthorhombic aa-phase in
PbTiO$_3$. This mechanism also plays a role in the more general
phenomenon of epitaxial stabilization, discussed, for example, in
\onlinecite{Gorbenko2002}.

\begin{figure}
\centerline{\epsfig{file=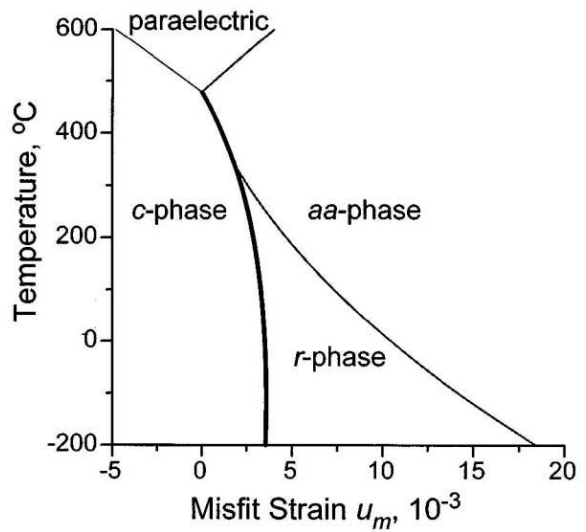,width=3.00in,angle=0}}
\caption{ Phase diagram of a (001) single-domain PbTiO3  thin films
epitaxially grown on different cubic substrates providing various
misfit strains um in the heterostructures. The second- and
first-order phase transitions are shown by thin and thick lines,
respectively.
 From
 \protect\onlinecite{Pertsev1998}.}
\label{fig:Pertsev1998}
\end{figure}

\begin{figure}
\centerline{\epsfig{file=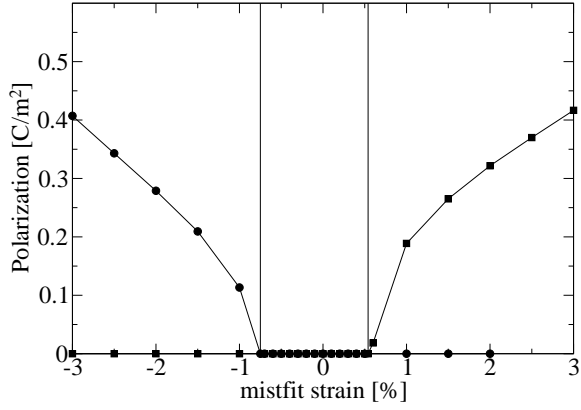,width=3.00in,angle=0}}
\caption{Polarization of SrTiO$3$ as a function of in-plane strain.
Solid circles and squares denote polarization along [001] and [110],
respectively.
 From
 \protect\onlinecite{Antons2004}.}
\label{fig:Antons2004}
\end{figure}

\begin{figure}
\centerline{\epsfig{file=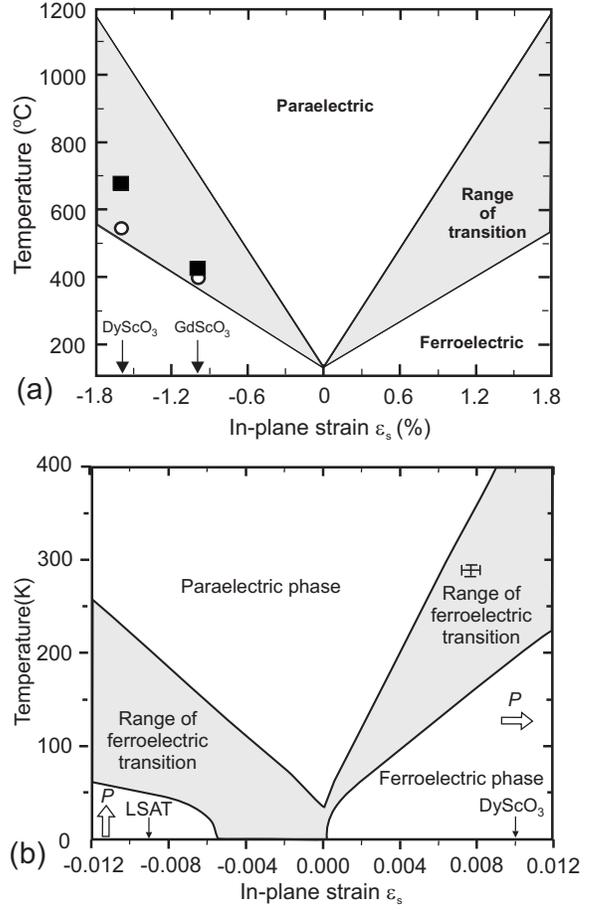,width=3.00in,angle=0}}
\caption{Expected T$_c$ of (a) (001) BaTiO$_3$ (from
\textcite{Choi2004}) and (b) (001) SrTiO$_3$ (from
\textcite{Haeni2004}) based on thermodynamic analysis. The range of
transition represents the uncertainty in the predicted T$_c$
resulting from the spread in reported property coefficients. }
\label{fig:btststrain}
\end{figure}

As discussed in Section \ref{subsection:lessons}, theoretical
analysis of the in-plane strain phase diagrams has focused on
isolating the effects of strain by computing bulk single-crystal
monodomain phase diagrams under the epitaxial constraint and zero
macroscopic electric field, using phenomenological Landau theory or
first principles methods. Phenomenological analysis based on
Landau-Devonshire theory for a number of perovskite oxides has been
presented by Pertsev and co-workers \cite{
Pertsev1998,Pertsev1999,Pertsev2000,PertsevKoukhar2000,Koukhar2001,Pertsev2003},
with temperature-strain diagrams for BaTiO$_3$, PbTiO$_3$ (see Fig.
\ref{fig:Pertsev1998}), SrTiO$_3$, and Pb(Zr,Ti)O$_3$ (PZT), the
latter generalized to include nonzero stress.
First-principles methods have been used to construct a
temperature-strain diagram for BaTiO$_3$ \cite{Dieguez2004} and a
zero-temperature strain diagram for PbTiO$_3$ and ordered PZT
\cite{Bungaro2004a} and SrTiO$_3$ \cite{Antons2004}. These
theoretical phase diagrams have some notable features. In
particular, compressive in-plane strain is found to elevate the
ferroelectric(c)-paraelectric transition temperatures in BaTiO$_3$
and PbTiO$_3$, and tensile in-plane strain elevates the
ferroelectric (aa)-paraelectric transition temperatures. In both
cases, the transition is second order, in contrast to the
first-order transition in bulk. To eliminate a possible source of
confusion, we comment that zero misfit strain as defined in
\cite{Pertsev1998} is not equivalent to an unconstrained film (the
low-temperature bulk phases are in general not cubic, and the
constrained dielectric and piezoelectric responses are clamped, as
will be discussed further below). The nearly vertical morphotropic
phase boundary characteristic of bulk PZT is substantially modified
\cite{LiChoudhury2003,Pertsev2003}. In SrTiO$_3$, ferroelectricity
is found to be induced by both sufficiently compressive and tensile
strains, with a corresponding direction for the spontaneous
polarization (c-type and aa-type) \cite{Pertsev2000,Antons2004}, as
shown in Fig. \ref{fig:Antons2004}.
The enhancement of the polarization in the c-phase by compressive
in-plane strain has been noted for BaTiO$_3$ \cite{Neaton2003} and
PZT \cite{Pertsev2003}. For both strained SrTiO$_3$ and strained
BaTiO$_3$, the ferroelectric T$_c$'s are predicted to increase as
the strain magnitudes increase (Choi 2004, Haeni 2004), as shown in
Fig. \ref{fig:btststrain}.

The use of phenomenological bulk Landau parameters yields a very
accurate description for small strains near the bulk T$_c$. However,
different parameter sets have been shown, for example in the case of
BaTiO$_3$, to extrapolate to qualitatively different phase diagrams
at low temperatures \cite{Pertsev1998,Pertsev1999}. Quantitatively,
the uncertainty in predicted phase boundaries, produced by the
fitting of the Landau theory parameter increases with increasing
misfit strain as shown in Fig. \ref{fig:btststrain} \cite{Choi2004}.
The Landau analysis is thus well-complemented by first-principles
calculations, which can provide very accurate results in the limit
of zero temperature. In the case of BaTiO$_3$, the ambiguity in the
low-temperature phase diagram \cite{Pertsev1998,Pertsev1999} has
been resolved in this way \cite{Dieguez2004}, in the case of
PbTiO$_3$, the phenomenological result is confirmed
\cite{Bungaro2004a}.

The epitaxial-strain induced changes in structure and polarization
are also expected to have a substantial effect on the dielectric and
piezoelectric responses. While overall, the dielectric and
piezoelectric responses should be reduced by clamping to the
substrate \cite{Li2001,Canedy2000}, these responses will tend to
diverge near second-order phase boundaries
\cite{Pertsev2003,Bungaro2004a}
\begin{figure}
\centerline{\epsfig{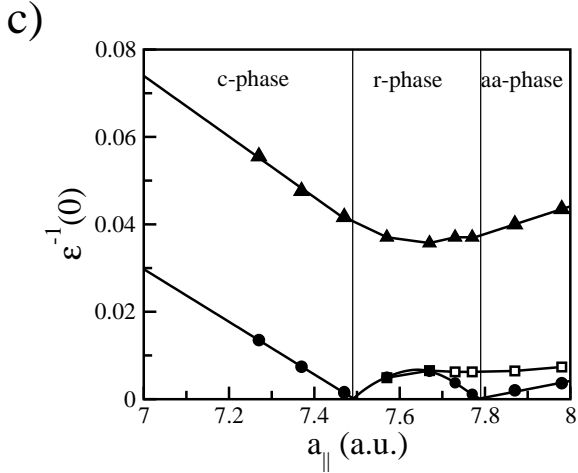}}
\caption{ Inverses of the eigenvalues of the phonon contribution to
the dielectric tensor as a function of the in-plane lattice constant
for the [001]-(PbTiO3)1(PbZrO3)1 superlattice. The lines are a guide
to the eye. From  \protect\onlinecite{Bungaro2004a}.}
\label{fig:Bungaro2004}
\end{figure}
Responses will also be large in phases, such as the r-phase, in
which the direction of the polarization is not fixed by symmetry, so
that an applied field or stress can rotate the polarization
\cite{Wu2003}. This polarization rotation has been identified as a
key mechanism in the colossal piezoresponse of single-crystal
relaxors \cite{Fu2000,Park1997}. The sensitivity of the zero-field
responses should also be reflected in the nonlinear response; thus
the electric-field tunability of the dielectric response can be
adjusted by changing the misfit strains \cite{Chen1997}.

\begin{figure}
\centerline{\epsfig{file=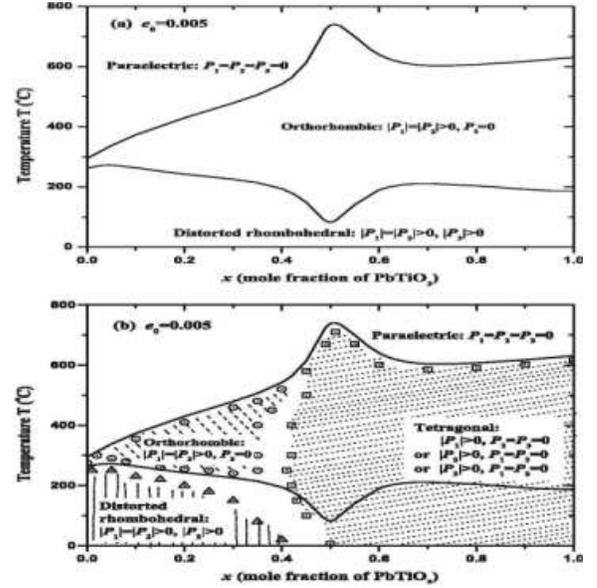,width=3.00in,angle=0}}
\caption{(a)Phase diagram of PZT film under in-plane tensile strain
of 0.005 obtained using thermodynamic calculations assuming a single
- domain state. There are only two stable ferroelectric phases. The
solid lines represent the boundaries separating the stability fields
of the paraelectric and ferroelectric phases, or the ferroelectric
orthorhombic and distorted rhombohedral phases. (b) Superposition of
the phase diagram of a PZT film under in-plane tensile strain of
0.005  from the phase-field approach (scattered symbols) and from
thermodynamic calculations assuming a single domain (solid lines).
There are three stable ferroelectric phases:  tetragonal-''square,''
orthorhombic-''circle,'' and distorted rhombohedral-''triangle''
according to the phase-field simulations. The scattered symbols
represent the ferroelectric domain state obtained at the end of a
phase-field simulation. The shaded portion surrounded by the
scattered symbols label the stability regions of a single
ferroelectric phase, and the nonshaded region shows a mixture of two
or three ferroelectric phases. From
\protect\onlinecite{LiChoudhury2003}.} \label{fig:LiChoudhury2003}
\end{figure}

While the phase diagrams thus derived are quite rich, even the
optimal single-crystal monodomain structure may be unfavorable with
respect to formation of polydomain
\cite{Speck1994,Bratkovsky2001,Roytburd2001} or multiphase
structures, which allow strain relaxation on average and reduce
elastic energy. The evaluation of the energies of polydomain
structures requires taking both strain and depolarization fields
into account. A recent discussion of PbTiO$_3$ using a phase-field
analysis \cite{LiChoudhury2003} suggested that including the
possibility of polydomain structure formation significantly affects
the phase diagram for experimentally relevant misfit strains and
temperatures, as shown in Fig. \ref{fig:LiChoudhury2003}.
For very thin films, additional effects associated with the
interface between the film and substrate are also expected to
contribute significantly. For example, although polydomain formation
can accommodate misfit strain averaged over different variants, each
domain will be mismatched to the substrate at the atomic level, with
a corresponding increase in interface energy. Furthermore, the
energy of a domain wall perpendicular to the substrate will be
higher than the energy of the corresponding wall in the bulk, due to
the geometrical constraint on the allowed shifts of the atomic
planes across the domain wall (as found for the bulk in
\onlinecite{Meyer2002}) imposed by the planar interface. Different
domain walls will in general be affected differently by the
constraint, possibly changing the relative energy of different
polydomain configurations.

With recent advances in thin film synthesis, it is possible to grow
and characterize high-quality films that are sufficiently thin to be
coherent or partially relaxed. Here, we give a few examples of
experimental observations of changes in structure and polarization
in very thin films. T$_c$ elevation in strained PbTiO$_3$ films has
been reported and analyzed in \textcite{Rossetti91} and
\textcite{Streiffer2002}. The strain-induced r-phase has been
observed in PZT films thinner than 150 nm on Ir-electroded Si wafers
\cite{Kelman2002}. An antiferroelectric to ferroelectric transition
has been observed in thin films of PbZrO$_3$ \cite{UPZT}, though
whether it is induced by strain or some other thin-film related
effect is considered an open question. Large polarization
enhancements have been observed in epitaxially strained BaTiO$_3$
films \cite{Choi2004}. Most dramatically, room-temperature
ferroelectricity has been achieved for SrTiO$_3$ under biaxial
tensile strain induced by a DyScO$_3$ substrate \cite{Haeni2004}.
For thicker films, observation of polydomain structures is reported
in \onlinecite{Roytburd2001}. It has been observed that domain
formation may be suppressed by rapid cooling through the transition
\cite{Ramesh1993}.

\begin{figure}
\centerline{\epsfig{file=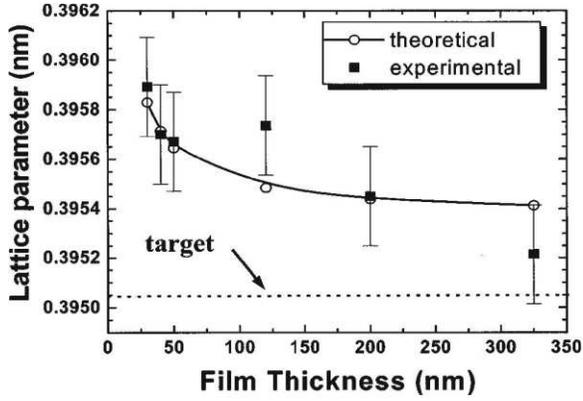,width=3.00in,angle=0}}
\caption{Evolution of a$^\perp$ as a function of film thickness for
Ba$_{0.6}$Sr$_{0.4}$TiO$_3$  thin films grown on
0.29(LaAlO$_3$):0.35(Sr$_2$TaAlO$_6$) substrates. .Also shown is the
theoretical curve, given by the open circles. The straight dashed
line represents the lattice parameter of the ceramic target (a =
0.39505 nm). From  \protect\onlinecite{Canedy2000}.}
\label{fig:Canedy2000}
\end{figure}

\begin{figure}
\centerline{\epsfig{file=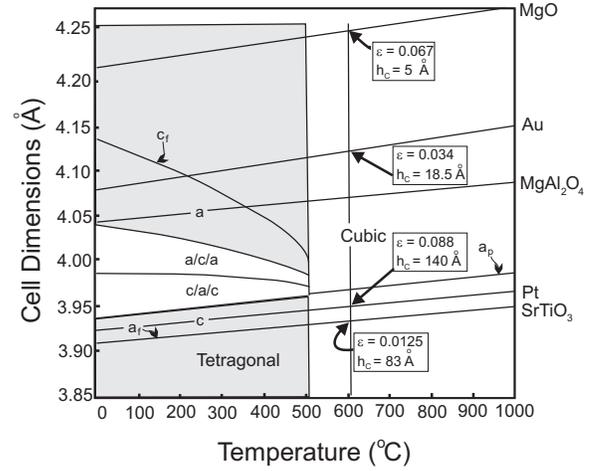,width=3.00in,angle=0}}
\caption{ Coherent temperature dependent domain stability map for
PbTiO$_3$ including the cubic lattice parameter for several common
single crystal oxide substrates. The misfit strains for epitaxial
growth of PbTiO$_3$ at 600$^o$ C are included in the insets along
with the critical thickness h for misfit dislocation formation. From
\protect\onlinecite{Speck1994}.} \label{fig:Speck1994}
\end{figure}

For thicker films grown at high temperature, misfit dislocations
will form at the growth temperature partially or completely to relax
misfit strain. The degree of relaxation increases with increasing
thickness, until, for thick enough films, the epitaxial strain is
negligible. This behavior has been studied theoretically
\cite{Matthews1974} and observed experimentally, for example as in
Fig. \ref{fig:Canedy2000}.
Additional strain can arise during cooling from the growth
temperature if there is differential thermal expansion between the
film and the substrate and the formation of misfit dislocations is
kinetically inhibited. A detailed theoretical study of strain
relaxation in epitaxial ferroelectric films, with discussion of the
interplay of misfit dislocations, mixed domain formation and
depolarizing energy, was undertaken by Speck and Pompe
\cite{Speck1994}. It was assumed that for rapid cooling from the
growth temperature, the effect of misfit dislocations can be
incorporated by using an effective substrate lattice parameter,
while in the limit of slow cooling, the system optimally
accommodates misfit strain with dislocations. (This assumption is
valid for films with thickness of order 1 $\mu$, while the treatment
needs to be slightly modified for intermediate thicknesses where the
equilibrium concentration of misfit dislocations leads to only
partial strain relaxation). Elastic domains form to relax any
residual strain. Below T$_c$, depolarizing energy can change the
relative energetics of different arrangements of polarized domains
and misfit dislocations. It was suggested that the electrostatic
energy should be more of a factor for smaller tetragonality systems
(BaTiO$_3$ vs PbTiO$_3$) where the strain energy is less, though
this could at least be partially balanced by the fact that the
polarization is smaller as well. A typical coherent diagram is shown
in Fig. \ref{fig:Speck1994}.

With transmission electron microscopy (TEM) it is possible to make
detailed studies of the types and arrangements of misfit
dislocations in perovskite thin films. Recent studies of
high-quality films include  \onlinecite{Suzuki1999} and
\onlinecite{Sun2004}. While strain-relaxing defects, such as misfit
dislocations, reduce or eliminate the elastic energy associated with
homogeneous strain, these and other defects prevalent in films do
generate inhomogeneous strains. As mentioned at the beginning of
this section, the inhomogeneous strains couple strongly to the
polarization, and it has been shown by phenomenological analysis
\cite{Balzar2002,Balzar2004} that their effects on T$_c$ can be
significant. They have also been argued to contribute to the
degradation of the dielectric response in thin films relative to
bulk values \cite{Canedy2000}.

Strains and their coupling to polarization are also central to the
properties exhibited by short-period superlattices of
lattice-mismatched constituents. As the result of recent work on
artificial superlattices of ferroelectric materials, there are some
indications that improved ferroelectric properties and/or very large
dielectric constants can be achieved. The most studied system at
present is BaTiO$_{3}$/SrTiO$_{3}$
\cite{Tabata1994,Ishibashi2000,Nakagawara2000,Shimuta2002,Neaton2003,Jiang2003,Rios2003,Panpreprint}.
In BaTiO$_{3}$/SrTiO$_{3}$ superlattices lattice-matched to a
SrTiO$_{3}$ substrate, the compressive in-plane strain on the
BaTiO$_{3}$ layer substantially raises its polarization. Theoretical
studies suggest that the SrTiO$_{3}$ layer is polarized (and the
polarization in the BaTiO$_{3}$ layer is reduced) by electrostatic
energy considerations, which favor continuity of the component of
the polarization along the normal. Overall the polarization is
enhanced above that of bulk BaTiO$_{3}$, though not as high as that
of a pure coherent BaTiO$_{3}$ film if it were possible to suppress
the formation of strain-relaxing defects. While the natural lattice
constant of BaTiO$_{3}$/SrTiO$_{3}$ is intermediate between the two
endpoints, so that on a SrTiO$_{3}$ substrate the superlattice is
under compressive in-plane stress, it has been suggested that the
multilayer structure tends to inhibit the formation of misfit
dislocations so that a thicker layer of coherent superlattice
material can be grown. As the superlattice material thickness
increases, there will be strain relaxation via misfit dislocations
and the in-plane lattice constant should increase, putting the
SrTiO$_{3}$ layer under in-plane tensile strain. In this case the
SrTiO$_{3}$ layer is observed to have a component of polarization
along [110] \cite{Jiang2003,Rios2003}, consistent with theoretical
studies of epitaxially strained SrTiO$_{3}$ \cite{Pertsev2000,
Antons2004} and of the BaTiO$_{3}$/SrTiO$_{3}$ superlattice with
expanded in-plane lattice constant \cite{Johnstonpreprint}.

The real appeal of short-periodicity ferroelectric multilayers is
the potential to make ``new" artificially structured materials with
properties that could open the door to substantial improvements in
device performance or even radically new types of devices.
Perovskites are particularly promising, as individual materials
possess a wide variety of structural, magnetic, and electronic
properties, while their common structure allows matching at the
interface to grow superlattices. Beyond the prototypical example of
BaTiO$_{3}$/SrTiO$_{3}$ discussed in the previous paragraph, there
has been work on other combinations such as KNbO$_{3}$/KTaO$_{3}$
\cite{Christen1996,Sigman2002,Sepliarsky2001,Sepliarsky2002},
PbTiO$_{3}$/SrTiO$_{3}$ \cite{Jiang1999},
PbTiO$_{3}$/PbZrO$_{3}$\cite{Bungaro2002,Bungaro2004a},
La$_{0.6}$Sr$_{0.4}$MnO$_3$/La$_{0.6}$Sr$_{0.4}$FeO$_3$\cite{Izumi1999},
CaMnO$_3$/CaRuO$_3$\cite{Takahashi2001},
LaCrO$_3$-LaFeO$_3$\cite{Ueda1998,Ueda1999a},and
LaFeO$_3$-LaMnO$_3$\cite{Ueda1999b}. In nearly all cases, strain
plays an important role in understanding the aggregate properties of
these short-period multilayers and superlattices. In addition to
lattice mismatch, the layers also interact through the mismatch in
polarization along the layer normal, which leads to mutual
influences governed by considerations of electrostatic energy and
nonzero macroscopic electric fields. With three or more
constituents, it is possible to break inversion symmetry to obtain
superlattice materials with possibly favorable piezoelectric
properties. This idea was first proposed theoretically
\cite{Sai2000}, leading to experimental studies of
CaTiO$_3$/SrTiO$_3$/BaTiO$_3$ \cite{Eckstein2003} and
LaAlO$_3$/(La,Sr)MnO$_3$/SrTiO$_3$
\cite{Yamada2002,Ogawa2003,Kimoto2004}. Also, perovskite
superlattices combining ferroelectric and ferromagnetic layers offer
a path to the development of multiferroic materials. The
identification, synthesis and characterization of further
combinations remains the subject of active research interest.

\section{NANOSCALE FERROELECTRICS}

\subsection{Quantum confinement energies}

Confinement energies are a trendy topic in nanoscale semiconductor
microelectronics devices.\cite{Petroff01}  The basic idea is that in
a system in which the electron mean free path is long with respect
to the lateral dimension(s) of the device, a quantum-mechanical
increase in energy (and of the bandgap) in the semiconductor will
occur. In general confinement energies exist only in the ballistic
regime of conduction electrons, that is, where the electron mean
free path exceeds the dimensions of the crystal.  This usually
requires a high-mobility semiconductor at ultra-low temperatures.
Such effects are both interesting and important in conventional
semiconductors such as Si or Ge, GaAs and other III-Vs, and perhaps
in II-VIs.  However, despite the fact that the commonly used oxide
ferroelectrics are wide-bandgap p-type semiconductors (3.0 eV $<$
E$_{g}$ $<$ 4.5 eV),\cite{Waser96} neither their electron nor hole
mean free paths are sufficiently long for any confinement energies
to be measured. Typically the electron mean free path in an
ABO$_{3}$ ferroelectric perovskite is 0.1 to 1.0 nm,\cite{Dekker54}
depending on applied electric field E, whereas the device size d is
at least 20 nm. Therefore any confinement energy (which scales as
d$^{-2}$) might be a meV or two, virtually unmeasurable, despite a
few published claims \cite{Yu97},\cite{Kohiki00},\cite{Scott00qce}
reporting extraordinarily large effects. In the case of
Bi$_{2}$O$_{3}$ and SrBi$_{2}$Ta$_{2}$O$_{9}$ (SBT) these effects
may arise from two-phase regions at the sample
surfaces.\cite{Zhou92},\cite{Switzer99} This is theoretically
interesting and very important from an engineering device point of
view; if it were not true the contact potential at the electrode
interface in a 1T-1C device, or at the ferroelectric-Si interface in
a ferroelectric-gate FET, would depend critically on the cell size,
which would add a very undesirable complication to device design.

\subsection{Coercive fields in nanodevices}

\begin{figure}[h]
 \includegraphics[width=7cm]{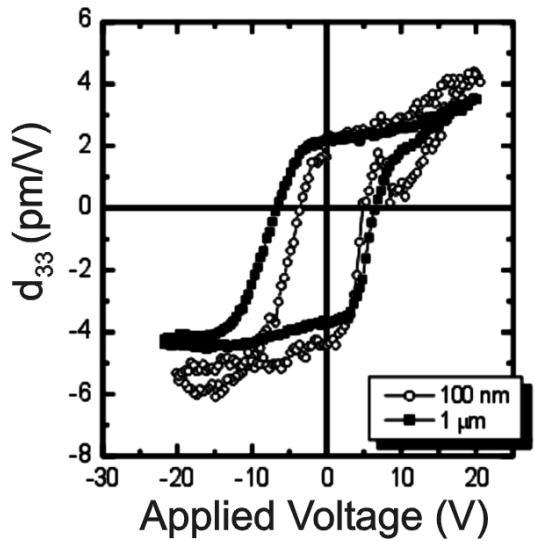}
 \caption{\textit{Lack of significant dependence of coercive field on lateral area in nanoscale ferroelectrics \textcite{Alexe99}}\label{fig:coercivearea}}
\end{figure}

One of the most pleasant surprises in the research on small-area
ferroelectrics is the observation, shown in Fig.
\ref{fig:coercivearea} , that the coercive field is independent of
lateral area.\cite{Alexe99} Coercive fields in nanophase
ferroelectric cells have generally been measured via atomic force
microscopy (AFM).\cite{Gruverman96} Domain structures, polarization
and coercive fields of nanoscale particles of BaTiO$_3$ have been
studied theoretically using interatomic potentials
\cite{Stachiotti2004} and a first-principles effective Hamiltonian
\cite{Fu2003}

\subsection{Self Patterned nanoscale ferroelectrics}

\begin{figure}[h]
\epsfig{figure=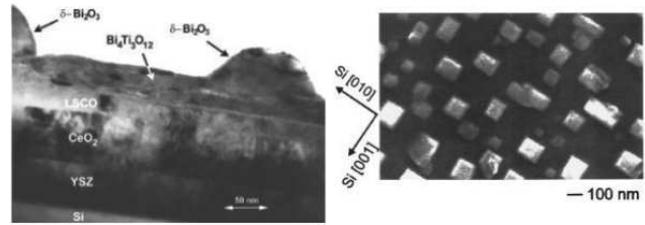,width=8.5cm}\caption{\textit{Sample of
Alexe et al\cite{Alexe98}, (a) TEM cross-section showing underlying
layers and bismuth oxide nanoeelectrodes, (b) Semi-registered array
of nanoelectrodes taking their orientation from the underlying Si
substrate}}\label{fig:alexesp}
\end{figure}

One approach to producing nanoscale ferroelectrics is to attempt to
produce self patterned arrays of nanocrystals, in which ordering is
produced by interactions between islands through the substrate. This
approach could be used to produce arrays of metallic nanoelectrodes
on top of a ferroelectric film or alternatively arrays of crystals
from the ferroelectric materials themselves. The first scheme was
suggested by Alexe et al. \cite{Alexe98} who found that a bismuth
oxide wetting layer on top of a bismuth titanate film formed an
array of metallic bismuth oxide nanocrystals on top of the film,
which were partially registered along the crystallographic
directions of the underlying substrate (Fig. \ref{fig:alexesp}).
These nanocrystals were used successfully as electrodes to switch
regions of the film\cite{Alexe99}. In the second approach one might
use a material such as PbTiO$_{3}$ on a SrTiO$_{3}$ substrate, which
was first demonstrated to form islands when grown epitaxially at
very thin film thicknesses by Seifert et al. \cite{Seifert96} In the
context of self patterning of oxide materials a recent work by Vasco
et al studies the growth of self organised SrRuO$_{3}$ crystals on
LaAlO$_{3}$ \cite{Vasco03}.

When small amounts of materials are deposited on substrates where
there is some degree of mismatch between the two materials, islands
form and the repulsive interactions between them are mediated via
strain fields in the substrate as first suggested by
Andreev.\cite{Andreev81} This idea has been developed into a
detailed theory by Shchukin and Bimberg;\cite{Shchukin99} however
this theory is a zero-temperature theory, whereas a thermodynamic
theory is required to describe the crystallization processes which
occur at quite high temperatures. An extension of the theory to
finite temperatures has been carried out by Williams and
co-workers.\cite{Williams00}$^{,}$\cite{Rudd03} The chief result of
this theory are the prediction of three different kinds of
structures (pyramids, domes and superdomes), a volume distribution
for a particular species of structure, and a shape map to describe
relative populations of structures as a function of coverage and
crystallization temperature. One interesting result from experiment
is that similar shaped structures are observed in both the
Volmer-Weber (VW) and Stranski-Krastanow (SK) growth modes, but on
different size scales. In the work of Williams the thickness above
which dome populations occur is of the order of 4-5 monolayers,
corresponding to the critical thickness for misfit dislocations for
Ge on Si(100). On the other hand Capellini et al. \cite{Capellini97}
studied via atomic force microscopy the growth of Ge on Si(100) in
the SK growth mode and found a much larger critical structure height
of 50 nm at which dislocations were introduced and the structures
changed from being pyramidal in geometry to domelike. The large
increase in critical thickness is due to a substantial part of the
misfit strain being taken up by the substrate in the SK growth mode,
as described by Eaglesham and Cerrulo \cite{Eaglesham90}. The
description of self patterned ferroelectric nanocrystals by the
models of Schukin and Williams has recently been undertaken by
\textcite{DawberSP}.

\begin{figure}[h]
 \includegraphics[width=6cm]{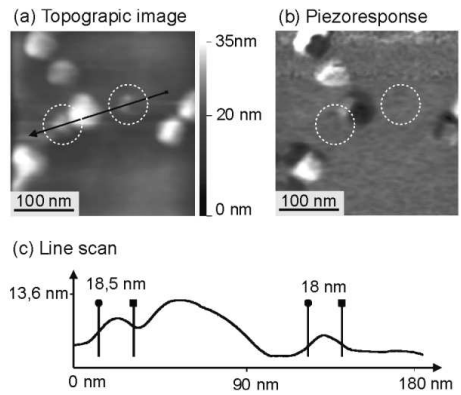}
 \caption{\textit{\textcite{Roelofs03}(a) topographic image of grains from 100 nm down to 20 nm in lateral size (b) piezoresponse image of same grains showing the absence of peizoresponse for the grains below 20nm}}\label{fig:roelofsfig}
\end{figure}

Prior to this two groups have grown PbTiO$_3$ nanocrystals on
Pt/Si(111) substrates to measure size effects in ferroelectricity.
\cite{Roelofs03}$^{,}$\cite{Shimizu03} These works both show a lack
of piezoresponse in structures below ~20nm in lateral size (Fig.
\ref{fig:roelofsfig}), though we expect that this is connected to
mechanical constraints rather than any fundamental limiting size for
ferroelectric systems. \textcite{Chu04}  have highlighted the role
that misfit dislocations can play in hampering ferroelectricity in
small structures. Interestingly in the work of \textcite{Roelofs03}
and \textcite{Shimizu03} because of the (111) orientation of their
substrates, instead of square-based pyramids they obtain triangular
based structures that display hexagonal rather than cubic
registration (an analogous result is observed when Ge is grown on
Si(111). \cite{Capellini99}). The growth and analysis of PZT
nanocrystals on SrTiO$_{3}$ has been carried out by
\textcite{Szafraniak03}). A review on size effects in ferroelectric
nanocrystals is currently in preparation by \textcite{Ruedigerrev}.

Although there is potential to produce self patterned arrays with
greater registration by better choice of materials and processing
conditions our general conclusion is that highly registered memory
arrays will not occur spontaneously in the absence of a
pre-patterned field.

\subsection{Non-planar geometries:Ferroelectric nanotubes}

Almost all recent work on ferroelectric oxide films have involved
planar geometries.  However, from both a device engineering point of
view and from theoretical considerations, it is now appropriate to
analyze carefully non-planar geometries, especially nanotubes.

\begin{figure}[h]
 \includegraphics[width=8.5cm]{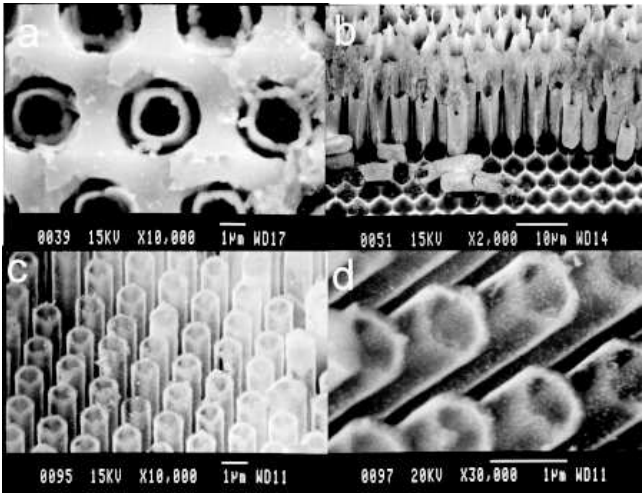}
 \caption{\textit{SEM micrograph indicating a plan view of a regular array of SBT tubes in host silicon substrate with diameter ca. 2 $\mu m$ and wall thickness ca. 200 nm (a).   SBT tubes in cross sectional view indicating coating to bottom of pore (b).   Micrograph of free-standing array of tubes with diameter ca. 800 nm (c) and wall thickness $<$ 100 nm (d).}\label{fig:ferronanotubes}}
\end{figure}

Nanotubes made of oxide insulators have a variety of applications
for pyroelectric detectors, piezoelectric ink-jet printers, and
memory capacitors that cannot be filled by other nanotubes
\cite{Herzog85},\cite{Sakamaki01},\cite{Sajeev02},\cite{Gnade00},\cite{Averdung01}.
In the drive for increased storage density in FRAM and DRAM devices,
complicated stacking geometries, 3D structures and trenches with
high aspect ratios are also being investigated to increase the
dielectric surface area. The integration of ferroelectric nanotubes
into Si substrates is particularly important in construction of 3D
memory devices beyond the present stacking and trenching designs,
which according to the international ULSI schedule
\footnote{International Technology Roadmap for Semiconductors (ITRS)
2002 (available at
http://public.itrs.net/Files/2002Update/Home.pdf)} must be achieved
by 2008. Template synthesis of nanotubes and wires is a versatile
and inexpensive technique for producing nanostructures. The size,
shape and structural properties of the assembly are simply
controlled by the template used.  Using carbon nanotubes as
templates, tubular forms of a number of oxides including
V$_{2}$O$_{5}$, SiO$_{2}$, Al$_{2}$O$_{3}$ and ZrO$_{2}$ have been
generated \cite{Patzke02}.   Much larger ($>$20-micron diameter)
ferroelectric micro-tubes have been made by sputter deposition
around polyester fibres \cite{Fox95},\cite{Pokropivny01} - Fox has
made them from ZnO and PZT, with 23 $\mu$m inside diameter, about
1000x larger than the smallest nano-tubes reported in the present
paper. Porous sacrificial templates as opposed to fibres have also
been used. Porous anodic alumina has a polycrystalline structure
with ordered domains of diameter 1-3 $\mu$m, containing
self-organised 2D hexagonal tubular pore arrays with an interpore
distance of 50-420 nm \cite{Li98}. This nano-channel material can
therefore be used as a template for individual nanotubes but is not
suitable for making an ordered array of tubes over length scales
greater than a few mm.   Many oxide nanotubes, such as TiO$_{2}$,
In$_{2}$O$_3$, Ga$_{2}$O$_{3}$, BaTiO$_{3}$ and PbTiO$_{3}$, as well
as nanorods of MnO$_{2}$, Co$_{3}$O$_{4}$ and TiO$_{2}$, have been
made using porous alumina membranes as templates \cite{Patzke02}].
\textcite{Hernandez02}, used a sol-gel template synthesis route to
prepare BaTiO$_{3}$ and PbTiO$_{3}$ nanotube bundles by dipping
alumina membranes with 200 nm pores into the appropriate sol. The
BaTiO$_{3}$ and PbTiO$_{3}$ nanotubes were shown to be cubic
(paraelectric) and tetragonal (ferroelectric) by x-ray diffraction,
although Raman studies indicated some non-centrosymmetric phase on a
local scale in the BaTiO$_{3}$. Porous silicon materials are also
available as suitable templates. \textcite{Mishina02} used a sol-gel
dipping technique to fill nanoporous silicon with a
PbZr$_{1-x}$Ti$_{x}$O$_{3}$ (PZT) sol producing nanograins and
nanorods 10-20 nm in diameter. The presence of the ferroelectric PZT
phase was shown by second harmonic generation (SHG) measurements. In
this instance the porous silicon does not have a periodic array of
pores \cite{Smith92} and as in the case for those produced by
Hernandez et al, we emphasise that those nanotubes are not ordered
arrays, but instead spaghetti-like tangles of nano-tubes that cannot
be used for the Si device embodiments. A second type of porous Si
templates, however, consist of a very regular periodic array of
pores with very high aspect ratios.  By a combination of
photolithography and electrochemical etching hexagonal or orthogonal
arrays of pores with diameters 400 nm to a few mm and up to 100
$\mu$m deep can be formed in single crystal Si wafers
\cite{Schilling01},\cite{Ottow96}. These crystals were originally
developed for application as 2D photonic crystals, but also find
applications as substrates for templated growth and integration of
oxides nanostructures with Si technology. \textcite{Luo03} recently
used such crystals to produce individual, free standing PZT and
BaTiO$_{3}$ ferroelectric nanotubes by a polymeric wetting
technique. \textcite{Morrison03} described the use of liquid source
misted chemical deposition (LSMCD) to fill such photonic Si crystals
with SBT precursor.  During deposition, the SBT precursor was shown
to coat the inside of the pores. After etching of the photonic
crystal with pore diameter 2 $\mu$m for 30 seconds with aq.
HF/HNO$_{3}$ the interface between the Si substrate and SBT coating
is dissolved, exposing the uniform SBT tube, Fig.
\ref{fig:ferronanotubes}a. The tube walls are very uniform with a
thickness of ca. 200 nm. The same sample is shown in cross-sectional
view after complete removal of the host Si walls between pores, Fig.
\ref{fig:ferronanotubes}b. The result is a regular array of tubes
attached to the host Si matrix only at the tube base. Although these
tubes have suffered damage during handling, it is clear that the
pores have been filled uniformly to the bottom, a depth of ca 100
$\mu$m.

The second photonic crystal with pore diameter 800 nm underwent
fewer depositions and after etching revealed a regular array of
uniform tubes of diameter 800 nm, Fig. \ref{fig:ferronanotubes}c.
The wall thickness is uniform and $<$ 100 nm, Fig.
\ref{fig:ferronanotubes}d.  The tubes are ca. 100 $\mu m$ long,
completely discrete and are still attached to the host Si matrix,
creating a perfectly registered hexagonal array. Free-standing tubes
may be produced by completely dissolving the host Si matrix. As yet,
no one has applied cylindrical electrodes to the tubes; however,
\textcite{Steinhart03} recently used porous anodic alumina templates
to grow palladium nanotubes. Using a similar method it may be
possible to alternately deposit Pd or Pt and SBT to produce a
concentric electrode/FE/elctrode structure in each nanotube.  The
use of the photonic crystal template with a regular array of pores
has significant benefits over other porous substrates in that the
coatings/tubes produced are also in a registered array ordered over
several mm's or even cm's. This facilitates addressing of such an
array for device applications. DRAMs utilise high surface area
dielectrics, and high aspect ratio SBT coatings such as these
embedded in Si could increase storage density.  Current
state-of-the-art deep trenched capacitors are 0.1 mm diameter by 6
mm deep, aspect ratio 60:1. Using SBT (or other FE oxide) nanotubes
of wall thickness $<$100 nm, a trench (or array of trenches) of 0.1
$\mu$m diameter and 100 $\mu$m deep, an aspect ratio of $>$ 1000:1
is possible.  Applying and addressing electrodes to an array of FE
nanotubes could generate 3D FRAM structures offering high storage
density with improved read/write characteristics compared to
conventional planar stacks. On removal of the Si walls, the
piezoelectric response (expansion/contraction under an applied
field) of such an array of nanotubes could be utilized for a number
of MEMs applications. These could include: (1) ink-jet printing -
delivery of sub-picolitre droplets for lithography free printing of
submicron circuits; (2) biomedical applications - nanosyringes,
inert drug delivery implants; (3) micropositioners or movement
sensors.

Almost no theoretical work has been published on the physics of
ferroelectric nanotubes. Analytical solutions for the effects on the
$d_{ij}$ piezoelectric coefficients of hollow tubes have been given
for both the case in which polarization P is along the length z
\cite{Ebenezer03} and for P radial \cite{Ebenezer02}, they did not
however solve the azimuthal case where polarization goes around the
tube. It it this latter case which been measured as hysteresis by
\textcite{Luo03} with a tube lying on a bottom electrode with a
semicircular sputtered top electrode. Important matters such as the
dependence of T$_{C}$ upon tube diameter have also not been
examined.

\section{Conclusions}
\label{sec:conclusions}

In this review we have sought to cover the important advances in
recent years in the physics of thin film ferroelectric oxides. At
the present point of time ferroelectric thin films memory devices
have reached a point of maturity where they are beginning to appear
in real commercial devices. At the same time new directions such as
the drive to faster, smaller, nanoscale devices and non-planar
geometries are evolving and new levels of physical understanding
will be required. Over the next years it is expected that first
principles computational approaches will continue to develop,
suggesting a new synergy between the computational modeling and
experimental realizations of ferroelectric systems with new and
exciting properties.

\end{document}